\documentclass[lettersize,journal]{IEEEtran}
\usepackage{amsmath,amsfonts}
\usepackage{algorithm}
\usepackage{array}
\usepackage{textcomp}
\usepackage{stfloats}
\usepackage{url}
\usepackage{verbatim}
\usepackage{graphicx}
\usepackage{cite}
\usepackage{siunitx}
\usepackage{amssymb}
\usepackage{booktabs}
\usepackage{colortbl}
\usepackage{color}
\usepackage{changepage}
\usepackage{algcompatible}
\usepackage{algpseudocode}
\usepackage{tikz}
\usetikzlibrary{positioning}
\usepackage{subfigure}
\usepackage[section]{placeins}
\usepackage[T1]{fontenc}
\usepackage{multirow}
\usepackage[pagebackref,breaklinks,colorlinks]{hyperref}
\usepackage[pages=some,placement=bottom]{background}

\usepackage[capitalize]{cleveref}
\crefname{section}{Sec.}{Secs.}
\Crefname{section}{Section}{Sections}
\Crefname{table}{Table}{Tables}
\crefname{table}{Tab.}{Tabs.}

\newcommand\copyrighttext{%
  \footnotesize \textcopyright 2022 IEEE. Personal use of this material is permitted. Permission from IEEE must be obtained for all other uses, in any current or future media, including reprinting/republishing this material for advertising or promotional purposes, creating new collective works, for resale or redistribution to servers or lists, or reuse of any copyrighted component of this work in other works.
  DOI: \href{https://doi.org/10.1109/TCI.2022.3233188}{10.1109/TCI.2022.3233188}}
\newcommand\copyrightnotice{%
\backgroundsetup{contents={\fbox{\parbox{\dimexpr\textwidth-2\fboxsep-2\fboxrule\relax}{\copyrighttext}}},opacity=1,color=black,scale=1,vshift=4pt}
\BgThispage
}

\newif\ifIncludeSupplementary%
\IncludeSupplementarytrue%

\newif\ifArxiv%
\Arxivtrue

\newcommand{\argmin}{\mathop{\text{argmin}}}

\definecolor{hlbest}{rgb}{0.9,0.9,0.9}
\newcommand{\hlbest}[1]{\hspace*{-1pt}\colorbox{hlbest}{\hspace{1pt}\strut{}#1\hspace{1pt}}\hspace*{-1pt}}

\newcommand{\draftgraphics}{false}

\newcommand{\email}[1]{\nolinkurl{#1}}

\begin{document}

\title{An Educated Warm Start For Deep Image Prior-Based Micro CT Reconstruction
}

\author{Riccardo~Barbano$^{\dagger}$, Johannes~Leuschner$^{\dagger}$, Maximilian~Schmidt, Alexander~Denker, Andreas~Hauptmann, Peter~Maass and Bangti~Jin%
\thanks{R.~Barbano is with the Department of Computer Science, University College London, UK (e-mail: \email{riccardo.barbano.19@ucl.ac.uk}).}%
\thanks{J.~Leuschner, M.~Schmidt, A.~Denker and P.~Maass are with the Center for Industrial Mathematics, University of Bremen, Germany (e-mail: \email{{jleuschn,maximilian.schmidt,adenker,pmaass}@uni-bremen.de}).}%
\thanks{A.~Hauptmann is with the Research Unit of Mathematical Sciences, University of Oulu, Finland, and also with the Department of Computer Science, University College London, UK (e-mail: \email{andreas.hauptmann@oulu.fi}).} \thanks{B. Jin is with Department of Mathematics, The Chinese University of Hong Kong, Shatin, N.T., Hong Kong (e-mail: \email{btjin@math.cuhk.edu.hk,bangti.jin@gmail.com})\\ $\dagger$ equal contributors.} }%

\ifArxiv%
\else%
\markboth{Journal of \LaTeX\ Class Files,~Vol.~14, No.~8, August~2021}%
{Shell \MakeLowercase{\textit{et al.}}: A Sample Article Using IEEEtran.cls for IEEE Journals}
\fi

\maketitle%
\copyrightnotice%

\begin{abstract}
Deep image prior (DIP) was recently introduced as an effective unsupervised approach for image restoration tasks.
DIP represents the image to be recovered as the output of a deep convolutional neural network, and learns the network's parameters such that the model output matches the corrupted observation.
Despite its impressive reconstructive properties, the approach is slow when compared to supervisedly learned, or traditional reconstruction techniques.
To address the computational challenge, we bestow DIP with a two-stage learning paradigm: (i) perform a supervised pretraining of the network on a simulated dataset; (ii) fine-tune the network's parameters to adapt to the target reconstruction task.
We provide a thorough empirical analysis to shed insights into the impacts of pretraining in the context of image reconstruction.
We showcase that pretraining considerably speeds up and stabilizes the subsequent reconstruction task from real-measured 2D and 3D micro computed tomography data of biological specimens.
The code and additional experimental materials are available at \href{https://educateddip.github.io/docs.educated_deep_image_prior/}{educateddip.github.io/docs.educated\_deep\_image\_prior/}.
\end{abstract}

\section{Introduction}\label{sec:intro}

Inverse problems in imaging center around recovering an unknown image $x\in\mathbb{R}^n$ of interest from the noisy measurement $y_\delta = Ax + \eta$, where $y_{\delta} \in \mathbb{R}^{m}$ is
the noisy measurement data, $A$ the linear forward operator, and $\eta$ an i.i.d.\ noise (e.g.\ Gaussian noise $\eta \sim \mathcal{N}(0, \sigma^{2}I)$). Due to the inherent ill-posedness of the problem, suitable regularization is crucial and is key for a successful recovery of $x$ \cite{EnglHankeNeubauer:1996,scherzer2009variational,ItoJin:2015}.

Over the last years, deep learning methods have been successfully applied to solve all types of imaging problems, with supervised training being the dominant paradigm \cite{ongie2020deep, arridge2019solving_ip_data_driven}.
That means, a deep neural network is trained to restore the image from noisy data using a set of paired training data. A large number of such high-quality paired training data may be needed \cite{baguer2020diptv}. Except simulated data, these are usually not obtainable, or too expensive to collect. Further challenges arise from the distributional shifts of the test data (e.g.\ change of image class, noise level or forward operator at test time). Ideally, the trained model should be robust to these changes, and transfer its reconstructive properties from one domain to another using as little additional data as possible \cite{gilton2021model,barbano2021unsupervised,lunz2021learned}. Unfortunately, this is often not the case.

An effective solution to these challenges is deep image prior (DIP) \cite{Ulyanov:2018}, which represents a new approach to regularize image restoration.
Rather than taking the supervised route, DIP learns to reconstruct without reference data, by assuming that a natural image can be well represented by a convolutional neural network (CNN).
This is achieved by training the network's parameters to generate an image that fits the data $y_\delta$ (often equipped with suitable early stopping).
The method is very attractive for imaging tasks with scarce training data. DIP
has received enormous attention in the imaging community, and delivered state-of-the-art performance for unsupervised methods on a number of imaging tasks, including computed tomography (CT) \cite{baguer2020diptv,KnoppGrosser:2022}, magnetic resonance imaging (MRI) \cite{DarestaniHeckel:2021}, positron emission tomography (PET) \cite{GongLi:2019,cui2021populational,cui2019pet} and compressive ptychography \cite{BarutcuGursoyKatsaggelos:2022}, closely matching its supervised counterparts.

While DIP has been shown to be effective, it is not free from drawbacks.
Notably, it requires ``fresh training'' each time it is deployed, which
leads to high computational overhead and demanding VRAM
requirements at test time when compared to supervised counterparts \cite{Monga:2021,ongie2020deep, hauptmann2018model}; the latter ones only require one feed-forward pass through the network and thus computationally cheap.
This inefficiency is considerably exacerbated by the fact that DIP requires a lengthy (and unstable) optimization process \cite{leuschner2021quantitative, baguer2020diptv}.
For example, reconstructing a single image of resolution $(\SI{501}{px})^2$ requires approximately \num{30}-\num{50}k iterations to reach the early-stopping point, which translates to \num{3}-\num{5}\,\si{h} of computing-time on NVIDIA GeForce RTX 2080Ti/1080Ti.
It gets even worse in the 3D setting: a $(\SI{167}{px})^3$ reconstruction takes approximately one day on NVIDIA GeForce RTX 3090 using mixed precision!
This hinders its applicability to solve imaging inverse problems, especially when fast reconstruction is critical.
These observations motivate us to explore the following:

\begin{adjustwidth}{9pt}{9pt}
\vspace{0.2em}
\textit{
Can DIP benefit from pretraining for accelerating subsequent reconstructive tasks? If so, can we easily construct an informative dataset to warm-start DIP?
How do inductive biases of pretraining impact the reconstructive task?}
\end{adjustwidth}

Pretraining is one well-established paradigm to address data scarcity in supervised learning \cite{azizi2021big,howard2018universal}. Models are often pretrained using large-scale datasets, and fine-tuned on target tasks that have less training data \cite{DonahueJia:2014}.
However, the idea of pretraining has not received the attention it deserves for DIP, and presents a new challenge.
The challenge is to learn (via supervised pretraining) feature representations that are transferable and generalizable to subsequent fine-tuning.

To overcome the computational challenge, we systematically explore a supervised pretraining strategy for accelerating DIP-based $\mu$CT reconstruction, and introduce an effective two-stage learning paradigm.
Our contributions can be summarized as follows.
We develop an effective strategy to greatly {accelerate} the convergence of DIP for $\mu$CT reconstruction, by recasting DIP within the ``{supervised pretraining + unsupervised fine-tuning}'' paradigm.
We show that carefully designed {pretraining} with {simulated data} from a synthetic image class can considerably speed up and stabilize DIP-based $\mu$CT reconstruction with real-measured data, including computationally demanding 3D tasks, for which we develop a specialized U-Net architecture to perform DIP-based $\mu$CT reconstruction under the constraint of \num{24}\,GB VRAM. To the best of our knowledge, this is the first successful 3D $\mu$CT reconstruction using DIP.
The experiment results show that despite its simplicity, it can be highly effective. Further, we conduct a thorough {experimental study} to shed {insights} into the mechanism of knowledge {transfer} between the supervised pretraining and unsupervised fine-tuning stages, including a novel {linear analysis} of pretraining, which exhibits sparsity-promoting in the parameters' bases.

The paper is organized as follows. We describe the standard DIP in Section \ref{sec:background} and related works in Section \ref{sec:relatedwork}. In Section \ref{sec:method}, we present the two-stage framework for DIP. We give the experimental details and results in Sections \ref{sec:data} and \ref{sec:experiment}, and analyze the impact of pretraining in Section \ref{sec:analysis}.

\section{Deep Image Prior}\label{sec:background}

The idea of DIP \cite{Ulyanov:2018} is to find a minimizer of the fidelity $\|Ax - y_\delta\|^2$, by representing the unknown $x$ as the output of a CNN, $x = \varphi_\theta(z)$, where
$z\in\mathbb{R}^n$ is a fixed random vector (often pixel-wise i.i.d.\ samples of random noise), and $\theta\in\mathbb{R}^p$ denotes the network's
parameters to be learned. A U-Net \cite{ronneberger2015u} like architecture is commonly used for the network. DIP solves
\begin{equation*}
\theta^*\in\argmin_\theta \|A\varphi_\theta(z) - y_\delta\|^2,
\end{equation*}
and presents $\varphi_{\theta^*}(z)$ as the reconstruction.
Note that the training of the network parameters $\theta$ coincides with the recovery process, and has to be repeated for each measurement. The procedure is unsupervised, and
guided by the principle of matching the forward projected network output $A\varphi_\theta(z)$ to the measurement data $y_\delta$.
Due to the overparameterization of the neural networks used in DIP,
a direct minimization of the loss can suffer from overfitting. DIP often uses early-stopping to deliver a satisfactory
reconstruction: the update of $\theta$ is stopped early to avoid overfitting to the noise \cite{Ulyanov:2018}. This has motivated
developing automated rules for early stopping \cite{JoChunChoi:2021,WangSun:2021}.

\section{Related Works}\label{sec:relatedwork}

\paragraph{Deep Image Prior}

Since the first proposal in \cite{Ulyanov:2018}, there have been several important developments on DIP.
Heckel et al.\ \cite{HeckelHand:2019} propose deep decoder, using under-parameterized networks to
ease the need for early-stopping. Dittmer et al.\ \cite{DittmerKluthMaass:2020} study DIP
through the lens of regularization theory \cite{EnglHankeNeubauer:1996,ItoJin:2015,scherzer2009variational},
and Cheng et al.\ \cite{Cheng_2019_CVPR} discuss its connection with Gaussian processes as the number of
architecture channels grows to infinity, and propose the use of Bayesian learning. There are several efforts
to combine DIP with explicit regularization to improve the reconstruction quality. \cite{LiuSunXuKamilov:2019,baguer2020diptv}
propose the use of total variation penalty for stabilizing the learning process,
and \cite{Mataev_2019_ICCV} combines DIP with regularization by denoising. Besides, the use of explicit regularization significantly relaxes the need of early
stopping. Jo et al.\ \cite{JoChunChoi:2021} propose to penalize the
complexity of the reconstruction using Stein's unbiased risk estimator. 
See also
\cite{WangSun:2021} for a stopping criterion based on monitoring the running variance of iterate sequence and references therein for further discussions.
\cite{baguer2020diptv} suggests at test-time to start optimizing a randomly initialized DIP to match a reconstruction, produced by another method.
Heckel and Soltanolkotabi \cite{HeckelSoltanolkotabi:2020} prove that for compressed sensing, an untrained CNN can approximately reconstruct signals and images that are sufficiently structured, from a near minimal number of random measurements.
The very recent work \cite{Arndt:2022} establishes the equivalence of "analytic" DIP with the
standard Tikhonov regularization, and several basic properties in the lens of classical
regularization theory.
This work complements and expands on these existing studies by
addressing the computational challenge associated with regularized DIP, especially the works
\cite{baguer2020diptv,LiuSunXuKamilov:2019}, where the regularized DIP was proposed and
empirically demonstrated.

\paragraph{Advances in Pretraining}

Supervised pretraining on ImageNet has been established as a common practice in computer vision. Neural networks are pretrained to solve image classification, and transferred to downstream tasks (e.g.\ object detection \cite{RedmonDivvala:2016,RenHe:2015} and semantic segmentation \cite{LongShelhamerDarrell:2015}). However, pretraining on ImageNet does not necessarily improve the accuracy of the downstream task \cite{HeGirshick:2019}, and similar observations about pretraining on ImageNet are made about medical image classification \cite{RaghuZhangBengio:2019}.
Within tomographic imaging, several works \cite{HanYe:2018,DarCukur:2020} employ transfer learning to adapt a trained neural network from one task setting to another. Our work shares similarities with these works in adapting to changes of the image distribution.
These works focus on supervised end-to-end fine-tuning, whereas we focus on an unsupervised learning framework: we study pretraining with a synthetic dataset as a means for accelerating DIP reconstruction on measured $\mu$CT data, and provide a detailed analysis of its acceleration mechanism. Very recently, Gilton et al \cite{gilton2021model} proposes to fine-tune the pretrained model so as to accommodate model errors, but unlike this work, the image distribution is unchanged. Inspired by an early version of this paper, Knopp and Grosser \cite{KnoppGrosser:2022} also demonstrated the potential of warm-starting DIP for dynamic tomography.

\section{Proposed Method}\label{sec:method}
The TV-regularized DIP approach obtains $x^{\ast}$ by
\begin{gather}
    \begin{aligned}
    \theta^{\ast}_\mathrm{t} \in\argmin_\theta \Bigl\{l_{{\rm t}}(\theta)\!:=\Vert A\,\varphi_{\theta}(z) - y_\delta\Vert^2+
    \gamma\,\text{TV}\left(\varphi_{\theta}(z)\right)&\Bigr\},
    \end{aligned}\nonumber\\
     x^* = \varphi_{\theta^{\ast}_{\rm t}}(z),
    \label{eqn:diptv}
\end{gather}
where $\varphi_\theta$ is a CNN, and $\gamma\geq 0$ balances the data consistency with the regularization term ${\rm TV}(\varphi_\theta(z))$, which denotes the total variation seminorm on the network output $\varphi_\theta(z)$, defined as $ \text{TV}(x) = \Vert \nabla_h x \Vert_1 + \Vert \nabla_v x \Vert_1$, where $\nabla_h$ and $\nabla_v$ denote the derivative in the horizontal and vertical directions. Several studies \cite{LiuSunXuKamilov:2019,baguer2020diptv} found that incorporating the total variation penalty is beneficial to DIP.
The loss $l_{{\rm t}}$ in \eqref{eqn:diptv} is optimized with Adam \cite{kingma2014adam}, by randomly initializing $\theta$.
The learning is performed as (single-batch) test time adaptation to $y_\delta$.

\begin{figure}
\centering%
\definecolor{pretrainingcolor}{HTML}{A42A2E}%
\definecolor{finetuningcolor}{HTML}{3D78B2}%
\resizebox{\columnwidth}{!}{%
\begin{tikzpicture}[phase/.style={rectangle, draw}, sep/.style={dotted, draw=black!50}, knowledge/.style={rectangle, fill=gray!25, draw=gray!60}, information/.style={draw=gray!60, line width=1.5pt}]
    \node (dip_formula) {\shortstack[l]{\\[-0.26em]$\hspace{-2.5em}{\color{pretrainingcolor}\theta^{\ast}_\mathrm{s}}\to\theta^\mathrm{init}$\\$\displaystyle{\color{finetuningcolor}\theta^{\ast}_\mathrm{t}} \in\argmin_\theta l_{{\rm t}}(\theta; {\color{finetuningcolor}y_\delta})$\\$x^* = \varphi_{\theta^{\ast}_{\rm t}}(z)$}};  %
    \node[left=1em of dip_formula.north west, anchor=north east] (pretraining_formula) {$\displaystyle \argmin_{\theta} l_{{\rm s}}(\theta; \{({\color{pretrainingcolor}x^n, A^\dagger y_{\delta}^n})\}_{n=1}^{N})\ni\hspace{1.1em}$};
    \node[phase, fill=finetuningcolor!25, draw=none, %
        above=2em of dip_formula.north west, anchor=west] (dip) {fine-tuning};
    \node[phase, fill=pretrainingcolor!25, draw=none, %
        above=2em of pretraining_formula.north east, anchor=east] (pretraining) {pretraining};
    \path (dip.north west) -- (pretraining.north east) node[midway] (top) {};
    \path (dip_formula.south west) -- (pretraining_formula.south east) node[midway] (bottom) {};
    \node[above=0em of top] (top_outer) {};
    \node[below=0.5em of bottom] (bottom_outer) {};
    \draw[sep] (top_outer) -- (bottom_outer);
    \node[knowledge, below=1em of bottom] (op) {$A$}; %
    \node[right=-3.7em of dip_formula] (dip_loss_x) {};
    \node[below=-2.5em of dip_formula] (dip_loss_y) {};
    \coordinate (dip_loss) at (dip_loss_y -| dip_loss_x);
    \node[right=-8.1em of pretraining_formula] (pretraining_loss_x) {};
    \node[below=-1.2em of pretraining_formula] (pretraining_loss_y) {};
    \coordinate (pretraining_loss) at (pretraining_loss_y -| pretraining_loss_x);
    \draw[information, ->] (op) -- (op-|dip_loss) -- (dip_loss) node[midway, right] (tv) {}; %
    \draw[information, ->] (op) -- (op-|pretraining_loss) -- (pretraining_loss);
    \draw[information, ->] (op) -- (op-|pretraining_loss) -- (pretraining_loss);
    \node[below left=.1em of pretraining_loss] {\color{pretrainingcolor}synthetic};
    \node[below right=.1em of pretraining_loss] {\color{pretrainingcolor}simulated};
\end{tikzpicture}
}
\caption{A two-stage learning paradigm.
The parameters $\theta$ of the U-Net are first optimized on a dataset comprising ordered pairs of synthetic ground truth images $x^{n}$ and simulated measurement $y_{\delta}^{n}$. The optimal configuration $\theta_{{\rm s}}^{\ast}$ is then used to warm-start the unsupervised fine-tuning on real $\mu$CT data.
}
\label{fig:approach-diagram}
\end{figure}
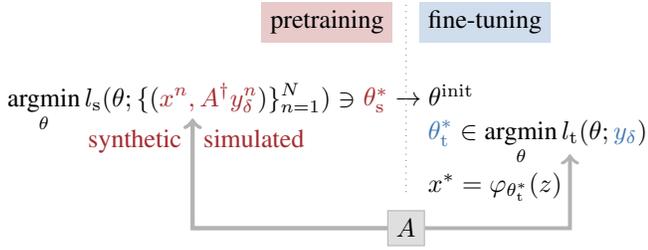

In this work, we recast DIP into the ``supervised pretraining + unsupervised fine-tuning'' paradigm as a two-stage process, called educated DIP (EDIP); see Fig.~\ref{fig:approach-diagram} for a schematic illustration of the framework. In the first stage, we pretrain the network $\varphi_\theta(A^\dag y_\delta)$, where $A^\dagger$ is an approximate inverse operator (e.g.\ filtered back-projection (FBP) for CT \cite{PanSidky:2009}). The training is carried out on a synthetic dataset $\mathcal{D}=\{(x^n,y_\delta^n)\}_{n=1}^N$, composed of $N$ pairs drawn from the joint distribution of ground truth $x^n$ and corresponding simulated measurement $y_\delta^n$. This step is tailored to the target reconstruction task in \eqref{eqn:diptv}, and learns the optimal parameters $\theta^*_{\rm s}$ via supervised training,
\begin{equation}\label{eqn:dipsup}
    \theta^{\ast}_\mathrm{s} \in \argmin_{\theta} \Bigl\{l_{{\rm s}}(\theta)\!:=\dfrac{1}{N}\!\!\!\!\!\!\sum_{\;\;(x^n, y_{\delta}^n)\in\mathcal{D} }\!\!\!\!\!\Vert\varphi_\theta(A^\dagger y_{\delta}^n) - x^n\Vert^2\Bigl\}.
\end{equation}
Note that $\varphi_\theta$ receives $A^\dag y_\delta^n$ as its input (instead of the random noise in \cite{Ulyanov:2018}), serving as a post-processing reconstructor \cite{JinUnser:2017}.
The objective of this stage is to enforce ``benignant'' inductive biases via supervised learning. This educates DIP with knowledge contained in the dataset $\mathcal{D}$, which is then exploited, but still needs to be amended, in solving the reconstruction task in \eqref{eqn:diptv}.

In the second stage, for a given new query measurement $y_\delta$, we use the optimal parameters $\theta^{\ast}_\mathrm{s}$ obtained in the pretraining stage to initialize the network $\varphi_\theta(A^\dag y_\delta)$ in \eqref{eqn:diptv} so as to get DIP up to speed in handling target tasks on real-measured data.
That is, we regard the DIP optimization as a self-adaptation step, where the parameters $\theta$ are fine-tuned unsupervisedly, with their drift conditioned on $\theta^{\ast}_{\rm s}$. Note that the robustness of this method at test time does not rely solely on how well the pretraining stage anticipates distributional shifts. The model makes a good use of pretraining --- the supervised pretraining stage sets and constrains the stage --- but adapts to distributional shifts at test time, and reserves its right to amend the received supervision.

There are several possible variants of the basic framework.
U-Net consists of two parts, a decoder with parameters $\theta_{\rm dec}$, and an encoder with parameters $\theta_{\rm enc}$.
A direct variant of EDIP is to fine-tune only the decoder parameters $\theta_{{\rm dec}}$, but fixing the encoder parameters to the educated guess $\theta^{\ast}_{\rm s, {\rm enc}}$, which are regarded as a shared (between stages) feature extractor.
At test time, we solve \eqref{eqn:diptv} only with respect to $\theta_{{\rm dec}}$ and rely on the pretraining to construct a suitable ``universal'' encoding.
Thus, the learned reconstructor $\varphi_{\theta^{\ast}}$ recovers from the measurement data with $\theta^{\ast} = (\theta^{\ast}_{{\rm s}, {\rm enc}}, \theta^{\ast}_{{\rm t}, {\rm dec}})$. This variant with the fixed encoder (FE) is termed as EDIP-FE.

\section{Datasets}\label{sec:data}

\subsection{Synthetic Training Dataset}

\begin{figure}[ht]
\centering%
  \includegraphics[
    draft=\draftgraphics,
    width=\linewidth]{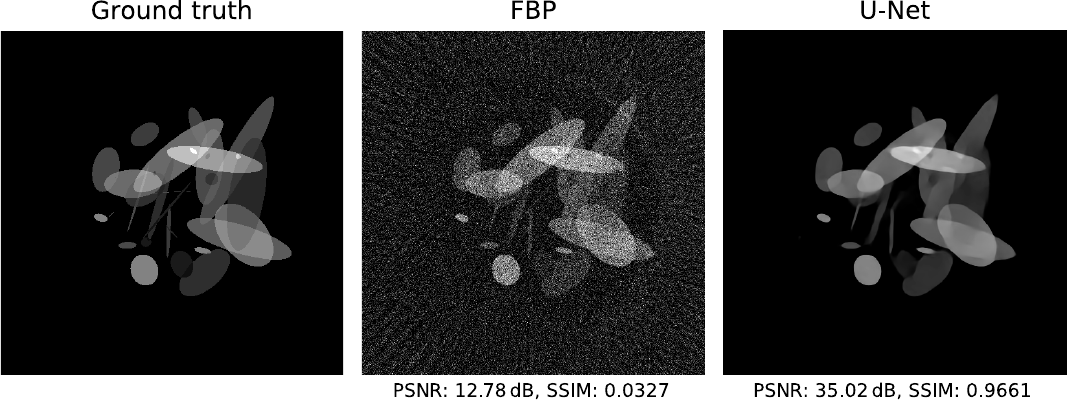}
  \caption{
  An exemplary ground truth image used in the pretraining stage.
  The FBP and U-Net reconstructions are also shown.
  The measurement data $y_\delta$ is simulated using the Walnut \texttt{Sparse 120} setting, adding $5\%$ white noise.
  }
  \label{fig:pretraining_samples_ellipses_walnut}
\end{figure}

We pretrain on a synthetic training dataset of images composed of ellipses or ellipsoids with random position, shape, orientation and intensity values, which
are commonly used to train and evaluate learned reconstruction methods.
This image class encompasses basic building blocks of more complex images, while favoring piece-wise smoothness.
Synthetic data is particularly useful when it is infeasible to collect high-quality ground truth images reassembling the image class of the target reconstructive task, while enabling the learning of features tailored to the inversion of the forward operator $A$. In the experiments,
we use datasets of \num{32000} training and \num{3200} validation images generated on-the-fly using ODL \cite{adler2018odl}.
The image resolution and the distribution of the ellipses / ellipsoids can be easily adapted to match different target data.
The synthetic projection data is computed by forward projecting the ground truth images and adding \num{5}\,\% white noise. Fig.~\ref{fig:pretraining_samples_ellipses_walnut} shows an exemplary ground truth image and reconstructions obtained by the FBP and U-Net from the simulated noisy data.

\subsection{Real ${\mu}$CT Measurement Data}
\label{ssec:measurement_data}

\begin{figure}[ht]
  \centering%
  \includegraphics[
    draft=\draftgraphics,
    width=0.6\linewidth]{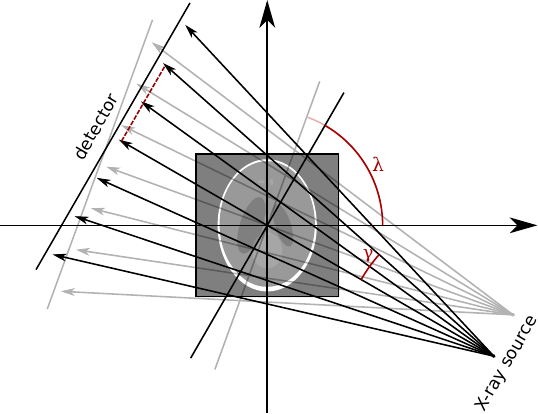}%
  \caption{ {Diagram of the 2D cone-beam geometry (a.k.a. fan-beam geometry).
  }}
  \label{fig:fan_beam}
\end{figure}

We evaluate our approach on two real $\mu$CT datasets to showcase the effectiveness of the approach. The forward operator $A$ is a ray transform matching a 2D or 3D cone-beam geometry (cf.~Fig.~\ref{fig:fan_beam} for 2D). %
The scanner rotates around the object (or, equivalently, the object is rotated inside the scanner), taking projections from different source angles $\lambda$. Within each projection, each detector pixel (e.g.\ parameterized by $\gamma$) measures the intensity for a specific line, attenuated by the object.

\label{sec:2d_cone_beam_geometry}

\paragraph{X-ray Lotus Root Dataset}

$\mu$CT measurements of a Lotus root slice filled with different materials are available from \cite{bubba2016lotus_paper}.
The dataset contains fan-beam measurements corresponding to a 2D volume slice, with \num{120} projections at angles equally distributed over $[0, 360^\circ)$ and \num{429} detector pixel values each.
The sparse matrix modeling the forward operator for an image resolution $(\SI{128}{px})^2$ is used.
In the evaluation, we consider the setting of \texttt{Sparse 20}: a \num{6}-fold angular sub-sampling, \num{20} angles, equally distributed over $[0, 360^\circ)$. We use a TV-regularized reconstruction from all \num{120} projection angles, obtained by Adam, as the reference solution.

\paragraph{X-ray Walnut Dataset}

A collection of cone-beam $\mu$CT measurement data from \num{42} Walnuts was provided in \cite{der_sarkissian2019walnuts_data}.
For each walnut, a set of three 3D cone-beam measurements is included, each obtained with a different source position. Projections are acquired at \num{1200} angles equally distributed over $[0, 360^\circ)$, with a resolution of \num{972} detector rows and \num{768} detector columns.
A volume resolution of $(\SI{501}{px})^3$ is used.
We consider reconstructing a single 2D slice from a suitable subset of detector pixel measurements, and 3D reconstruction with a downscaled image resolution of $(\SI{167}{px})^3$.
For the 2D task, we use the setting of \texttt{Sparse 120}: a $10$-fold angular sub-sampling with \num{120} angles, equally distributed over $[0, 360^\circ)$; for 3D, we consider the settings \texttt{3D Sparse 20} and \texttt{3D Sparse 60} with \num{20} and \num{60} equally distributed angles, and sub-sample the projection rows and columns by a factor of \num{3}. The 3D settings are chosen to mimic industrial applications, where a high degree of sparsity is often desired.
The approximations $A^\dagger y_\delta$ are computed via the Feldkamp-Davis-Kress (FDK) algorithm \cite{feldkamp1984fdk}.
FDK is an FBP-based algorithm with a weighting step for cone-beam measurements, and is still denoted as ``FBP''. To achieve accurate automatic differentiation of the forward projection operator in %
2D, we utilize its sparse matrix representation.
In 3D, we opt for forward and backward projection routines of ASTRA via tomosipo \cite{hendriksen2021tomosipo}.
We use the ground truth provided with the dataset \cite{der_sarkissian2019walnuts_data}, which was obtained with accelerated gradient descent using the measurements from all \num{1200} projection angles and all three source positions.

\section{Experiments and Results}\label{sec:experiment}

Throughout, we denote the type of the network input $z$ used for a method in brackets: for example, ``DIP (noise)'' refers to the standard DIP with noise input, while ``EDIP (FBP)'' stands for the educated DIP with FBP input.

\subsection{Neural Network Architecture}

\begin{figure}[h!]
    \centering
    \includegraphics[width=\linewidth]{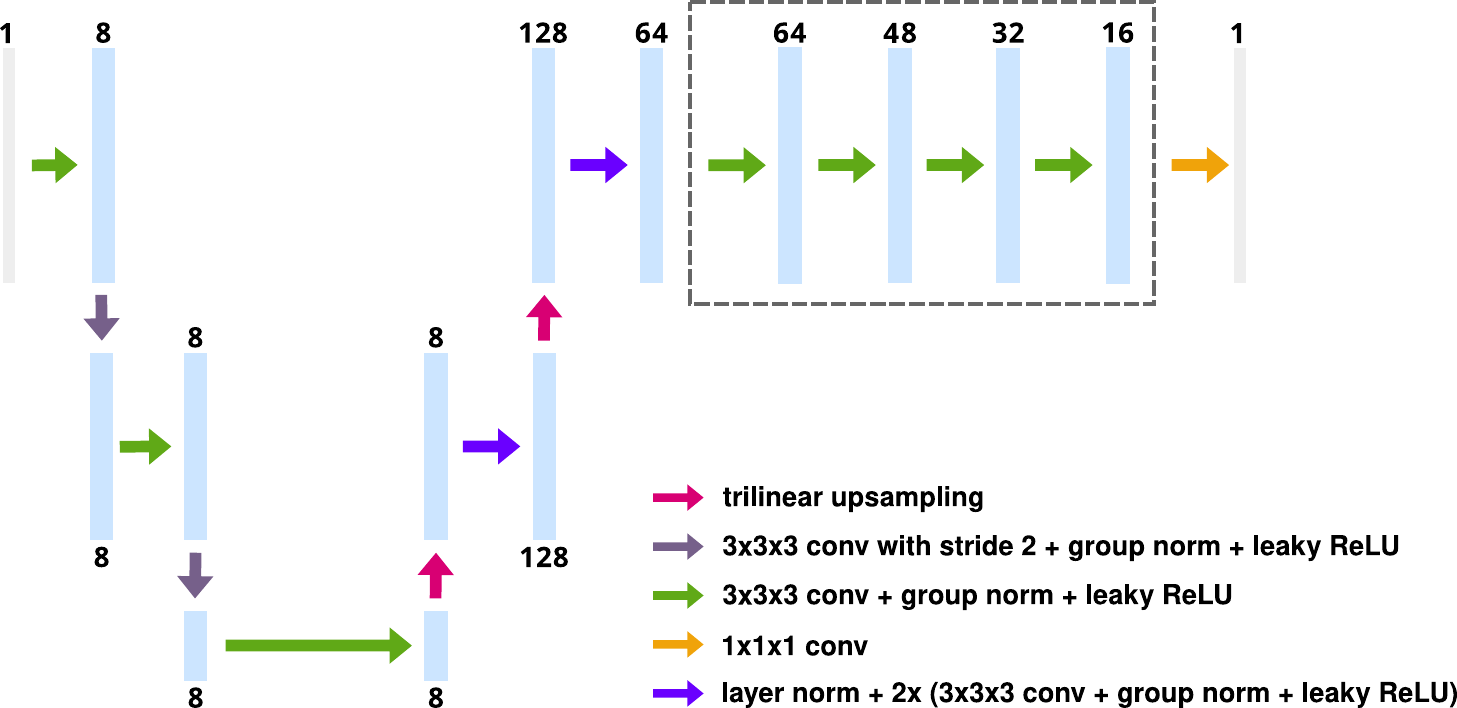}
    \caption{The architecture of the proposed 3D U-Net. Each light-blue bar corresponds to a multi-channel feature map. Arrows denote the different operations.}
    \label{fig:architecture_of_3D_unet}
\end{figure}

For 2D settings, we adopt the U-Net proposed by \cite{baguer2020diptv}, but replace batch-normalization layers with group-normalization layers. For 3D $\mu$CT reconstructions, we fine-tune the architecture, cf.\ Fig.~\ref{fig:architecture_of_3D_unet}, since the standard 3D U-Net --- originally introduced for segmentation \cite{cciccek20163d} --- does not meet our memory constraint, and a naively reduced version leads to sub-optimal reconstructions. We modify
the U-Net architecture as follows: (i) reduce the numbers of channels per convolutional layer in the encoder;
(ii) increase the expressivity of the decoder by chaining subsequent convolutional layers with decreasing number of channels;
(iii) remove skip connections. Due to memory constraints (i.e.\ 24\,GB VRAM), we use a 3-scale 3D U-Net. 

\subsection{Evaluation Metrics}

We measure the reconstruction quality via peak signal-to-noise ratio (PSNR), and
include structural similarity index measure (SSIM) \cite{wang2004image} for reconstructions.
To assess the convergence speed, we employ two metrics: steady PSNR and rise time (denoted by $\star$ in the figures).
The steady PSNR is the median PSNR over the last \num{5}k iterations.
The rise time is the iteration number at which we reach the baseline PSNR (i.e.\ DIP's steady PSNR) up to a threshold \SI{0.1}{dB}.
In addition, we always consider the iteration-wise median PSNR over repeated runs of the same experiment (with varying seeds) for these metrics; we use \num{5} runs for 2D and \num{3} runs for 3D.
The variability between runs does arise not only from random initialization of the network parameters or noise input, but also from numerical effects in parallel computations on GPU. The optimal reconstruction $\varphi_{\theta_{\text{min-loss}}}(z)$ is taken from the iteration with minimum loss value $l_{{\rm t}}(\theta_{\text{min-loss}})= \min_{i\in 0\dots N} l_{{\rm t}}(\theta^{[i]})$.
This remedies non-monotonous loss minimization, yet the (E)DIP optimization plots and steady PSNR computations use the actual iterate to facilitate a direct analysis.

\begin{figure*}[h!]
\begin{minipage}{0.475\textwidth}
  \centering%
  \includegraphics[
    trim={0 0 0 0}, clip,
    draft=\draftgraphics, width=\textwidth]{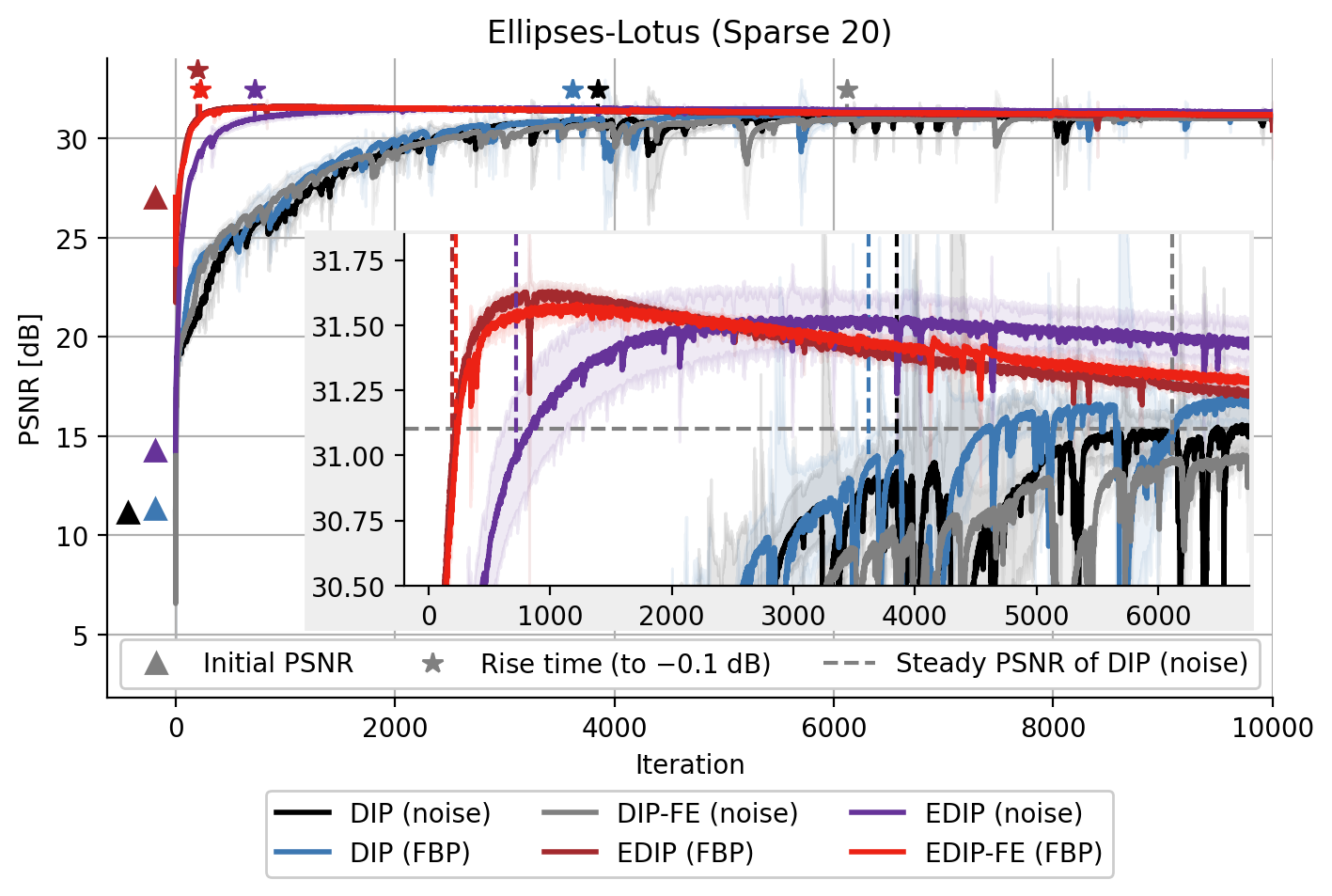}
  \caption{The optimization of EDIP versus DIP on Lotus \texttt{Sparse 20}. The symbols $\star$ and $\blacktriangle$
  denote initial PSNR and rise time, respectively, and the horizontal dashed line indicates the steady PSNR of DIP (noise).
  }
  \label{fig:comp_ellipses_lotus_20-main}
 \end{minipage}\hfill
\begin{minipage}{0.475\textwidth}
  \centering%
  \includegraphics[
    draft=\draftgraphics,width=0.85\textwidth]{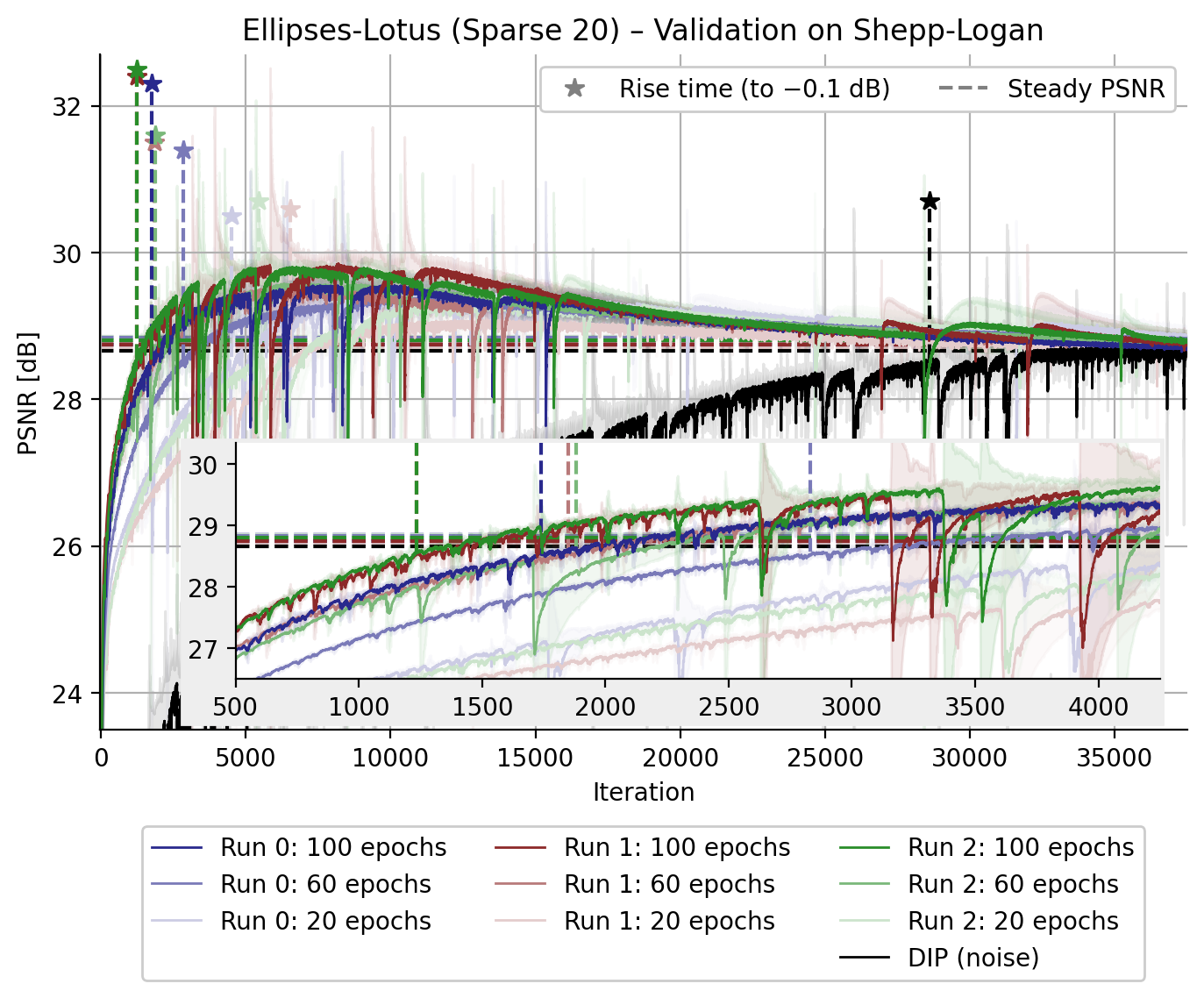}
  \caption{ {Checkpoint selection} on the Shepp-Logan phantom for the initial EDIP (FBP) model parameters for the Lotus \texttt{Sparse 20} setting.
  }
  \label{fig:validation_lotus_20-main}
\end{minipage}
\end{figure*}
\subsection{Hyperparameter's Selection}

The learning rate and regularization parameter $\gamma$ are fine-tuned for standard DIP.
For EDIP, we do not conduct additional hyperparameter search, but use the values identified for DIP. These hyperparameters values also perform well for EDIP, which saves us from performing an individual search for each pretraining checkpoint (to be defined next).

\subsection{Selection of the Checkpoints}\label{sec:evaluation}

Multiple parameters' configurations (i.e.\ $\theta_{{\rm s}}^{\ast}$) may be obtained from the pretraining stage: repeated runs (varying random initializations) and multiple checkpoints along the optimization trajectory of each run.
From a set of checkpoints, one needs to identify solutions maximizing the speed-up at test time.
More broadly, this is an open question. 
In the experiments below, the selection strategy is based on assessing the performance on the Shepp-Logan phantom \cite{shepp1974fourier}, a standard test image within the medical imaging community.
The checkpoint leading to the shortest rise time is then selected, among those with a steady PSNR that is at most \SI{0.25}{dB} lower than the maximum steady PSNR of any checkpoint.
This selection is carried out for 2D reconstruction settings; 3D runs use the best performing checkpoint for computational reasons.
For Lotus \texttt{Sparse 20}, we repeat the pretraining \num{3} times (varying the seed) and collect checkpoints after every \num{20} epochs, training for a maximum of \num{100} epochs.
For the Walnut \texttt{Sparse 120}, we pretrain for \num{20} epochs, and retain the minimum validation loss checkpoint of each run. For the 3D Walnut settings, we pretrain for a maximum of \num{2} epochs, and retain checkpoints every \num{0.125} epochs (i.e.\ \num{4}k gradient updates).

\subsection{The Lotus Root}
\label{sec:lotus}

Table~\ref{tab:numerical_eval_results_lotus-main} shows the convergence properties of EDIP %
and DIP for Lotus \texttt{Sparse 20}.
We include in our analysis cases where the FBP  $A^{\dagger}y_{\delta}$ is fed as the input (instead of noise) when solving \eqref{eqn:diptv} for DIP, and inputting noise for EDIP.
EDIP significantly outperforms DIP in terms of the convergence speed for either a fixed noise image or FBP.

\begin{table}[h!]
  \caption{
        {Quantitative evaluation} for the Lotus \texttt{Sparse 20}. %
        }
  \small%
  {\centering%
  \setlength\fboxsep{0pt}%
  \resizebox{0.85\columnwidth}{!}{%
  \begin{tabular}{l@{\extracolsep{4pt}}ccc}
    \strut{}Ellipses-Lotus \texttt{Sparse 20}\hspace*{-10em}\\
  \cline{1-1}\cline{2-4}\\[-0.7em]
   & \shortstack{\strut{}Rise time} & \shortstack{\strut{} (Max PSNR; iters)} & \shortstack{\strut{}Steady PSNR} \\
   \midrule
  DIP (noise) & \num{3848} & (\num{31.17}; \num{8846}) & \num{31.10}\\
  DIP (FBP) & \num{3622} & (\num{31.25}; \num{8813}) & \num{31.17} \\
  DIP-FE (noise) & \num{6118} & (\num{31.10}; \num{9818}) & \num{31.00}\\
  EDIP (FBP) & \hphantom{0}\hlbest{\num{195}} & (\hlbest{\num{31.65}}; \hphantom{\num{0}}\hlbest{\num{981}}) & \num{31.21}\\
  EDIP (noise) & \hphantom{0}\num{723} & (\num{31.53}; \num{3548}) & \hlbest{\num{31.39}}\\
  EDIP-FE (FBP) & \hphantom{0}\hlbest{\num{226}} & (\hlbest{\num{31.59}}; \hlbest{\num{1421}}) & \num{31.26} \\
  TV & -- & -- & \num{30.73} \\
  \bottomrule
  \end{tabular}%
  }\\}
  \label{tab:numerical_eval_results_lotus-main}
\end{table}

\begin{table}[b]
  \caption{
        Checkpoints' comparison from the pretraining stage for EDIP (FBP) on Lotus \texttt{Sparse 20}. The checkpoint from run \num{2} after \num{100} epochs is selected using the Shepp-Logan data (cf.\ Fig.~\ref{fig:validation_lotus_20-main})}
  \small%
  {\centering%
  \setlength\fboxsep{0pt}%
  \resizebox{0.6\columnwidth}{!}{%
  \begin{tabular}{l@{\extracolsep{4pt}}rcc}
   & Epochs & \shortstack{\strut{}Rise time} & \shortstack{\strut{} (Max PSNR; iters)}\\ %
  \cline{1-2}\cline{3-4}\\[-0.7em]
  \multirow{ 3}{*}{Run \num{0}} & \num{100}  & \num{247} & (\num{31.49}; \num{1545}) \\ %
  &\hphantom{\num{0}}60 & \hlbest{\num{174}} & (\num{31.56}; \hphantom{\num{0}}\hlbest{\num{842}})  \\ %
  &\hphantom{\num{0}}20 & \num{291} & (\hlbest{\num{31.61}}; \num{1614}) \\[0.35em] %
  \multirow{ 3}{*}{Run \num{1}} & \num{100}  & \hlbest{\num{162}} & (\num{31.53}; \hphantom{\num{0}}\hlbest{\num{779}}) \\ %
  &\hphantom{\num{0}}60  & \num{243} & (\num{31.53}; \num{1755}) \\ %
  &\hphantom{\num{0}}20  & \num{390} & (\num{31.56}; \num{1518}) \\[0.35em] %
  \multirow{ 3}{*}{Run \num{2}} & \num{100}  & \hlbest{\num{195}} & (\hlbest{\num{31.65}}; \hphantom{\num{0}}\hlbest{\num{981}}) \\ %
  &\hphantom{\num{0}}60  & \hlbest{\num{194}} & (\hlbest{\num{31.58}}; \hlbest{\num{1083}}) \\ %
  &\hphantom{\num{0}}20  & \num{318} & (\num{31.51}; \num{1706}) \\ %
  \bottomrule
  \end{tabular}%
  }\\}
  \label{tab:numerical_test_checkpoints_ellipses_lotus_20-main}
\end{table}

\begin{figure*}[t]
  \centering%
\begin{minipage}{0.475\textwidth}
  \includegraphics[
    draft=\draftgraphics,
    width=\textwidth]{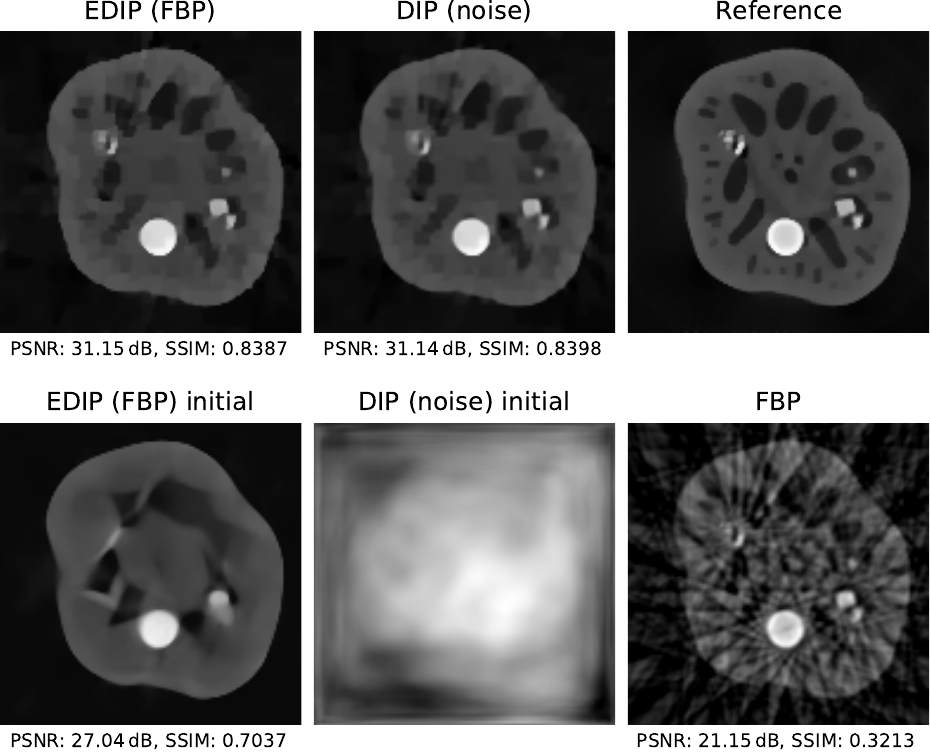}%
  \caption{ EDIP
  versus DIP reconstruction on Lotus \texttt{Sparse 20}.
  From the \num{5} runs (varying the seed), the one with the (closest to) median PSNR was selected for each method.
}
  \label{fig:reco_ellipses_lotus_20}
\end{minipage}\hfill
\begin{minipage}{0.475\textwidth}
  \includegraphics[
    draft=\draftgraphics,
    width=\textwidth]{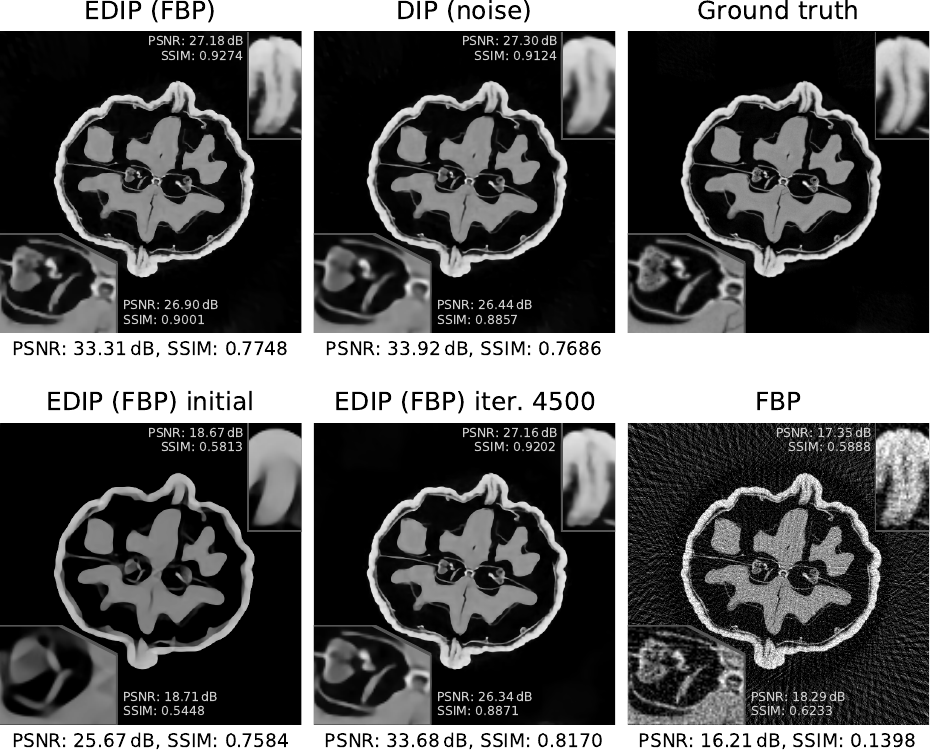}%
  \caption{ {EDIP versus DIP reconstruction of Walnut \texttt{sparse 120}}.
  }
\label{fig:reco_ellipses_walnut_120}
  \end{minipage}
\end{figure*}

EDIP only takes \num{195} (and \num{723} for noise input) iterations to reach \SI{-.1}{dB} of the baseline PSNR, against \num{4.1}k iterations needed for DIP. Thus, pretraining greatly accelerates the convergence. The optimization process is considerably more stable (cf.\ Fig.~\ref{fig:comp_ellipses_lotus_20-main}), implying a possibly much more favorable loss landscape for EDIP. Thus, pretraining stabilizes the optimization process of DIP, which is highly desirable in practice.
Note that EDIP-FE, which fixes the encoder parameters to the pretrained ones $\theta_{s,\rm enc}^*$, is as fast as EDIP, and the reconstruction quality of EDIP and EDIP-FE are largely comparable with each other. With fewer parameters to be updated, EDIP-FE is computationally lighter than EDIP (since backpropagation is only needed for the decoder, and the forward pass through the encoder can be pre-computed beforehand).
Fig.~\ref{fig:reco_ellipses_lotus_20} shows the reconstruction (along with the reference and FBP) for Lotus \texttt{Sparse 20}.
We observe that pretraining can also boost the performance of DIP: EDIP considerably overshoots the baseline PSNR, cf.\ Fig.~\ref{fig:comp_ellipses_lotus_20-main}. This suggests that pretraining, if coupled with proper early-stopping (approximately a few hundred iterations after the rise time), can lead to better reconstructions.

To maximize the speed-up, we select the warm-start configuration $\theta^{\ast}_{{\rm s}}$ on the Shepp-Logan phantom.
Fig.~\ref{fig:validation_lotus_20-main} shows the validation runs. We select $\theta^{\ast}_{{\rm s}}$ from  run \num{2} after \num{100} epochs since it results in the smallest rise time.
Interestingly, a substantial overshoot of baseline PSNR is observed on the Shepp-Logan phantom, possibly due to its in-distribution nature with respect to the ellipses.
Table~\ref{tab:numerical_test_checkpoints_ellipses_lotus_20-main} reports the rise time at test time for different checkpoints collected for each run. The results indicate that the checkpoint selection does impact the achievable acceleration factor, but not the maximum PSNR.

\subsection{The Walnut}
\label{sec:walnut}
\begin{table*}[t]
   \caption{
        {Quantitative evaluation} for the Walnut.}
  \small%
  {\centering%
  \setlength{\tabcolsep}{4.6pt}%
  \setlength\fboxsep{0pt}%
   \label{tab:tabular_results_overall_walnut}
  \resizebox{\textwidth}{!}{%
  \begin{tabular}{l@{\extracolsep{4pt}}cccccccccc}
  \strut{}Ellipses/Ellipsoids-Walnut \hspace{-10em} & \multicolumn{3}{c}{\texttt{Sparse 120}}&\multicolumn{3}{c}{\texttt{3D Sparse 20}} & \multicolumn{3}{c}{\texttt{3D Sparse 60}}\\
  \cline{1-1}\cline{2-4}\cline{5-7}\cline{8-10}\\[-0.7em]
  & \shortstack{\strut{}Rise time} & \shortstack{\strut{} (Max PSNR; iters)} &  \shortstack{\strut{}Steady PSNR} & \shortstack{\strut{}Rise time} & \shortstack{\strut{} (Max PSNR; iters)} &  \shortstack{\strut{}Steady PSNR} & \shortstack{\strut{}Rise time} & \shortstack{\strut{} (Max PSNR; iters)} &  \shortstack{\strut{}Steady PSNR} \\\midrule
  DIP (noise) & \num{20373} & (\num{34.02}; \num{25357}) & \num{33.87} &\num{17200} & (\num{30.68}; \num{23477}) & \num{30.37} & \num{49041} & (\num{34.05}; \num{58901}) & \num{33.93}\\
  DIP (FBP) & \num{13778} & (\num{34.07}; \num{28094}) &  \num{33.90} & \num{13016} & (\num{31.32}; \num{25063}) & \hlbest{\num{31.19}} & \num{27873} & (\hlbest{\num{34.37}}; \num{53731}) & \hlbest{\num{34.22}}\\
  EDIP (FBP) & \hphantom{0\,}\hlbest{\num{4496}} & (\num{33.92}; \hlbest{\num{13039}}) & \num{33.56} & \hphantom{0\,}\hlbest{\num{3739}} & (\hlbest{\num{31.48}}; \hlbest{\num{10689}}) & \num{30.94} & \hlbest{\num{11247}} & (\hlbest{\num{34.35}}; \hlbest{\num{40810}}) & \hlbest{\num{34.18}}\\
  EDIP-FE (FBP) & \hphantom{0\,}\hlbest{\num{4384}} & (\num{33.91}; \hlbest{\num{12540}}) & \num{33.70} & \hphantom{0\,}\hlbest{\num{2979}} & (\num{31.38};  \hlbest{\num{10749}}) & \num{30.93} & \hlbest{\num{14520}} & (\hlbest{\num{34.33}}; \hlbest{\num{45259}}) & \hlbest{\num{34.15}} \\
  TV & -- & -- & \num{31.67} & -- & -- & \num{28.89} & -- & -- & \num{33.35}\\
  \bottomrule
  \end{tabular}%
  }\\}
\end{table*}

Fig.~\ref{fig:reco_ellipses_walnut_120} shows the reconstructed Walnut slice; see Table~\ref{tab:tabular_results_overall_walnut} for quantitative results.
A speed-up is observed, similar to the Lotus root: EDIP takes about \SI{30}{min} at rise time (approximately \num{4.4}k iterations), whereas DIP (with noise input) takes \SI{2}{h} and \SI{30}{min} at rise time (approximately  \num{20.4}k iterations) with NVIDIA GeForce RTX 2080Ti.
A TV regularized reconstruction of the Walnut takes \SI{6}{min}, and requires \num{1.7}k gradient steps to converge to \SI{31.67}{dB}.
EDIP takes only \SI{3}{min} (after \num{421} iterations) to match \SI{31.67}{dB}.
In \SI{6}{min}, EDIP reaches \SI{32.80}{dB}, with a gain of \SI{1.1}{dB}.
Finally, DIP-FE~/~EDIP-FE report similar performances to DIP~/~EDIP.

\begin{figure*}[t]
  \centering%
  \includegraphics[
    draft=\draftgraphics,
    width=\linewidth]{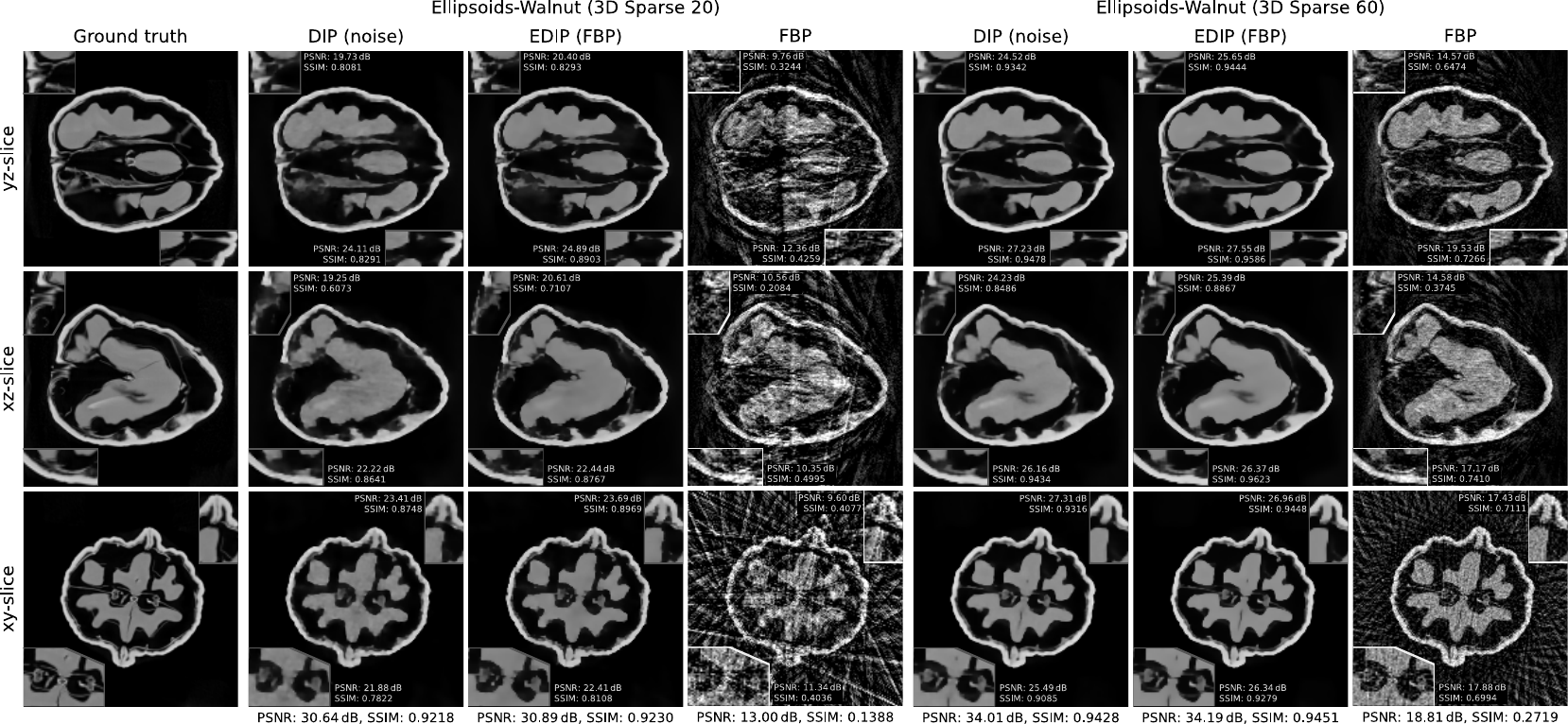}%
  \caption{ {3D Walnut reconstruction of EDIP} pretrained on ellipsoids dataset, compared to standard DIP, at three different slices.
  }
\label{fig:reco_ellipsoids_walnut_3d}
\vspace{-1em}
\end{figure*}

\begin{figure}
\centering
\noindent
\includegraphics[trim={0 1.95cm 0 0},clip,
width=0.90\columnwidth]{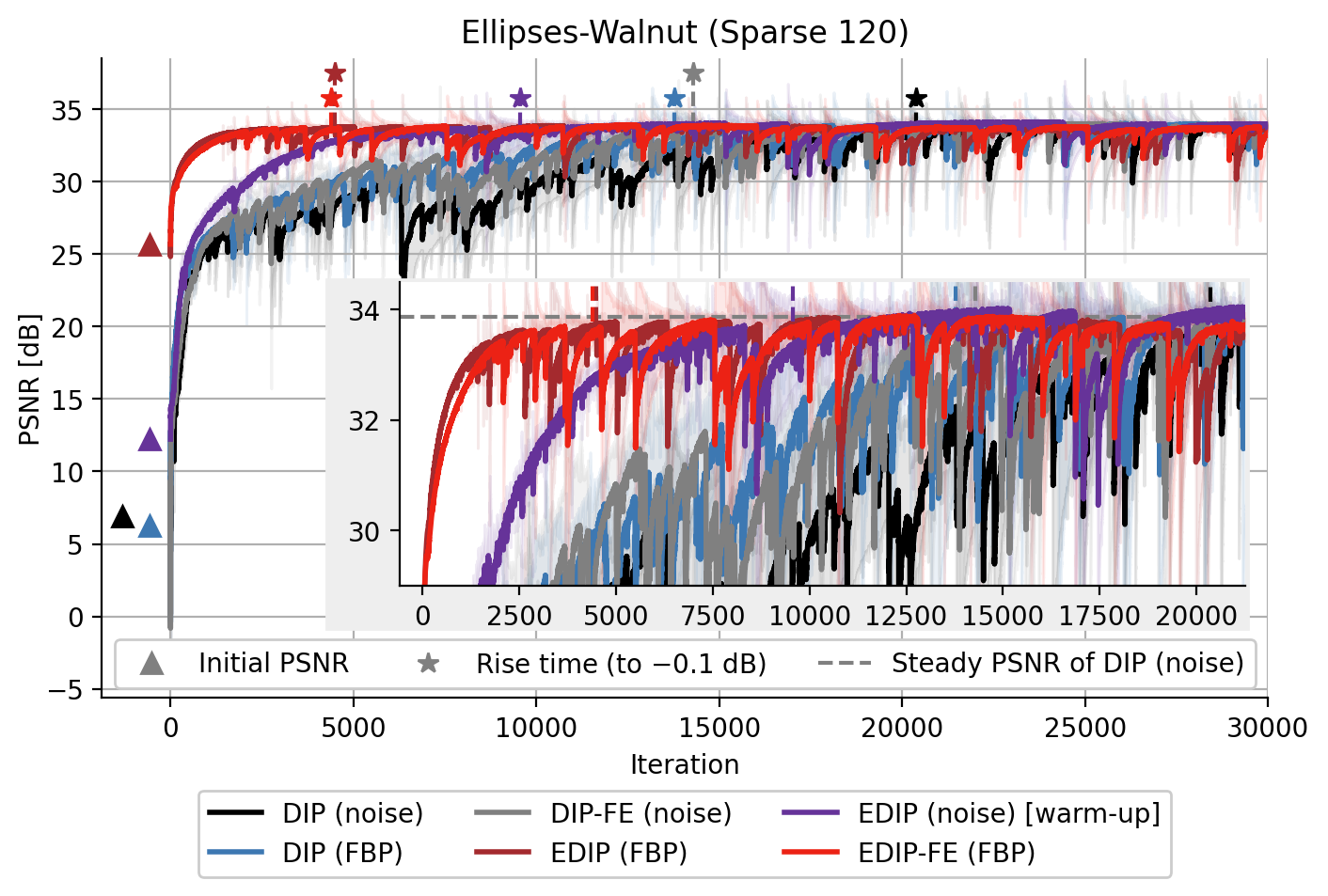}\\[0.2em]
\includegraphics[trim={0 1.4cm 0 0},clip, width=0.90\columnwidth]{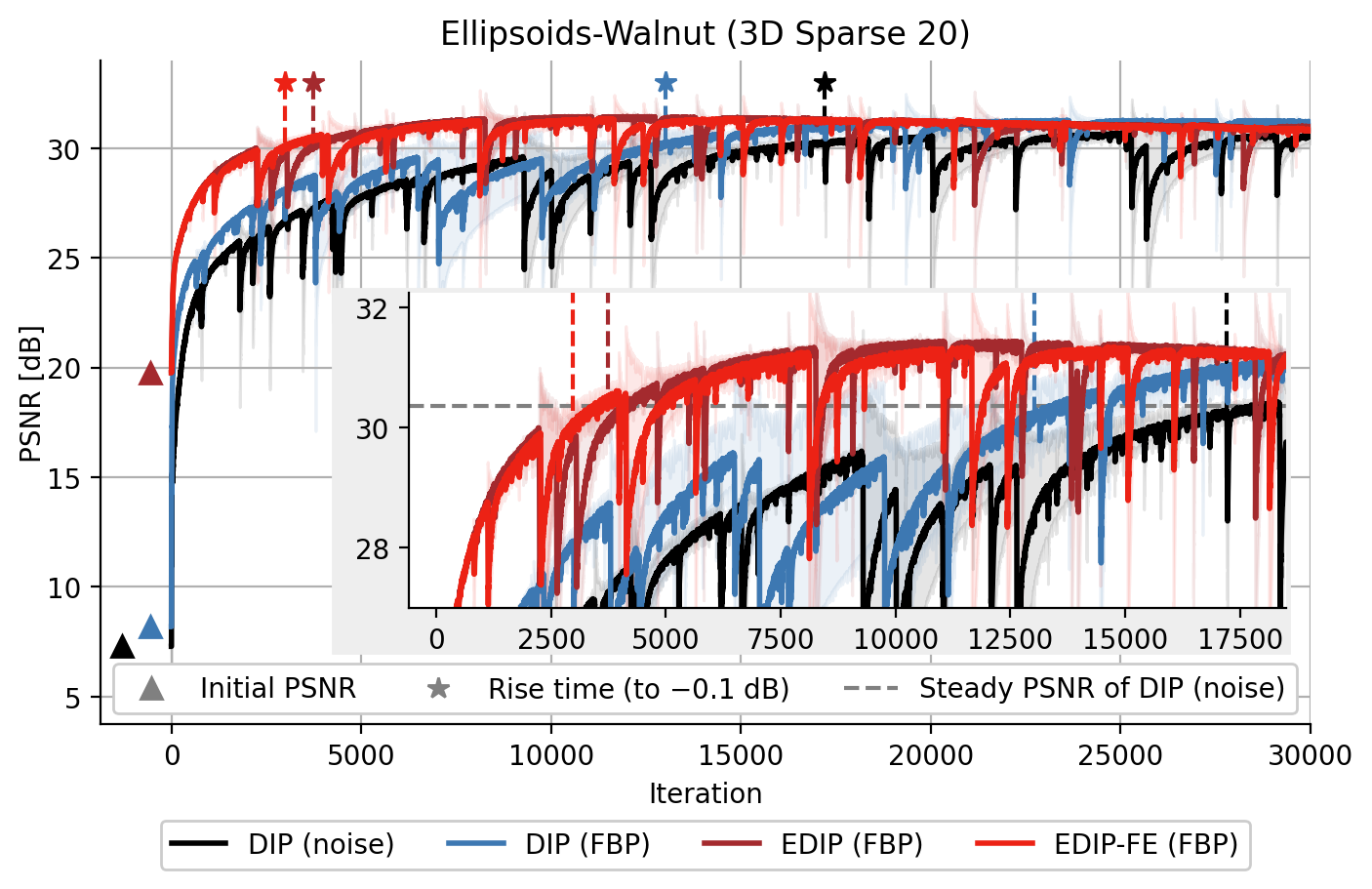}\\[0.2em]
\hspace{1.15em}
\includegraphics[width=0.65\columnwidth]{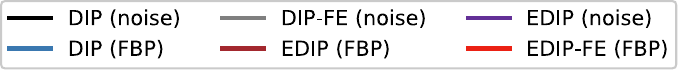}
\caption{The optimization of EDIP versus DIP for the Walnut reconstruction in 2D (top) and 3D (bottom). The symbols $\star$ and $\blacktriangle$
  denote initial PSNR and rise time, respectively, and the horizontal dashed line indicates the steady PSNR of DIP (noise).}
\label{fig:main-walnut-fig-optim}
\end{figure}

On a minor note, it is observed that EDIP better reconstructs finer structures (e.g.\ the wrinkled shell), and DIP suffers from over-smoothing artifacts. This concurs with the observation for Lotus \texttt{Sparse 20}: by incorporating the knowledge contained in the synthetic training data, pretraining can boost the performance of DIP.

Similar observations can be made for reconstructing the Walnut volume, cf.\ Fig.~\ref{fig:reco_ellipsoids_walnut_3d} for 3D reconstructions along the yz, xz, and xy axes and Table~\ref{tab:tabular_results_overall_walnut} for quantitative results. EDIP reconstruction from the \texttt{3D Sparse 20} data takes approx.\ \SI{1.5}{h} with a NVIDIA GeForce RTX 3090, and leads to \SI{33.77}{dB} in PSNR, compared to \SI{7.3}{h} and \SI{5.53}{h} for DIP (with noise / FBP as input).
EDIP matches the PSNR of a TV reconstruction in about \SI{30}{min}, gains \SI{1}{dB} over TV after additional \SI{20}{min}, and it takes \SI{2.3}{h} to observe a  \SI{2}{dB} gain.
The \texttt{3D Sparse 60} leads to similar speed-up.
It takes \SI{20}{h} for DIP with noise as input. Inputting the FBP results already in a considerable speed-up (about\ \SI{11}{h}), whereas EDIP requires only \SI{4}{h}.
In sum, pretraining on the synthetic ellipsoids dataset greatly accelerates the convergence of DIP for 3D $\mu$CT reconstruction.

\begin{figure}[h]
  \centering%
  \small%
  \includegraphics[
    draft=\draftgraphics,
    width=\linewidth]{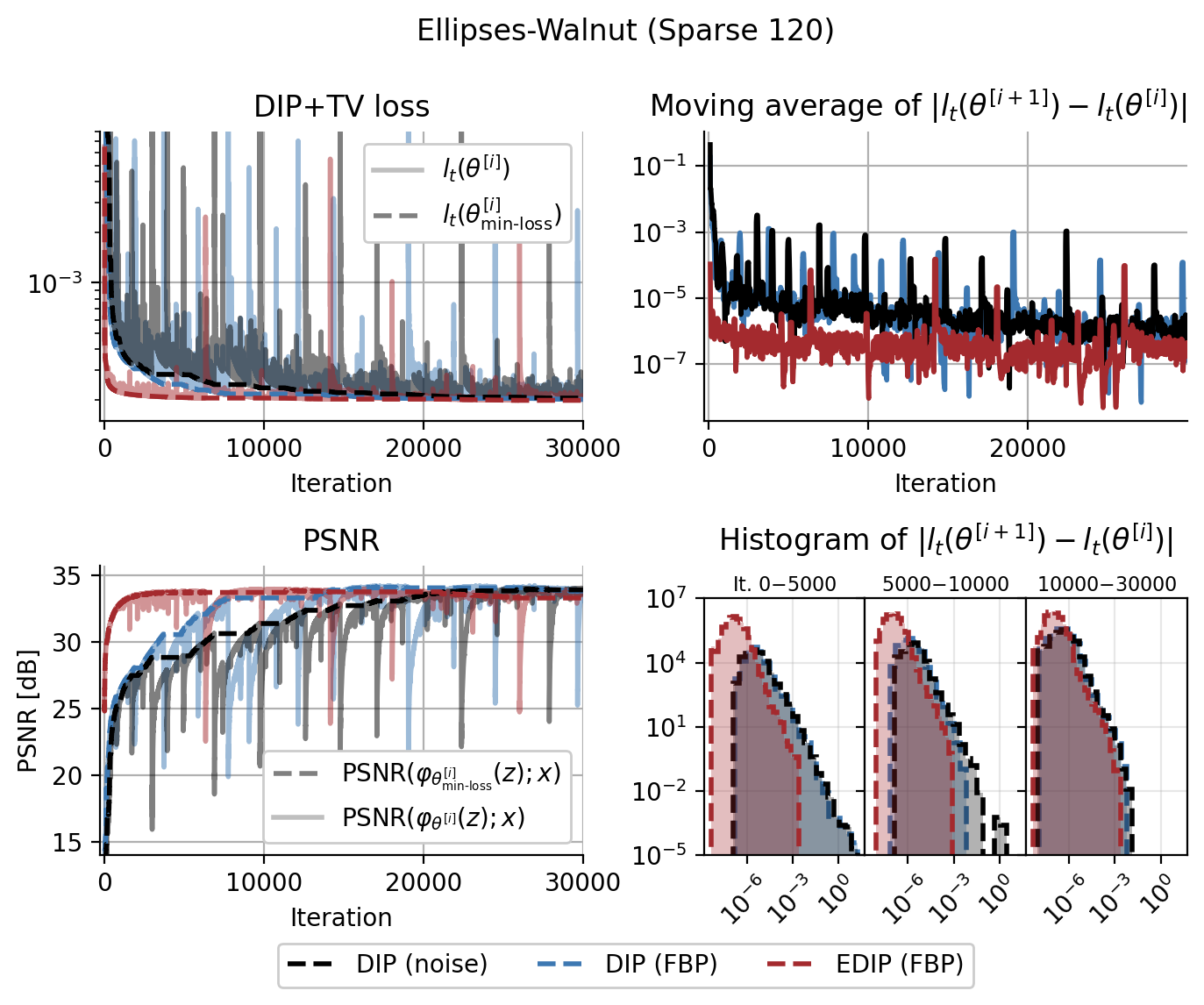}
  \caption{
  {The min-loss and PSNR computed with the min-loss output} for Walnut \texttt{Sparse 120} (left).
  Loss variation (i.e.\ $|l_t(\theta^{[i+1]}) - l_t(\theta^{[i]})|$) and respective histograms computed over three intervals (right).
  The moving average uses a window size of \num{100} iterations.}
  \label{fig:best_psnr_and_loss_ellipses_walnut_120}
\end{figure}

Last, we briefly comment on the convergence of the optimization process, cf.\ Fig.~\ref{fig:main-walnut-fig-optim}. The overall convergence behavior for 2D and 3D is similar to Lotus \texttt{Sparse 20}: pretraining stabilizes DIP optimization and greatly accelerates the convergence. Fig.~\ref{fig:best_psnr_and_loss_ellipses_walnut_120} shows the convergence and stability of the loss in \eqref{eqn:diptv}. The variation of the loss value is reduced if EDIP is used.
As a practical post-hoc strategy to overcome the instability of the DIP optimization scheme, the reconstructed image is taken as the network output at minimum loss.

\section{Investigation of the Role of Pretraining}\label{sec:analysis}

In this section, we first motivate why we use a standard pretraining strategy instead of resorting to more sophisticated schemes, and then we shed insight into the mechanism of knowledge transfer via pretraining, highlighting favorable as well as detrimental properties.

\paragraph{Standard, Adversarial, Meta?} 

In this work, we adopt the standard pretraining paradigm within our pretraining stage, as described in \cref{sec:method}. 
The choice is informed by comparing standard pretraining, adversarial pretraining \cite{goodfellow2014explaining,yi2021improved} and model agnostic meta-learning (MAML) \cite{FinnAbbeel:2017, raghu2019rapid} on the Lotus \texttt{Sparse 20}.
Adversarial pretraining uses a projected gradient descent attack (PGD-$L_2$ \cite{carlini2017towards, madry2017towards}).
MAML-based pretraining obtains a parameters' configuration training on six different tasks, comprising three different image classes: ellipses, rectangles \cite{barbano2022bayesian}, and natural images from the PASCAL VOC segmentation dataset \cite{everingham2010pascal}, as well as two different noise distributions: Gaussian and Poisson.
We do not vary the forward operator $A$, since the pretraining stage is tailored to a known acquisition geometry; varying the structure of $A$ (e.g., via sparsification) would only withhold from the model operator-specific knowledge, and introduce artifacts that are not expected to be found in the subsequent reconstruction tasks. 
We investigate whether the parameters' configurations, found with adversarial pretraining and MAML, lead to general representations adapting faster to the subsequent reconstruction problem. 
It is observed from \cref{tab:numerical_eval_pretrain_lotus-main} that all three pretraining strategies lead to parameters' configurations that adapt to the subsequent reconstruction task with approximately similar speed-up.
Even if the adaptation to the subsequent task shows on par properties, adversarial pretraining and MAML introduce a significant computational overhead, which we find unnecessary. The latter can be attributed to the facts that
adversarial pretraining requires the inclusion of an inner loop optimization to design the attack (adding 62h to the wall-clock time); MAML's outer loop updates $\theta_\mathrm{s}$, while the inner one (with one step of stochastic gradient descent) adapts $\theta_\mathrm{s}$ to a given task. MAML, instead, increases ($\times 5$) the overall VRAM required.

\begin{table}[h!]
  \caption{
        Quantitative Evaluation of Alternative Pretraining Strategies for the Lotus along with the wall-clock time recorded on a NVIDIA RTX 2080Ti. 
        }
  \small%
  {\centering%
  \setlength\fboxsep{0pt}%
  \resizebox{\columnwidth}{!}{%
  \begin{tabular}{l@{\extracolsep{4pt}}cccc}
    \strut{}Ellipses-Lotus \texttt{Sparse 20}\hspace*{-10em}\\
  \cline{1-1}\cline{2-5}\\[-0.7em]
   & \shortstack{\strut{}Rise time} & \shortstack{\strut{} (Max PSNR; iters)} & \shortstack{\strut{}(VRAM; batch size)} & \shortstack{\strut{} Time}\\
   \midrule
  EDIP (FBP) & \hphantom{0}{\num{195}} & ({\num{31.65}}; \hphantom{\num{0}}{\num{981}}) & (\num{5941}\,MiB; \num{32}) & 23h \\
  Adv.-$L_2$-EDIP (FBP) & \hphantom{0}{\num{143}} & ({\num{31.24}}; \hphantom{\num{0}}{\num{1175}}) & (\num{6093}\,MiB; \num{32}) & 85h\\
  MAML-EDIP (FBP) & \hphantom{0}{\num{545}} & ({\num{31.54}}; \hphantom{\num{0}}{\num{1512}}) & (\num{7949}\,MiB; \num{8}) & 31h\\
  \bottomrule
  \end{tabular}%
  }\\}
  \label{tab:numerical_eval_pretrain_lotus-main}
\end{table}

\paragraph{In Need to Amend}

Figs.~\ref{fig:reco_ellipses_lotus_20} and \ref{fig:reco_ellipses_walnut_120} show that the reconstructions obtained by directly deploying the pretrained network (i.e.\ $\varphi_{\theta_{{\rm s}}^{\ast}}$) on the FBP of the real-measured $\mu$CT data do enjoy good reconstructive properties, but the images tend to be overly-smooth and severely affected by ellipses-like artifacts, which are naturally present in the synthetic training dataset. Indeed, initializing the network's parameters to the pretrained configuration, on both Lotus and Walnut, shows a gain of \SI{5.8}{dB}, and of \SI{9.4}{dB} (\texttt{Sparse 120}), \SI{6.7}{dB} (\texttt{3D Sparse 20}), \SI{2}{dB} (\texttt{3D Sparse 60}) over the FBP.
The pretrained model enjoys high input-robustness, and feature reuse plays a very important role in the EDIP reconstruction.
However, the feature reuse mechanism leads to undesirable hallucinatory behaviors, as evidenced by the ellipses-like artifacts, which is a form of inductive biases induced by the synthetic image class. This also indicates the importance of properly designing the synthetic dataset used in the pretraining stage, from which the features are learned, and the strong dissimilarity between the synthetic training data and real test data may actually deteriorate the performance. In the supplementary materials, we showcase one potential pitfall of the ``supervised pretraining + unsupervised fine-tuning'' paradigm for DIP, resorting to synthetic data generated by a by far too specific and less diverse image class, i.e., human brain images for the supervised learning stage.

\begin{figure*}[ht]
  \centering%
  \begin{minipage}[b]{0.525\textwidth}
  \includegraphics[
    draft=\draftgraphics,width=0.494\textwidth]{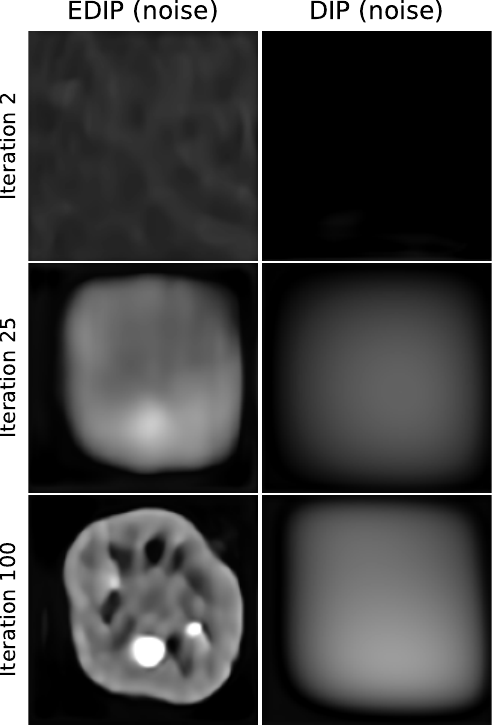}\;\hfill%
  \includegraphics[
    draft=\draftgraphics,width=0.494\textwidth]{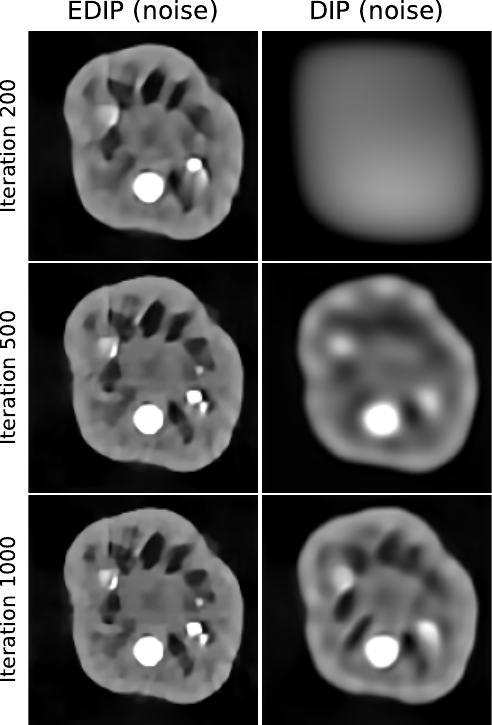}%
  \end{minipage}\hfill%
  \begin{minipage}[b]{0.425\textwidth}
  \centering\small
  \includegraphics[
    draft=\draftgraphics,width=.9\textwidth]{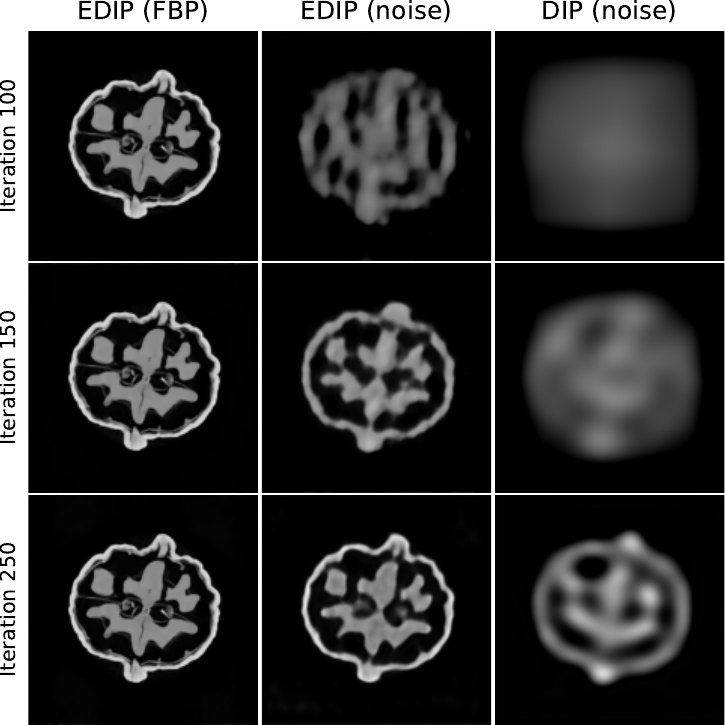}
  \end{minipage}
  \caption{ {Iterates collected throughout the EDIP/DIP reconstruction from Lotus \texttt{Sparse 20} (left) and Walnut \texttt{Sparse 120} (right)}, after different numbers of iterations. A video showing the reconstruction process is available at \href{https://educateddip.github.io/docs.educated_deep_image_prior/}{https://educateddip.github.io/docs.educated\_deep\_image\_prior/}.
    }
  \label{fig:iterates_noise}
\end{figure*}

The knowledge enforced via the synthetic dataset needs to be properly amended so that the reconstructed images recover a more realistic texture. This is achieved at the fine-tuning stage by enforcing the data consistency.
Amending the knowledge acquired via pretraining protects from hallucinations due to (inevitable) distributional shifts, thereby overcoming a well-known drawback of supervised learned reconstructors \cite{antun2020instabilities}.

\paragraph{Investigating Feature Reuse}

In a similar spirit to \cite{neyshabur2020tl}, we feed a noise image to EDIP (trained on pairs of FBP and ground truth image), which makes any visual features learned in the pretraining stage useless.
This allows us to disentangle influencing factors involved in the fine-tuning stage. We consistently observe faster convergence of EDIP with respect to the standard DIP for the Lotus dataset.
EDIP (fed with FBP) still results in faster convergence, which agree well with the intuition that decreasing feature reuse leads to diminishing benefits.
Fig.~\ref{fig:iterates_noise} (left) shows that EDIP remolds the noise image differently compared to the standard DIP.
The learned inductive biases prioritize reshaping the noise image as ellipse-like structures. The model makes an educated reconstruction. The features learned during pretraining are invariant of the input.
The pretrained model is then adapted by enforcing data-consistency via \eqref{eqn:diptv}.

On the Walnut, cf.\ Fig.~\ref{fig:iterates_noise} (right), the benefit of pretraining is less pronounced, if a noise image input is used. This might be due to the fact that the Walnut has a higher resolution and many more fine details, which are not present in the training dataset.
Nonetheless, pretraining can still remold noise input into a walnut faster than DIP, yet the FBP input (used in the pretraining) is even more effective.
These observations fully agree with that for Lotus.

\paragraph{Getting $\theta_{{\rm s}}^{\ast}$ Right}

\begin{figure}[ht]
  \centering%
  \includegraphics[
    draft=\draftgraphics,
    width=\linewidth]{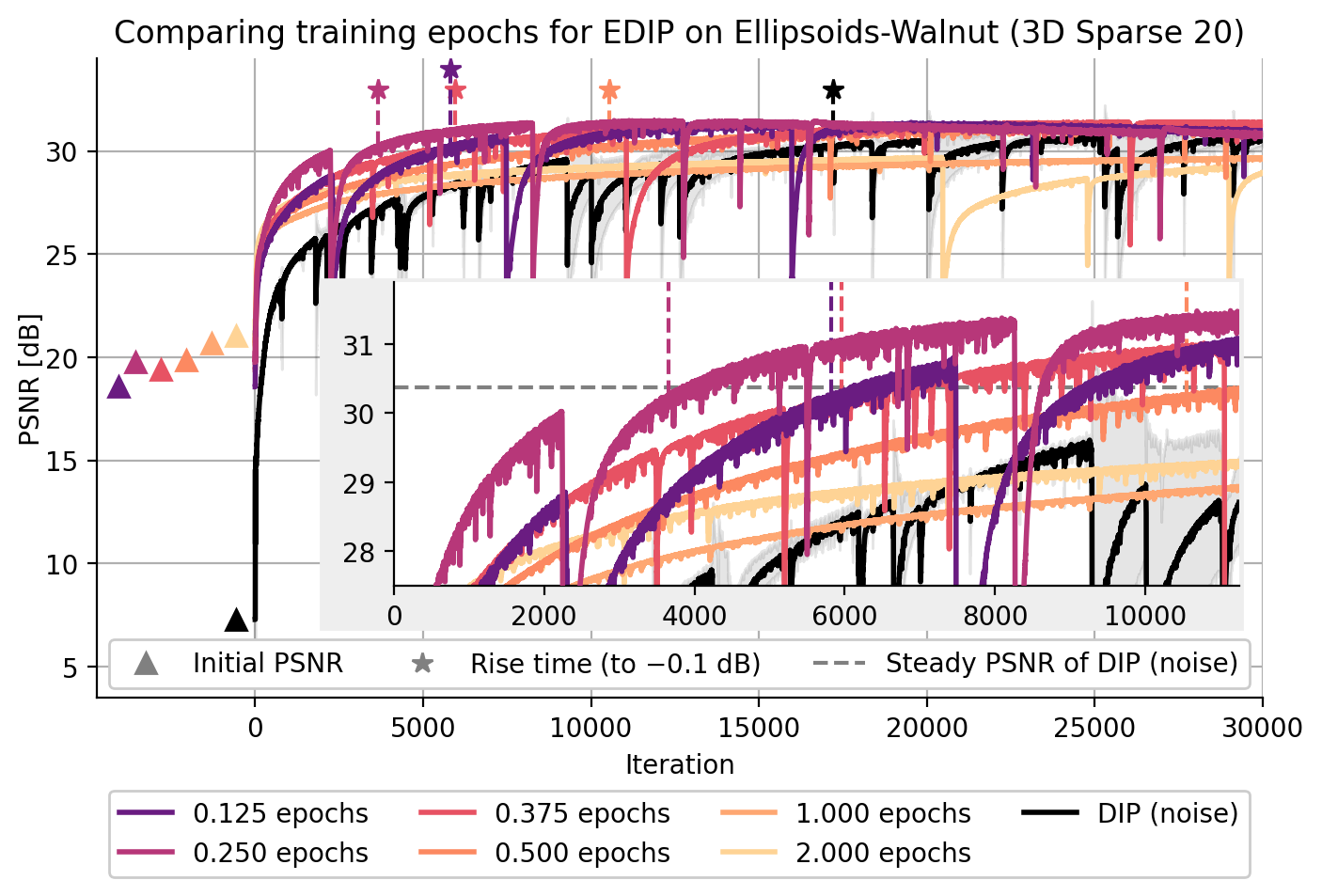}
  \caption{ {The optimization of EDIP using parameters from different checkpoints} for EDIP (FBP)
  on Walnut \texttt{3D Sparse 20}. The symbols $\star$ and $\blacktriangle$
  denote initial PSNR and rise time, respectively, and the horizontal dashed line indicates the steady PSNR of DIP (noise).}
  \label{fig:test_checkpoints_epochs_ellipsoids_walnut_3d}
\end{figure}

The starting point $\theta_{\rm s}^*$ of fine-tuning can impact the adaptation speed.
A selection procedure of $\theta_{\rm s}^*$ is desired to maximize transferable performance (e.g.\ speed-up).
On the 2D setting, pretraining for more epochs (\num{100} vs.\ \num{20}) leads to a faster adaptation.
This is clearly observed on the Lotus, possibly due to the in-distribution nature of the image class with respect to the ellipses dataset.
However, on more complex tasks (\texttt{3D Sparse 20} and \texttt{3D Sparse 60}), extensive pretraining leads to overfitting the image class, and enforcing dataset-specific knowledge appears detrimental to the transfer.
Fig.~\ref{fig:test_checkpoints_epochs_ellipsoids_walnut_3d} shows that extensively pretraining U-Net for \num{2} epochs (i.e.\ \num{64}k gradient updates with \num{32}k ellipsoid volumes), albeit yielding the highest initial PSNR, leads to a sub-optimal convergence: the network output is effectively constrained, as an over-trained $\phi_{\theta^{\ast}}$ after \num{2} epochs has little freedom to amend. This is also observed on the \texttt{3D Sparse 60} setting.

\paragraph{Spectral Evaluation}
We propose a spectral analysis to understand the ``education''  by
linearizing the non-linear forward map $F(\theta) = A \varphi_{\theta}(A^{\dagger}y_\delta)$ at $\theta_{0}$:
\begin{equation*}
  F(\theta) = F(\theta_0)+F'(\theta_0)(\theta-\theta_0),
\end{equation*}
with $F'(\theta_0) = A\varphi'_{\theta_0} \in \mathbb{R}^{m \times p}$ with $\varphi'_{\theta_0}=\partial \varphi_{\theta}/\partial \theta|_{\theta=\theta_{0}} \in \mathbb{R}^{n\times p}$ denoting the Jacobian of the network's output w.r.t.\ $\theta$.
We use the subspace spanned by leading right singular vectors $v_i$ of $F'(\theta_0)$ (i.e.\ with the largest singular values) as a faithful representation of the network's parameter space, which determines the dynamics of the learning process.
Due to the high-dimensionality of the output and parameter spaces, directly computing $\varphi_{\theta_0}'$ is intractable.
We approximate the first $\ell$ singular vectors of $F'(\theta_0)$ via randomized singular value decomposition (rSVD) \cite{halko2011finding, tropp2020randomized}, and proceed in two steps (cf.\ Algorithm \ref{algo:randomisedSVD}):

\begin{figure*}[hbt]
  \centering
  \includegraphics[
    draft=\draftgraphics,
    width=.9\linewidth]{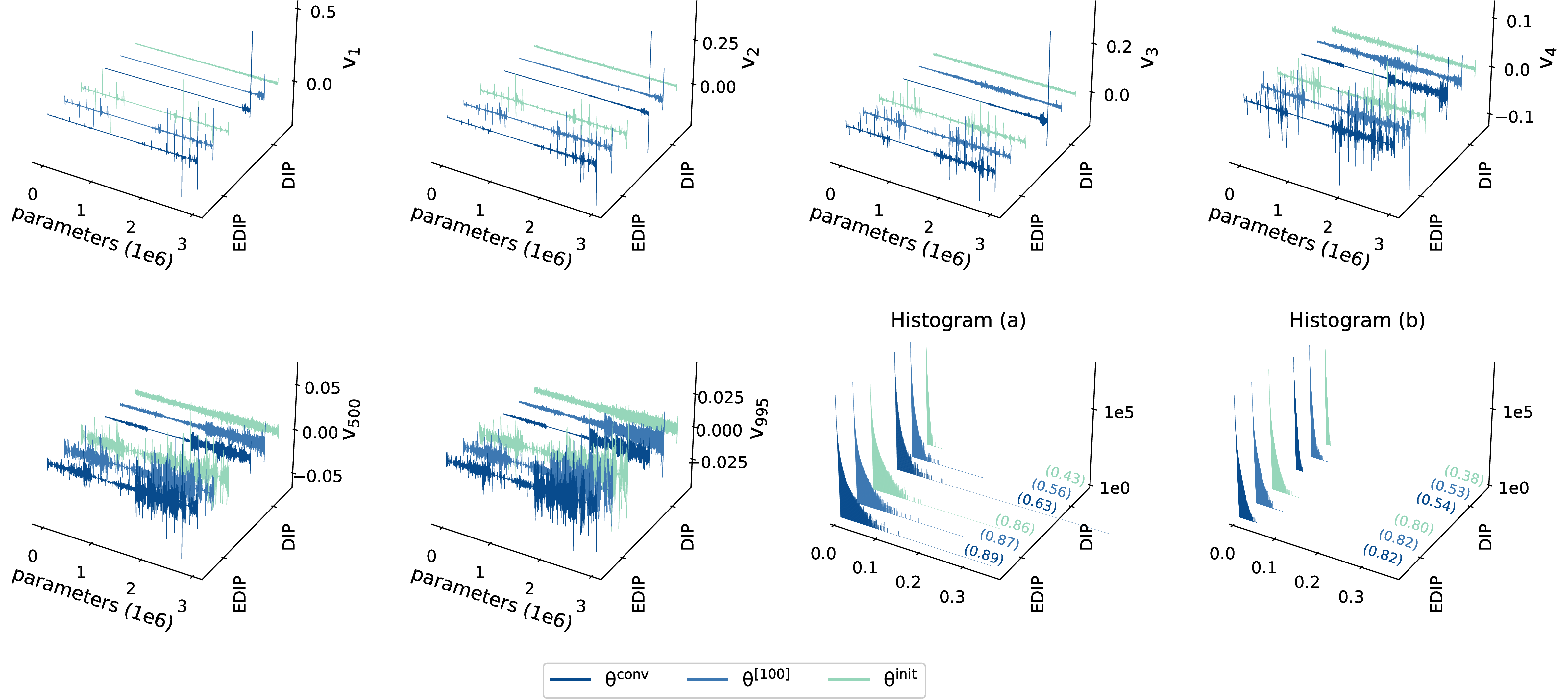}%
  \caption{{The evolution of right singular vectors} of the linearized forward map (i.e., the Jacobian) w.r.t.\ the network parameters $\theta$ for EDIP (FBP) versus DIP (FBP) on Lotus \texttt{Sparse 20} dataset. The parameters are ordered like they occur in the network, i.e.\ lower positions on the parameters axis refer to the encoder while higher positions refer to the decoder. (a) and (b) show mean histograms for the right singular vectors $v_1,\dots,v_{20}$ and $v_{976},\dots,v_{995}$, which represent the low-frequency and high-frequency bands of the singular vectors, respectively; the numbers in brackets denote Hoyer measure of sparsity \cite{hoyer2004sparse, hurley2009comparing}.}
  \label{fig:spct_analysis_vectors}
\end{figure*}

\textit{Stage \#1: Randomized Range Finder.} To construct a subspace capturing most of the action of $F'(\theta_0)$, we draw a Gaussian random matrix $\Omega\in\mathbb{R}^{p\times\ell}$ and form $\bar{F} =  F'(\theta_{0})\Omega \in\mathbb{R}^{m\times\ell}$.
To avoid the direct evaluation of $\varphi' _{\theta_{0}}$, for any column $\omega$ of $\Omega$, we use a finite difference approximation: $    \varphi_{\theta_0}'\omega = (\varphi_{\theta_0+\epsilon\omega}-\varphi_{\theta_0-\epsilon\omega})/(2\epsilon)$, where $\epsilon>0$ is a small constant.
Then we find an orthonormal matrix $Q\in\mathbb{R}^{m\times\ell}$ for the range of $\bar{F}$, using the standard QR factorization \cite{halko2011finding,tropp2020randomized}.

\textit{Stage \#2: Direct SVD.} Next we construct a low-rank matrix $B = Q^{\top}F' (\theta_{0})\in\mathbb{R}^{\ell\times p}$, or equivalently, $B^{\top} = {F' }(\theta_0)^{\top} Q$, which can be computed via backpropagation, and then approximate the singular values and the right singular vectors of $F'(\theta_{0})$ by that of $B \approx U \Sigma V^{\top}$ (with the last few discarded as oversampling: default choice \num{5}). Since the size of $B\in\mathbb{R}^{\ell\times p}$ is much smaller than that of $F'(\theta_0)$, a direct SVD computation is indeed feasible.

\begin{algorithm}[t]
  \caption{rSVD for Linearized Forward Map\label{algo:randomisedSVD}}
  \begin{algorithmic}[1]
    \Require the Jacobian matrix $F' (\theta_{0})$, the target rank $\kappa$, and oversampling parameter $o$
    \State  Draw a $p\times (\ell = \kappa + o)$ Gaussian random matrix $\Omega=(\omega_{ij})$
    \State  Form $\bar{F}=F'(\theta_{0})\Omega$
    \State  Construct an orthonormal basis $Q$ of $\mathrm{range}(\bar{F})$ using  QR decomposition
    \State  Form the matrix $B=Q^{\top}F'(\theta_0)$
    \State  Compute the SVD of $B=W\tilde \Sigma_\ell \tilde V_\ell$
    \State  {Return} $\tilde \Sigma_\kappa,\tilde V_\kappa$
  \end{algorithmic}
\end{algorithm}

\begin{figure}[t]
    \centering
    \includegraphics[
    draft=\draftgraphics,width=.75\linewidth, trim={0 0 0 0.85cm},clip
    ]{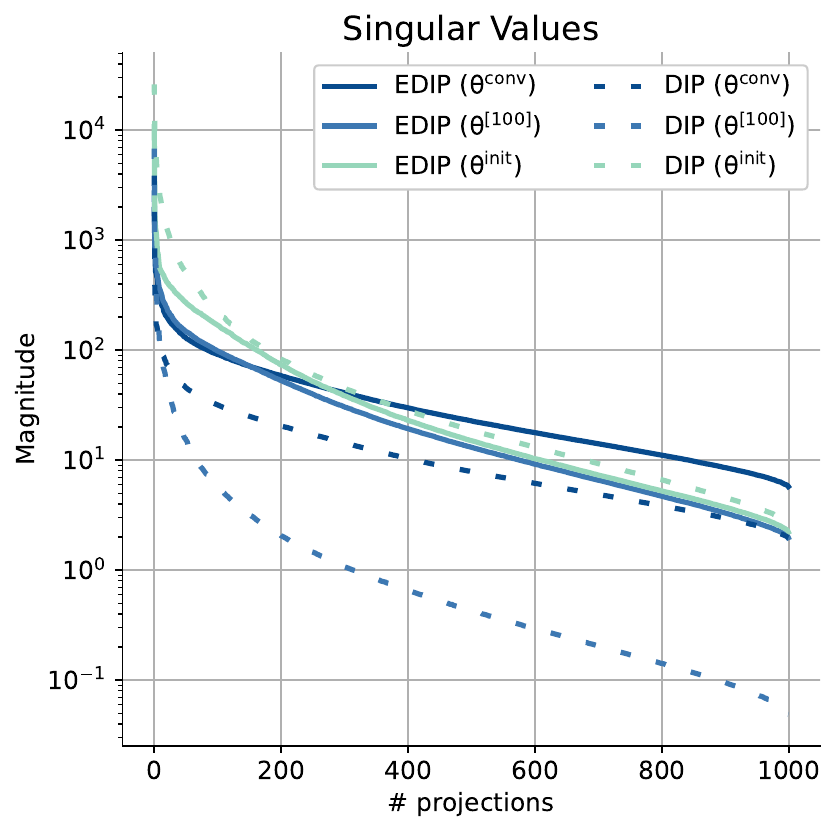}
    \caption{The {singular values} of the linearized forward map (i.e., the Jacobian) w.r.t. the network parameters $\theta$, at $\theta^{{\rm conv}}$ and $\theta^{{\rm init}}$ for EDIP (FBP) and DIP (noise) on Lotus \texttt{Sparse 20} data.}
    \label{fig:singular_values}
\end{figure}

In the analysis, we use the \num{995} leading singular values and the corresponding right singular vectors, which are used to represent the parameters.
We investigate EDIP and DIP, both receiving the FBP as the input, and respectively approximate the singular vectors of the Jacobian, evaluated at three checkpoints during the fine-tuning stage ($\theta^{ {\rm init} }$, $\theta^{{\rm [100]}}$, $\theta^{{\rm conv}}$). Fig.~\ref{fig:spct_analysis_vectors} summarizes our empirical findings, showing the right singular values component-wise plots and Hoyer measure of sparsity \cite{hoyer2004sparse, hurley2009comparing}. Hoyer measure takes a value $0$ if the vector is dense (i.e.\ all components are equal and non-zero) and $1$ if it is 1-sparse. The histogram is computed for the two sets of singular vectors, i.e., $\{v_1,\ldots,v_{20}\}$ and $\{v_{976},\ldots,v_{995}\}$, separately, in order to examine the behavior at the different frequency bands.
For DIP, the singular vectors are equally distributed throughout the parameter space (at $\theta^{{\rm init}}$) and across different singular values. During the fine-tuning stage, we observe a ``relevance shift'' towards the decoder's parameters (at $\theta^{[100]}$ and at $\theta^{{\rm conv}}$, respectively), which is attributed to the fact that the heavy-lifting of representing the target image is actually done by the decoder. This is also consistent with our experimental findings: EDIP-FE shows very similar reconstruction properties to EDIP.
For EDIP, pretraining enforces a hierarchical structure (i.e.\ a relevance shift towards the decoder's parameters), and again sparsity is clearly observed after pretraining. Pretraining strongly promotes sparsity in the basis of the parameter space, which is further promoted in the fine-tuning stage. This is observed in both low and high frequency bands. It is worth noting that even though individual singular vectors exhibit sparsity, the parameter vector $\theta$ does not necessary exhibit a very high level of sparsity, since the linear combination might spoil it. The emerging sparsity during pretraining may facilitate pruning the network, which however is still to be systematically explored.

Interestingly, pretraining also induces a shift in the singular values spectrum, and the overall behavior does not vary much during adaptation, cf.\ Fig.~\ref{fig:singular_values}. In contrast, for DIP, the shift is quite dramatic in terms of the magnitude, as well as the number of singular values larger than a given threshold.
This may offer an explanation to the very different dynamics of the optimization scheme for the pretrained model and the model trained from scratch: in the linearized regime, the singular value spectrum essentially determines the dynamics of gradient type algorithms (along with the learning rate), and the dramatic shift of the singular value spectrum of the DIP Jacobian may have contributed to the undesirable unsteady convergence behavior of DIP and indicates the necessity of carefully tuning the learning rate schedule in order to achieve a stable convergence behavior.

\section{Conclusions}

Our work advances unsupervised deep learning-based tomographic reconstruction.
We develop a two-stage learning paradigm for accelerating DIP in image reconstruction. It consists of a supervised pretraining stage on a simulated dataset to educate DIP and then a fine-tuning stage which adapts the network parameters to a single test image.
The extensive experimental evaluation clearly shows that pretraining on simulated data can significantly speed up, and stabilize DIP reconstruction for 2D / 3D real-measured sparse-view $\mu$CT.
The empirical study also indicates that the pretraining stage can facilitate learning a suitable feature representation, and that adapting only the decoder's parameters during the fine-tuning stage is sufficient to ensure good reconstruction accuracy.
The novel spectral analysis of the linearized model indicates a strong correlation of the sparsity pattern with the pretraining, and a drastically different shift of the singular values spectrum for the standard DIP and the educated version.

There are several avenues for further research. First, there are other techniques for learning a good 
initialization for neural networks, e.g., model-agnostic meta-learning (MAML) \cite{FinnAbbeel:2017} and adversarial pretraining \cite{goodfellow2014explaining,yi2021improved}. These strategies are also promising, but their full potentials are yet to be explored within the context of DIP reconstruction.
In the spirit of ANIL (Almost No Inner Loop) \cite{raghu2019rapid}, we would suggest using a variant that simplifies the inner loop optimization so as to improve the scalability of MAML.
Second, given the emerging sparsity pattern in singular vectors, it is natural to ask whether one can exploit for even faster adaptation, e.g., via pruning or optimizing in low-dimensional subspaces. Third, the proposal utilizes the specific forward operator in the pretraining stage, and hence the pretrained neural network is specialized, where specialization to the   target task is believed to be helpful. However, addressing multiple settings (e.g., different imaging modalities and multiple image classes) simultaneously is of course of interest.

\section*{Acknowledgment}

R.B. was supported by the i4health PhD studentship (UK EPSRC EP/S021930/1), and by The Alan Turing Institute under the UK EPSRC grant EP/N510129/1. J.L., M.S., and A.D. were funded by the German Research Foundation (DFG; GRK 2224/1). J.L. and M.S. additionally acknowledge support from the DELETO project funded by the Federal Ministry of Education and Research (BMBF, project number 05M20LBB). A.D. further acknowledges support from the Klaus Tschira Stiftung via the project MALDISTAR (project number 00.010.2019). A.H. acknowledges funding from the Academy of Finland projects 338408, 336796, 334817. P.M. was supported by the Sino-German Center for Research Promotion (CDZ) via the Mobility Programme 2021: Inverse Problems -- Theories, Methods and Implementations (IP--TMI). The research of B.J. is supported by UK EPSRC grants EP/T000864/1 and  EP/V026259/1.

\bibliographystyle{IEEEtran}
\bibliography{references}

\ifIncludeSupplementary
\cleardoublepage

\begin{appendices}

\section{$\mu$CT Measurement data}\label{app:data}

\subsection{Cone-Beam Geometry}

On the Lotus root, we employ the sparse matrix provided with the dataset.
For the 2D Walnut setting, a sparse matrix resembling the 2D cone-beam projection is constructed from the ASTRA geometry, by selecting a single volume slice, and a suitable subset of the 3D cone-beam projection lines. This is a non-standard 2D fan-beam setting: (i) the rotation axis is slightly tilted; (ii) the voxels / pixels are weighted according to the 3D projections, which differs from the 2D projection weighting. Specifically, in the integration of the beams for each detector ``pixel'', the contributing area / interval is spreading in two vs.\ one dimension(s) with increasing distance from the source, so the beam density decreases antiproportionally to the squared distance vs.\ antiproportionally to the distance.
For the 3D Walnut settings, ASTRA's direct projection routines are employed via tomosipo.
The backward gradients are approximated by back-projection.
The geometry definition has been adapted to match the sub-sampling applied to the volume and the measurements.

\subsection{X-ray Walnut Details}\label{sec:walnut_measurement_details}

From the collection of \num{42} Walnuts, we consider measurements of Walnut \num{1} taken with source position (or orbit) 2. The slice with offset $+\SI{3}{px}$ from the middle slice (i.e.\ zero-based index \num{253}) is selected for the 2D reconstruction task.
A subset of projection values is determined from the provided ASTRA geometry by computing the 3D forward projection of a mask, containing ones for the selected 2D slice and zeros for all other voxels.
We choose one single detector row per column and angle with maximum intensity. %
A sparse matrix representing the forward projection is constructed from the ASTRA forward projection routine for each unit vector, for which the transposed matrix gives an exact adjoint of the Jacobian, used in computing the gradient of \eqref{eqn:diptv}.
The more efficient ASTRA back-projection routine is not directly applicable due to the pseudo-2D geometry: some of the excluded detector rows close to the selected ones contribute to the selected 2D slice in the back-projection.
Another workaround (without matrix assembly) is to copy the measurement values from the selected rows to the neighboring rows (a.k.a.\ edge-mode padding); we use this to compute approximate FDK reconstructions.
For computing the gradient of the data fitting term in \eqref{eqn:diptv}, using the padding followed by the back-projection via ASTRA leads to degraded results, so we use the sparse matrix multiplication instead, which yields accurate gradients.

The implementation and the sparse matrix are available at {\href{https://educateddip.github.io/docs.educated_deep_image_prior/}{https://educateddip.github.io/docs.educated\_deep\_image\_prior/}}.

\newpage
\section{Methodology}

\subsection{2D Network architecture}

\begin{figure}[h]
  \centering%
  \small
  \includegraphics[
    draft=\draftgraphics,
    width=.95\linewidth]{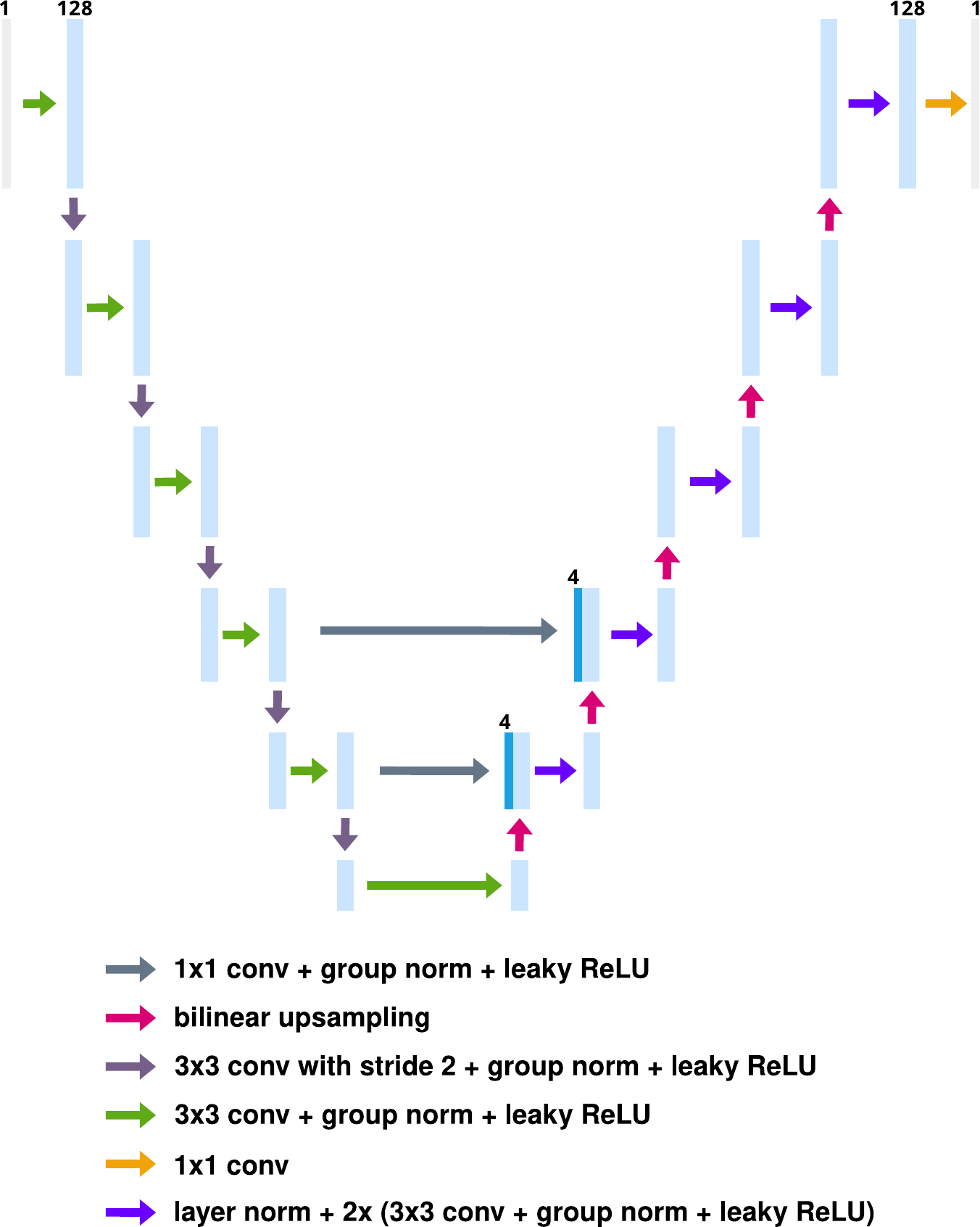}%
  \caption{The architecture of the U-Net used for the 2D experiments. Each light-blue bar corresponds to a multi-channel feature map. Arrows denote the different operations. The number of channels is set to \num{128} at every scale.}
  \label{fig:architecture_of_2D_unet}
\end{figure}

Figure~\ref{fig:architecture_of_2D_unet} shows the network architecture used.
We adopted the architecture proposed by \cite{baguer2020diptv}, with the only difference being that we replace batch-normalization layers with group-normalization layers.

See also Figure~\ref{fig:architecture_of_3D_unet} in the main text showing the U-Net architecture used for the 3D experiments.

\subsection{The Loss}

Our DIP implementation uses the loss function
\begin{align*}
    l'_{{\rm t}}(\theta) := \tfrac{1}{m} \Vert A \varphi_\theta(z) - y_\delta \Vert_2^2 + \gamma'\,\text{TV}(\varphi_\theta(z)),
\end{align*}
with the anisotropic total variation penalty $\text{TV}(x) = \Vert \nabla_h x \Vert_1 + \Vert \nabla_v x \Vert_1$,
where $m$ is the number of detector pixels (length of $y_\delta$) and $\nabla_h$ and $\nabla_v$ are the discrete difference operators in the horizontal and vertical directions, respectively.

\subsection{Hyperparameter Search}
\label{sec:hyperparams}

For each setting, suitable hyperparameters for DIP (noise) are selected by grid search.
While the learning rate \num{1e-4} is (near) optimal in all cases, the TV-regularization parameter $\gamma'$ varies both with the $\mu$CT geometry and between validation data (i.e.\ Shepp-Logan phantom, simulated data) and test data (i.e.\ Lotus or Walnut, real data).

\begin{table}[h]
  \caption{{Hyperparameters} for (E)DIP on {validation} and {test} data. }
  \centering%
  \sisetup{output-exponent-marker = \text{e}}
  \small%
  \begin{tabular}{lrrr}
  {Validation} & Learn. rate & $\gamma'$ & Iters. \\\toprule
  Lotus \texttt{Sparse 20} & \num{1e-4} & \num{4e-5} & \num{37500} \\
  Lotus \texttt{Limited 45} & \num{1e-4} & \num{1e-6} & \num{15000} \\
  $\hookrightarrow$ EDIP (FBP) & \num{1e-4} & \num{4e-6} & \num{10000} \\
  Walnut \texttt{Sparse 120} & \num{1e-4} & \num{2e-7} & \num{50000} \\[0.4em]
  {Test} & Learn. rate & $\gamma'$ & Iters. \\\toprule
  Lotus \texttt{Sparse 20} & \num{1e-4} & \num{1e-4} & \num{10000} \\
  Lotus \texttt{Limited 45} & \num{1e-4} & \num{6.5e-5} & \num{10000} \\
  Walnut \texttt{Sparse 120} & \num{1e-4} & \num{2e-7} & \num{30000} \\
  \shortstack[l]{$\hookrightarrow$ EDIP[-FE] (noise)\strut{}\\[-0.3em]\;\;\;pretrained on ellipses} & \shortstack{\num{5e-4} to \num{1e-4}\strut{}\\[0.5\baselineskip]} & \shortstack{\num{2e-7}\strut{}\\[0.5\baselineskip]} & \shortstack{\num{30000}\strut{}\\[0.5\baselineskip]} \\
  Walnut \texttt{3D Sparse 20} & \num{1e-4} & \num{1e-1} & \num{30000} \\
  Walnut \texttt{3D Sparse 60} & \num{5e-5} & \num{1e-1} & \num{60000} \\
  \end{tabular}
  \label{tab:hyperparams}
\end{table}

The hyperparameters used for DIP and EDIP are listed in Table~\ref{tab:hyperparams}.
The parameters are fine-tuned on DIP (noise), except for the override values specified in the rows starting with ``$\hookrightarrow$''.
For only two cases, we observe the hyperparameters that are optimal for DIP (noise) to be severely sub-optimal for EDIP.
For instance, no speed-up is observed for EDIP (noise), applied to the Walnut \texttt{Sparse 120}, after pretraining on the ellipses dataset,  if the default learning rate \num{1e-4} is used; while a higher learning rate leads to an unstable optimization.
A ``warm-up'' learning rate scheduling with an initial learning rate of \num{5e-4}, which is  linearly decreased to \num{1e-4} over the first \num{5}k iterations reveals a substantial speed-up.
We use the same learning rate scheduling with DIP (noise), but fail to observe any improvement.
Similarly, we observe that validating on the Shepp-Logan phantom for the Lotus \texttt{Limited 45} setting requires the regularization parameter $\gamma'$ to be increased to \num{4e-6} (instead of \num{1e-6}) for EDIP (FBP) to converge.

\begin{table}[h]
  \caption{{Hyperparameters} for {Lotus gold-standard reference} reconstruction.}%
  \centering%
  \sisetup{output-exponent-marker = \text{e}}
  \small%
  \begin{tabular}{lrrr}
  {Reference} & Learn. rate & $\gamma'$ & Iters. \\\toprule
  Lotus (full \num{120}) TV & 1e-3 & \num{5e-5} & \num{1000} \\
  \end{tabular}
  \label{tab:hyperparams_lotus_ground_truth}
\end{table}

\begin{table}[h]
  \caption{{Hyperparameters} for {TV baselines} on test data.}
  \centering%
  \sisetup{output-exponent-marker = \text{e}}
  \small%
  \begin{tabular}{lrrr}
  {Test} & Learn. rate & $\gamma'$ & Iters. \\\toprule
  Lotus \texttt{Sparse 20} TV & \num{5e-4} & \num{1e-4} & \num{5000} \\
  Lotus \texttt{Limited 45} TV & \num{5e-4} & \num{4e-5} & \num{5000} \\
  Walnut \texttt{Sparse 120} TV & \num{5e-4} & \num{4e-7} & \num{10000} \\
  Walnut \texttt{3D Sparse 20} TV & \num{5e-4} & \num{2e-1} & \num{5000} \\
  Walnut \texttt{3D Sparse 60} TV & \num{5e-4} & \num{1e-1} & \num{5000} \\
  \end{tabular}
  \label{tab:hyperparams_tv_baselines}
\end{table}

\section{Extended Experimental Results}

Here we report additional details about the experiments.

\subsection{The Lotus (Continued)}\label{sec:lotus_continued}

We also include a limited-view setting, named Lotus \texttt{Limited 45}: \num{45} angles, range $[0,135^\circ)$ in steps of $3^\circ$.
Fig.~\ref{fig:pretraining_samples_lotus} shows exemplary reconstructions on the test-fold of the synthetic datasets used for pretraining, for both \texttt{Sparse 20} and \texttt{Limited 45}.
The FBP suffers severe streak artifacts, but the trained U-Net can recover the shapes well.

\begin{figure}[ht]
  \centering%
  \small%
  Ellipses-Lotus \texttt{Sparse 20}\\[0.1em]
  \includegraphics[
    draft=\draftgraphics,
    width=\linewidth]{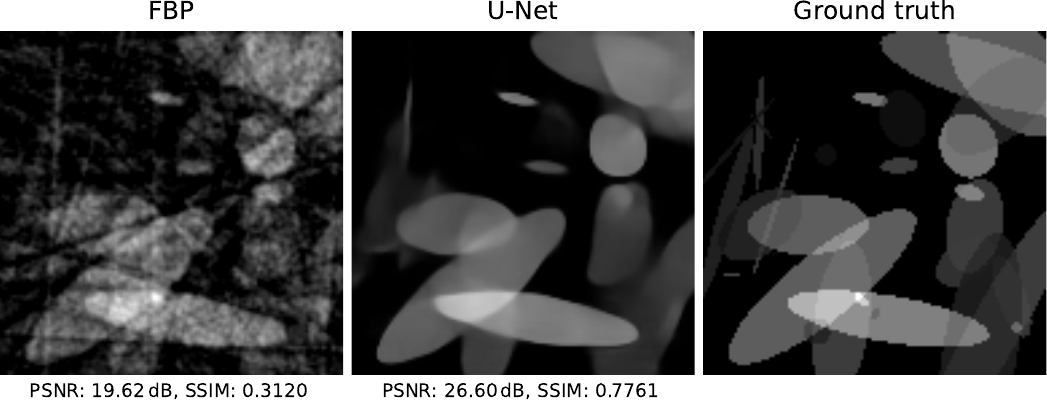}\\[0.1em]
  Ellipses-Lotus \texttt{Limited 45}\\[0.1em]
  \includegraphics[
    draft=\draftgraphics,
    width=\linewidth]{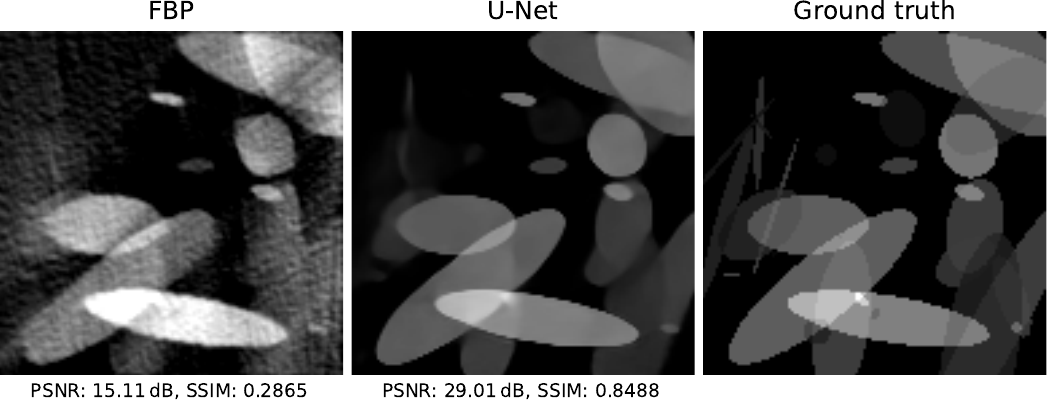}\\[0.1em]
  \caption{{Exemplary reconstructions} from the synthetic training datasets for Lotus \texttt{Sparse 20} and \texttt{Limited 45}.}
  \label{fig:pretraining_samples_lotus}
\end{figure}

The PSNR convergence of EDIP on Lotus root for the \texttt{Limited 45} setting is shown in Fig.~\ref{fig:comp_ellipses_lotus_limited_45-main}; the reconstructions are reported in Fig.~\ref{fig:reco_ellipses_lotus_limited_45}.
These numerical results indicate analogous conclusions as for the case of \texttt{Sparse 20}.

Table~\ref{tab:numerical_eval_results_lotus} reports overall tabular results for Lotus \texttt{Sparse 20} and Lotus \texttt{Limited 45}.
Rise time is defined to be the minimal number of iterations after which the PSNR reaches steady PSNR of DIP (noise) minus \SI{0.1}{dB}.
Both maximum PSNR and steady PSNR are computed using the iteration-wise median PSNR history over the \num{5} repeated runs (varying the random seed). For steady PSNR, the median value of the median PSNR history over the last \num{5}k iterations is considered. The convergence of TV is observed to be very stable, and we report the final PSNR. Initial PSNR is the mean value over the \num{5} repeated runs.

\begin{figure}[]
  \centering%
  \includegraphics[
    trim={0 0 0 0}, clip,
    draft=\draftgraphics, width=\linewidth]{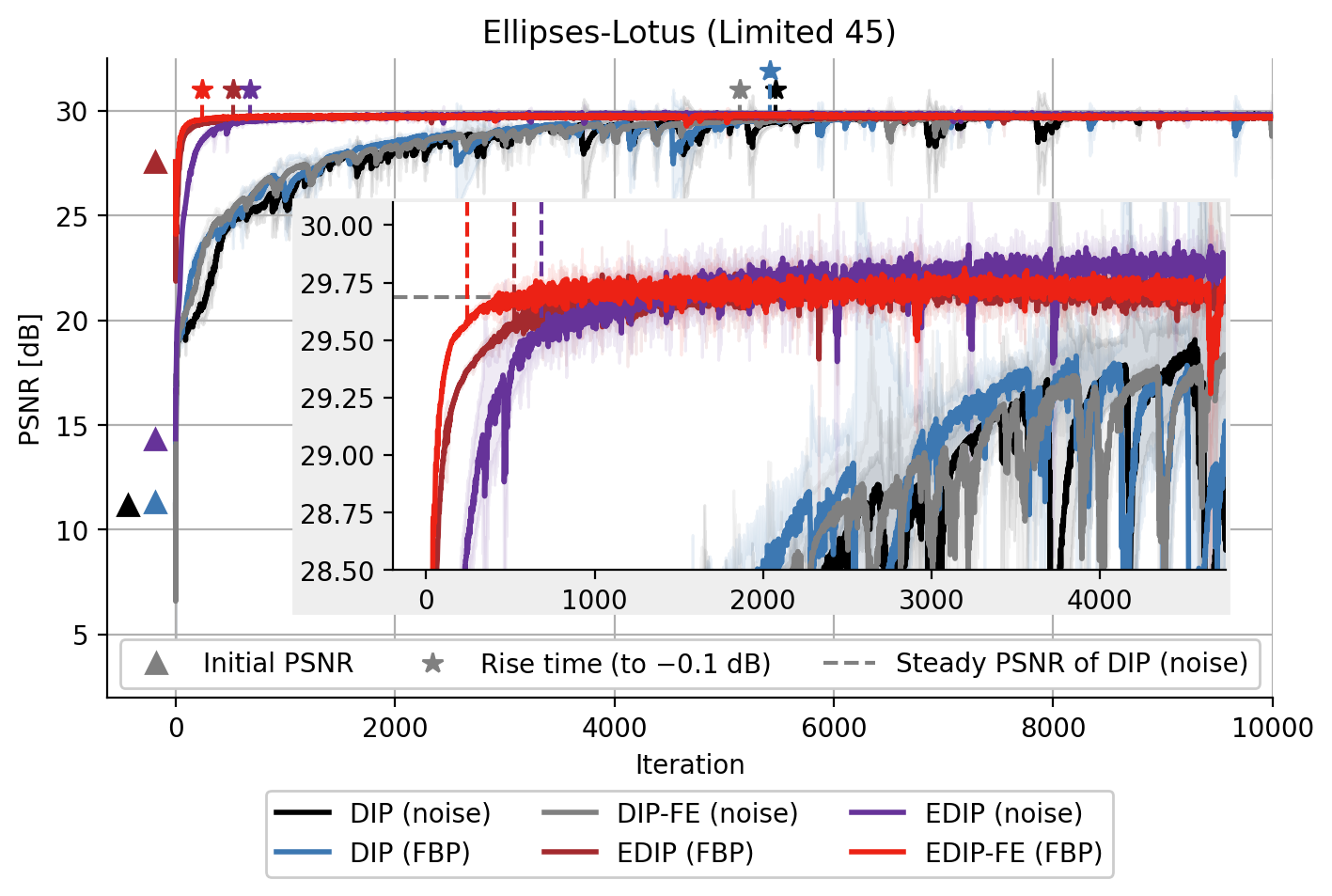}
  \caption{ {The optimization of EDIP}
  versus DIP on Lotus \texttt{Limited 45}.
  All traces are the mean PSNR of \num{5} runs (varying the seed). The notations $\blacktriangle$ and $\star$ denote the initial PSNR and rise time, respectively, and the horizontal dashed line indicates steady PSNR of DIP (noise).}
  \label{fig:comp_ellipses_lotus_limited_45-main}
\end{figure}

\begin{figure}[]
  \centering%
  \small
  \includegraphics[
    draft=\draftgraphics,
    width=.95\linewidth]{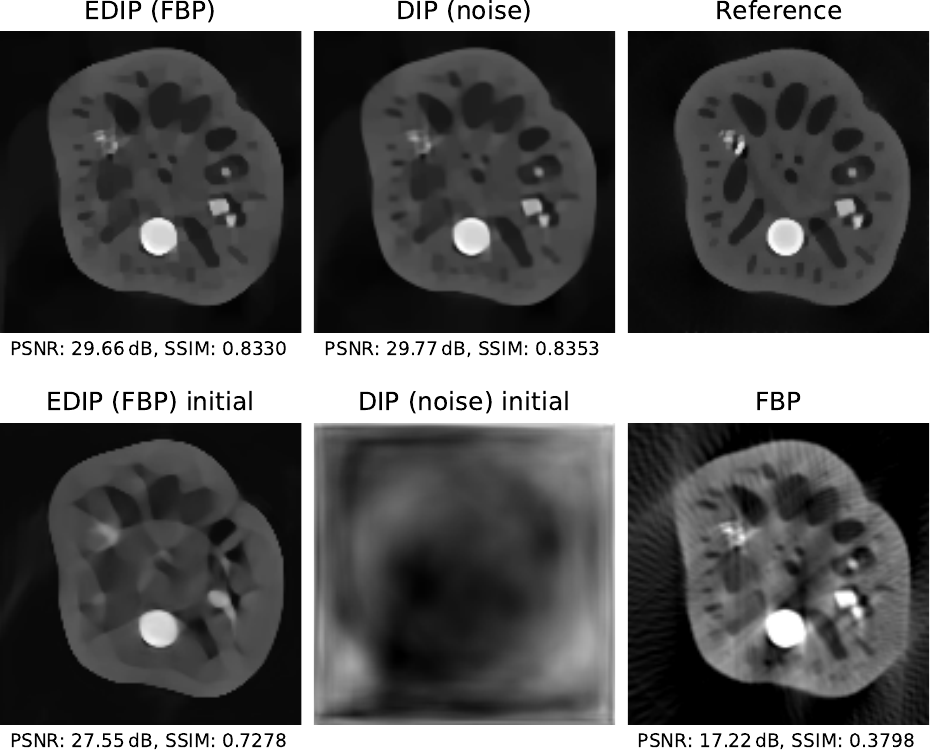}%
  \caption{ {Lotus reconstruction of EDIP} versus DIP on Lotus \texttt{Limited 45} data. From the \num{5} runs (varying the seed), the one with the (closest to) median PSNR was selected for each method.
  The reported reconstructions are the best reconstruction (i.e.\ reconstruction at the minimum loss value).}
  \label{fig:reco_ellipses_lotus_limited_45}
\end{figure}

It is observed that pretraining can substantially accelerate and stabilize the convergence of DIP.
The acceleration factor is more substantial, when considering the FBP as input.
The maximum PSNR (Max.\ PSNR) and steady PSNR suggest that pretraining also improves the reconstruction quality.
The performance of EDIP-FE is largely comparable to EDIP.

\begin{table*}[t]
  \caption{{Quantitative evaluation} for Lotus \texttt{Sparse 20} and Lotus \texttt{Limited 45} with EDIP being pretrained on ellipses data.
  }
  \small%
  {\centering%
  \setlength{\tabcolsep}{4.6pt}%
  \setlength\fboxsep{0pt}%
   \label{tab:numerical_eval_results_lotus}
  \resizebox{\textwidth}{!}{%
  \begin{tabular}{l@{\extracolsep{4pt}}ccccccccc}
  \strut{}Ellipses-Lotus \hspace{-10em} & \multicolumn{4}{c}{\texttt{Sparse 20}} & \multicolumn{4}{c}{\texttt{Limited 45}}\\
  \cline{1-1}\cline{2-5}\cline{6-9}\\[-0.7em]
  & \shortstack{\strut{}Rise time} & \shortstack{\strut{} (Max PSNR; iters)} &  \shortstack{\strut{}Steady PSNR} &  \shortstack{\strut{}Init PSNR} & \shortstack{\strut{}Rise time} & \shortstack{\strut{} (Max PSNR; iters)} &  \shortstack{\strut{}Steady PSNR} &  \shortstack{\strut{}Init PSNR} \\\midrule
  DIP (noise) & \num{3848} & (\num{31.17}; \num{8846}) & \num{31.10} & \num{11.17} & \num{5470} & (\num{29.85}; \num{9690}) & \num{29.69} & \num{11.17}\\
  DIP (FBP) & \num{3622} & (\num{31.25}; \num{8813}) & \num{31.17} & \num{11.33} & \num{5419} & (\num{29.84}; \num{8898}) & \num{29.69} & \num{11.32}\\
  DIP-FE (noise) & \num{6118} & (\num{31.10}; \num{9818}) & \num{31.00} & \num{11.17} & \num{5142} & (\num{29.82}; \num{8884}) & \num{29.69} & \num{11.17}\\
  DIP-FE (FBP) & \num{4516} & (\num{31.19}; \num{7677}) & \num{31.13} & \num{11.33} & \num{5056} & (\num{29.83}; \num{9891}) & \num{29.67} & \num{11.32}\\
  EDIP (FBP) & \hphantom{0}\hlbest{\num{195}} & (\hlbest{\num{31.65}}; \hphantom{\num{0}}\num{981}) & \num{31.21} & \hlbest{\num{27.04}} & \hphantom{0}\hlbest{\num{524}} & (\num{29.83}; \num{2734}) & \num{29.68} & \hlbest{\num{27.55}}\\
  EDIP (noise) & \hphantom{0}\num{723} & (\num{31.53}; \num{3548}) & \hlbest{\num{31.39}} & \num{14.28} & \hphantom{0}\num{682} & (\hlbest{\num{29.94}}; \num{4445}) & \hlbest{\num{29.80}} & \num{14.34}\\
  EDIP-FE (FBP) & \hphantom{0}\hlbest{\num{226}} & (\hlbest{\num{31.59}}; \num{1421}) & \num{31.26} & \hlbest{\num{27.04}} & \hphantom{0}\hlbest{\num{245}} & (\num{29.85}; \num{5533}) & \num{29.72} & \hlbest{\num{27.55}}\\
  EDIP-FE (noise) & \num{1414} & (\num{31.46}; \num{4278}) & \hlbest{\num{31.39}} & \num{14.28} & \num{1279} & (\hlbest{\num{29.95}}; \num{7095}) & \hlbest{\num{29.86}} & \num{14.34}\\
  TV & -- & -- & \num{30.73} & -- & -- & -- & \num{29.62} & -- \\
  \bottomrule
  \end{tabular}%
  }\\}
\end{table*}

Fig.~\ref{fig:best_psnr_and_loss_ellipses_lotus_20} shows the convergence of the loss in \eqref{eqn:diptv} and of the PSNR, where the PSNR is computed using the network output with minimum loss reached until the current iteration.
Using the minimum loss output is a practical way to overcome the instability of DIP optimization, clearly observed in the plots with the raw data in the main analysis.
Pretraining greatly accelerates and stabilizes subsequent unsupervised training of EDIP, when compared to the standard DIP.
This indicates a more favorable optimization landscape of EDIP~/~EDIP-FE than that of DIP.
A stable convergence in practice is important for designing stopping rules for DIP~/~EDIP.

\subsection{The Walnut (Continued)}\label{sec:walnut_continued}

The quantitative results in Table~\ref{tab:numerical_eval_results_walnut} validate our findings on the Lotus root.
See also Figs.~\ref{fig:test_checkpoints_epochs_ellipsoids_walnut_3d_60} and \ref{fig:pretraining_samples_ellipsoids_walnut_3d_sparse-20}--\ref{fig:pretraining_samples_ellipsoids_walnut_3d} for convergence behavior and exemplary reconstructions.

\begin{table*}[]
  \centering
  \small%
  \caption{{Quantitative evaluation} for Walnut \texttt{Sparse 120} with EDIP being pretrained on ellipses data. For the experiments marked with ``\textsuperscript{*}'' a higher initial learning rate was used (see Table~\ref{tab:hyperparams}).}
  \setlength{\tabcolsep}{4.6pt}%
  \setlength\fboxsep{0pt}%
  \resizebox{0.55\textwidth}{!}{%
  \begin{tabular}{l@{\extracolsep{4pt}}cccc}
   \strut{}Ellipses-Walnut \texttt{Sparse 120} \hspace{-20em} & & & &\\
   \cline{1-1}\cline{2-5}\\[-0.7em]
   & \shortstack{\strut{}Rise time} & \shortstack{\strut{} (Max PSNR; iters)} & \shortstack{\strut{}Steady PSNR} & \shortstack{\strut{}Init PSNR} \\\midrule
  DIP (noise) & \num{20373} & (\num{34.02}; \num{25357}) & \num{33.87} & \hphantom{0}\num{6.88}\\
  DIP (FBP) & \num{13778} & (\num{34.07}; \num{28094}) & \num{33.90} & \hphantom{0}\num{6.26}\\
  DIP-FE (noise) & \num{14289} & (\num{34.02}; \num{23573}) & \num{33.88} & \hphantom{0}\num{6.88}\\
  DIP-FE (FBP) & \num{13421} & (\hlbest{\num{34.19}}; \num{23266}) & \hlbest{\num{33.97}} & \hphantom{0}\num{6.26}\\
  EDIP (FBP) & \hphantom{0\,}\hlbest{\num{4496}} & (\num{33.92}; \num{13039}) & \num{33.56} & \hlbest{\num{25.67}}\\
  EDIP (noise) * & \hphantom{0\,}\num{9561} & (\hlbest{\num{34.12}}; \num{23352}) & \hlbest{\num{33.95}} & \num{12.22}\\
  EDIP-FE (FBP) & \hphantom{0\,}\hlbest{\num{4384}} & (\num{33.91}; \num{12540}) & \num{33.70} & \hlbest{\num{25.67}}\\
  EDIP-FE (noise) * & \num{21760} & (\num{33.89}; \num{29159}) & \num{33.75} & \num{12.22}\\
  TV & -- & -- & \num{31.67} & -- \\
  \bottomrule
  \end{tabular}}
    \label{tab:numerical_eval_results_walnut}
\end{table*}

\begin{table*}[]
  \small%
  {\centering%
     \caption{
        {Quantitative evaluation} for Walnut \texttt{3D Sparse 20} and \texttt{3D Sparse 60} with EDIP being pretrained on ellipsoids data. Both maximum PSNR and steady PSNR are computed using the iteration-wise median PSNR history over \num{3} repeated runs (varying the random seed). For steady PSNR, the median value of the median PSNR history over the last \num{5}k iterations is considered. The convergence of TV is very stable, and we report the final PSNR. Initial PSNR is the mean value over the \num{3} repeated runs. All PSNR values are in \si{dB}.}
  \setlength{\tabcolsep}{4.6pt}%
  \setlength\fboxsep{0pt}%
   \label{tab:tabular_results_3d_walnut_supp}
  \resizebox{\textwidth}{!}{%
  \begin{tabular}{l@{\extracolsep{4pt}}ccccccccc}
  \strut{}Ellipsoids-Walnut \hspace{-10em} & \multicolumn{4}{c}{\texttt{3D Sparse 20}} & \multicolumn{4}{c}{\texttt{3D Sparse 60}}\\
  \cline{1-1}\cline{2-5}\cline{6-9}\\[-0.7em]
  & \shortstack{\strut{}Rise time} & \shortstack{\strut{} (Max PSNR; iters)} &  \shortstack{\strut{}Steady PSNR} &  \shortstack{\strut{}Init PSNR} & \shortstack{\strut{}Rise time} & \shortstack{\strut{} (Max PSNR; iters)} &  \shortstack{\strut{}Steady PSNR} &  \shortstack{\strut{}Init PSNR} \\\midrule
  DIP (noise) & \num{17200} & (\num{30.68}; \num{23477}) & \num{30.37} & \hphantom{0}\num{7.29} & \num{49041} & (\num{34.05}; \num{58901}) & \num{33.93} & \hphantom{0}\num{7.29}\\
  DIP (FBP) & \num{13016} & (\num{31.32}; \num{25063}) & \hlbest{\num{31.19}} & \hphantom{0}\num{8.19} & \num{27873} & (\hlbest{\num{34.37}}; \num{53731}) & \hlbest{\num{34.22}} & \hphantom{0}\num{8.62}\\
  EDIP (FBP) & \hphantom{0\,}\num{3739} & (\hlbest{\num{31.48}}; \num{10689}) & \num{30.94} & \hlbest{\num{19.77}} & \hlbest{\num{11247}} & (\hlbest{\num{34.35}}; \num{40810}) & \hlbest{\num{34.18}} & \hlbest{\num{20.17}}\\
  EDIP-FE (FBP) & \hphantom{0\,}\hlbest{\num{2979}} & (\num{31.38}; \num{10749}) & \num{30.93} & \hlbest{\num{19.77}} & \hlbest{\num{14520}} & (\hlbest{\num{34.33}}; \num{45259}) & \hlbest{\num{34.15}} & \hlbest{\num{20.17}}\\
  TV & -- & -- & \num{28.89} & -- & -- & -- & \num{33.35} & -- \\
  \bottomrule
  \end{tabular}%
  }\\}
\end{table*}

\section{Validating Pretraining}\label{sec:validation}

Different checkpoints are obtained from multiple pretraining runs (varying the random seed), and by collecting checkpoints along the optimization trajectory from each run. We identify the parameters' configuration to be used at test time from these checkpoints by selecting the one with the best performance on a validation set. To this end, we design a reconstructive task based on the Shepp-Logan phantom, a standard test image created to assess reconstruction algorithms.
The phantom is by construction within the ellipses data manifold and shares the same noise distribution of ellipses measurements.
The checkpoint leading to the shortest rise time is selected, among those with a steady PSNR that is at most \SI{0.25}{dB} lower than the maximum reached steady PSNR.

\begin{figure}[]
\begin{minipage}{0.475\textwidth}
  \centering%
  \small%
  \includegraphics[
    draft=\draftgraphics,
    width=\linewidth]{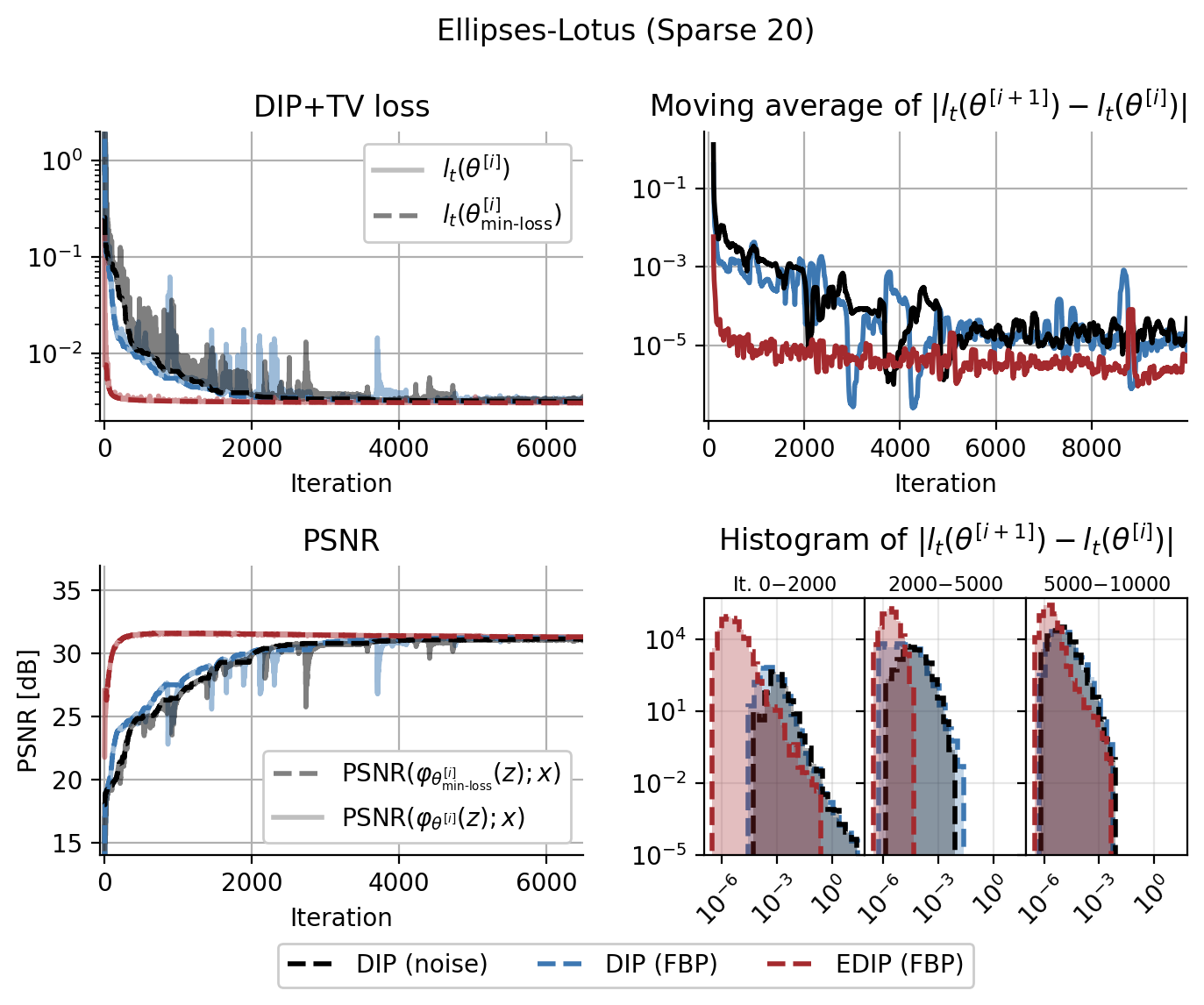}
  \caption{Min-loss and PSNR computed with the min-loss output for Lotus \texttt{Sparse 20} (left).
  Loss variation (i.e.\ $|l_t(\theta^{[i+1]}) - l_t(\theta^{[i]})|$) and respective histograms computed over three intervals (right). The moving average uses a window size of \num{100} iterations.
  }
  \label{fig:best_psnr_and_loss_ellipses_lotus_20}
\end{minipage}\\[1em]
\begin{minipage}{0.475\textwidth}
  \centering%
  \small
  \includegraphics[
    draft=\draftgraphics,
    width=\linewidth]{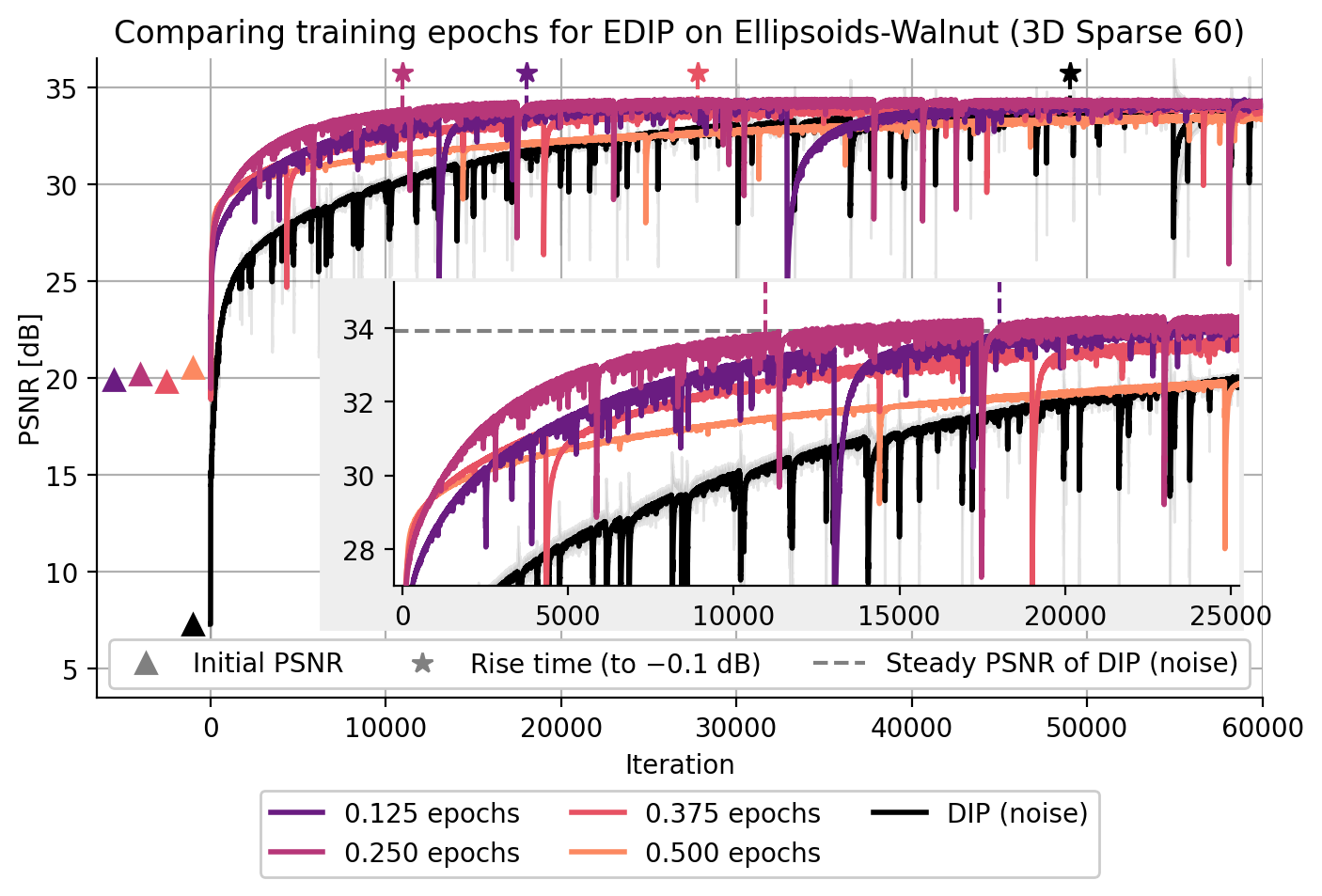}%
  \caption{{The optimization of EDIP}  using different checkpoints for EDIP (FBP) on Walnut \texttt{\texttt{3D Sparse 60}} data. The notations $\blacktriangle$ and $\star$ denote the initial PSNR and rise time, respectively, and the horizontal dashed line indicates steady PSNR of DIP (noise).}  \label{fig:test_checkpoints_epochs_ellipsoids_walnut_3d_60} %
  \end{minipage}
\end{figure}

\begin{figure}[]
\centering
  \begin{minipage}{0.475\textwidth}
    Ellipsoids-Walnut \texttt{3D Sparse 20}\\[0.1em]
  \includegraphics[
    draft=\draftgraphics,
    width=.95\linewidth]{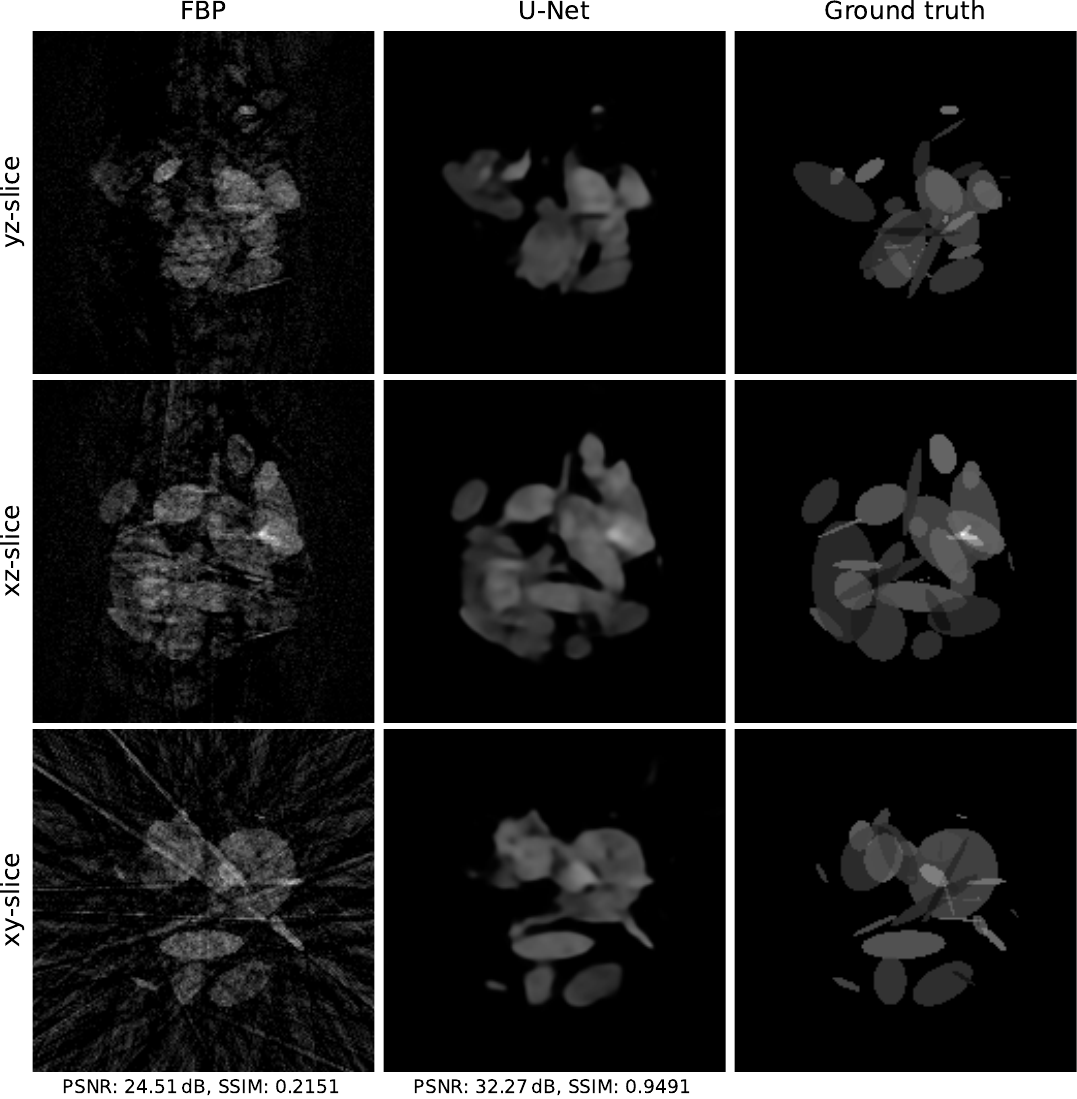}
  \caption{{Exemplary reconstructions} from the synthetic training dataset of ellipsoids images for Walnut \texttt{3D Sparse 20}.}\label{fig:pretraining_samples_ellipsoids_walnut_3d_sparse-20}
  \end{minipage}\\[2em]
  \begin{minipage}{0.475\textwidth}
    Ellipsoids-Walnut \texttt{3D Sparse 60}\\[0.1em]
  \includegraphics[
    draft=\draftgraphics,
    width=.95\linewidth]{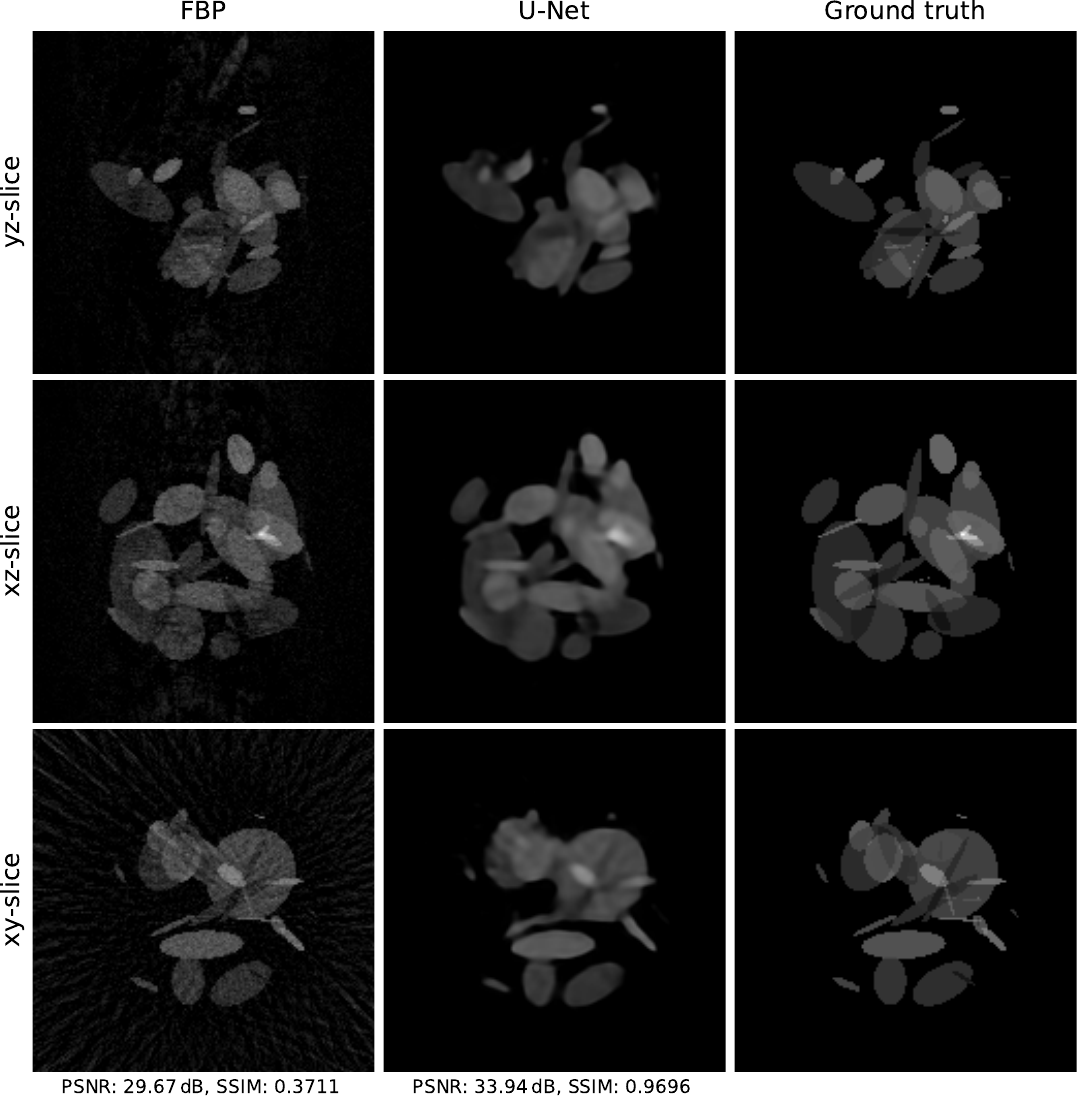}
  \caption{{Exemplary reconstructions} from the synthetic training dataset of ellipsoids images for Walnut \texttt{3D Sparse 60}.}  \label{fig:pretraining_samples_ellipsoids_walnut_3d}
  \end{minipage}

\end{figure}

We repeat the pretraining three times (varying the seed) and collect checkpoints after every \num{20} epochs for Lotus \texttt{Sparse 20} and Lotus \texttt{Limited 45}, training for a maximum of \num{100} epochs.
We also include the checkpoint for which the model shows minimum validation loss.
For Walnut \texttt{Sparse 120} we pretrain for \num{20} epochs, and retain only the minimum validation loss checkpoint.
Fig.~\ref{fig:pretraining_convergence} shows the convergence of the pretraining on the ellipses datasets for the Lotus and the Walnut settings, along with the learning rate scheduling.

\begin{figure*}[]
  \centering%
  \begin{minipage}{0.475\textwidth}
        \includegraphics[
            draft=\draftgraphics,
            width=0.975\linewidth]{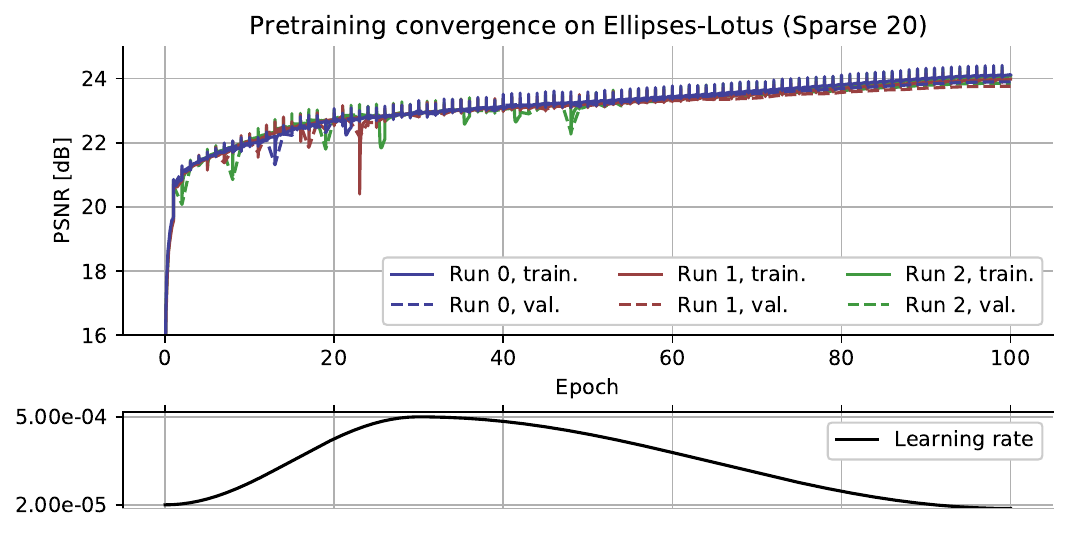}\\[0.8em]
    \end{minipage}
    \begin{minipage}{0.475\textwidth}
        \includegraphics[
            draft=\draftgraphics,
            width=0.975\linewidth]{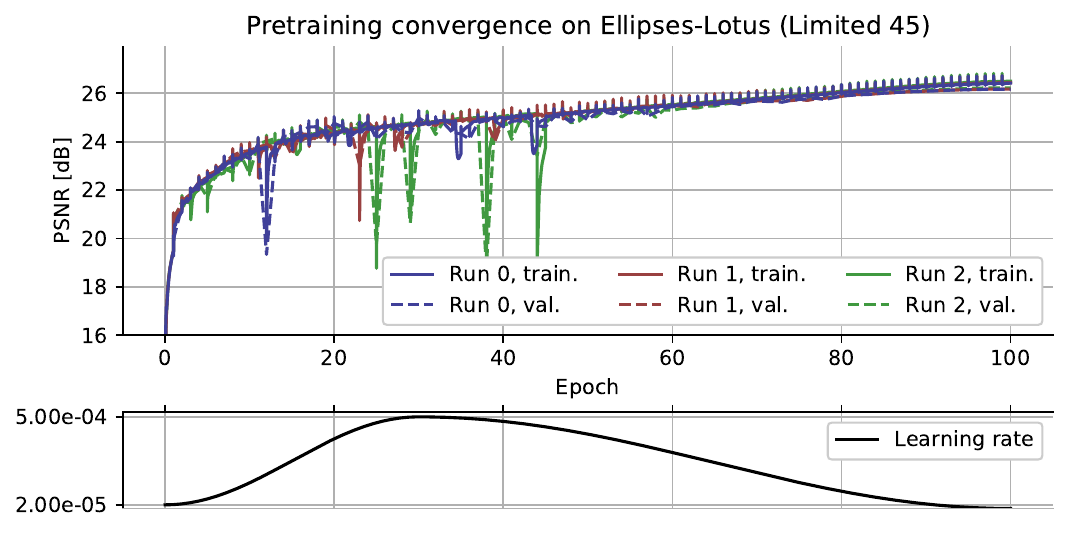}\\[0.8em]
    \end{minipage}
    \includegraphics[
            draft=\draftgraphics,
            width=0.5\linewidth]{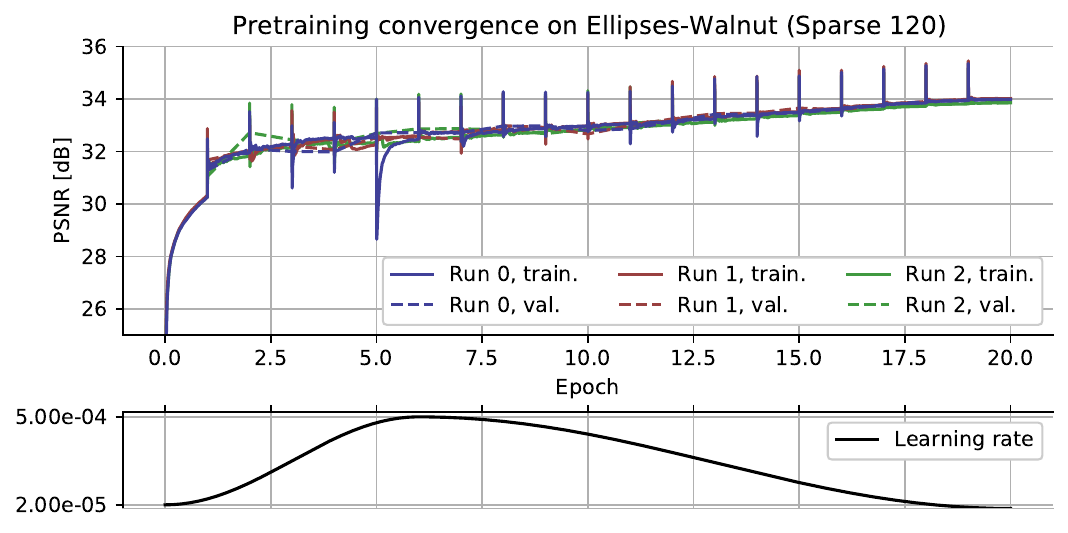}\\[0.8em]
    \begin{minipage}{0.475\textwidth}
        \includegraphics[
            draft=\draftgraphics,
            width=0.975\linewidth]{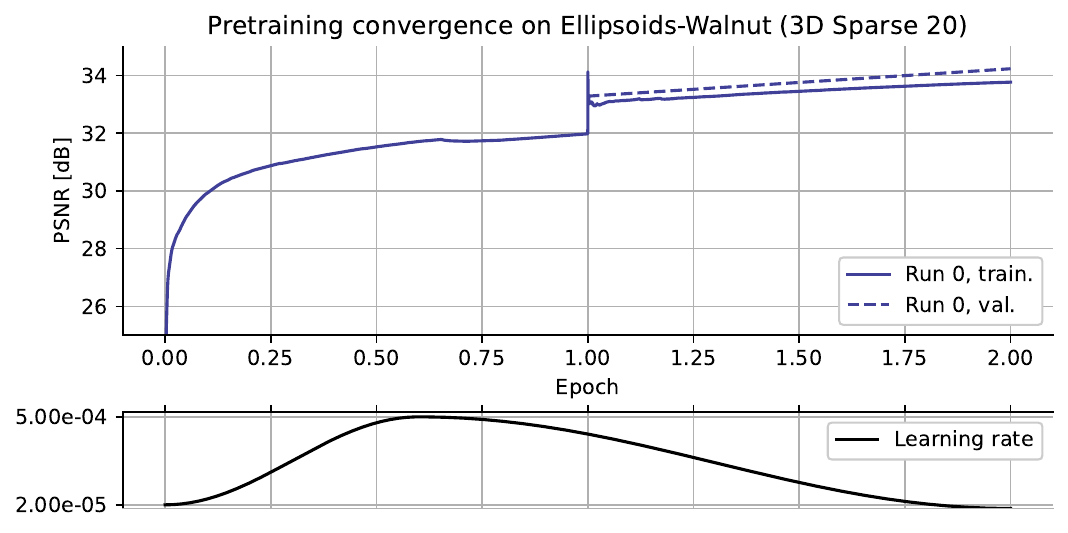}
    \end{minipage}
    \begin{minipage}{0.475\textwidth}
        \includegraphics[
            draft=\draftgraphics,
            width=0.975\linewidth]{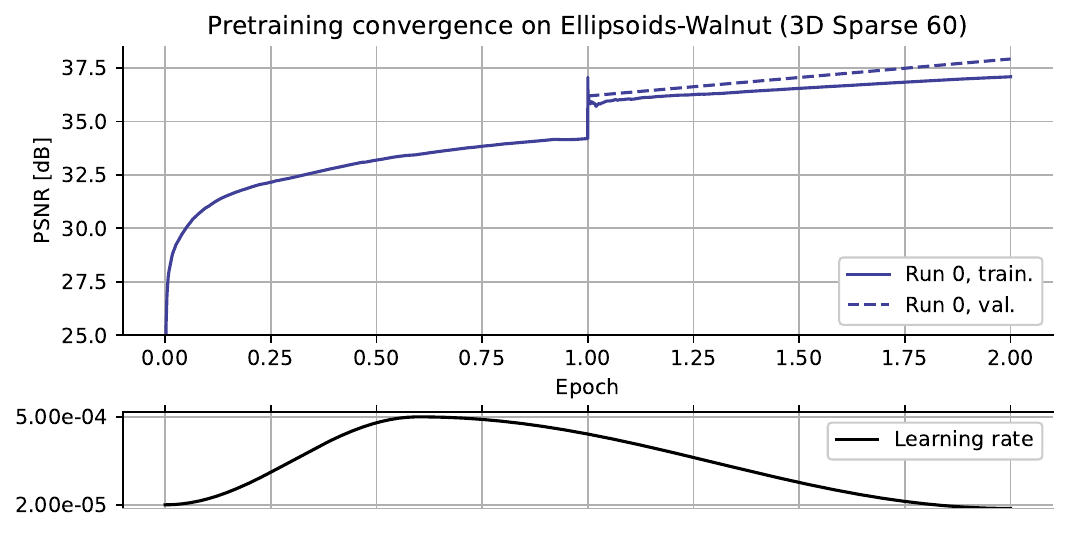}
    \end{minipage}
  \caption{ {Pretraining convergence}. Solid lines show the running mean of the training loss since the start of the respective epoch; dashed lines show the mean validation loss evaluated after each epoch (on a set of \num{3200} held-out images).}
  \label{fig:pretraining_convergence}
\end{figure*}

At the validation stage, each checkpoint is evaluated by performing EDIP fine-tuning on simulated data of the Shepp-Logan phantom.
The validation runs for Lotus \texttt{Sparse 20}, Lotus \texttt{Limited 45}, and Walnut \texttt{Sparse 120} are shown in Figs.~\ref{fig:validation_lotus_20}
and \ref{fig:validation_ellipses_walnut}, respectively.
In the Lotus settings, starting EDIP fine-tuning using checkpoints from a later epoch (e.g.\ \num{60}, \num{80}, \num{100}) is more beneficial.
Nonetheless, even pretraining for fewer epochs  (e.g.\ \num{20}) can already greatly benefit the EDIP fine-tuning, although to a lesser degree.
Pretraining considerably ameliorates the quality of the reconstruction of the Shepp-Logan phantom for both Lotus and Walnut settings. Especially for the Lotus \texttt{Limited 45} setting, it substantially increases the reconstruction quality.

\begin{figure*}[]
  \centering%
  \includegraphics[
    draft=\draftgraphics,
    width=0.4375\linewidth]{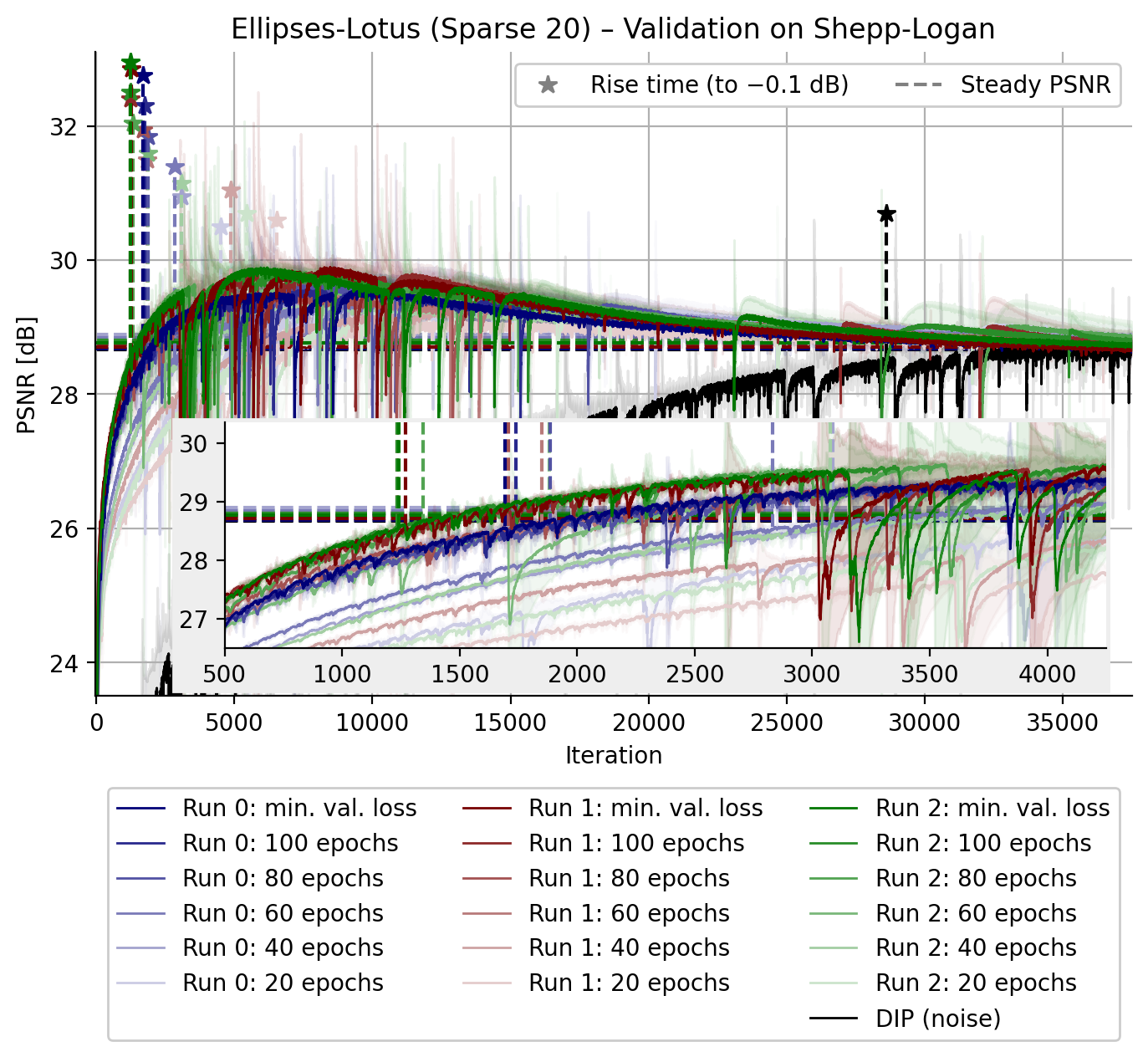}
     \includegraphics[
    draft=\draftgraphics,
    width=0.45\linewidth]{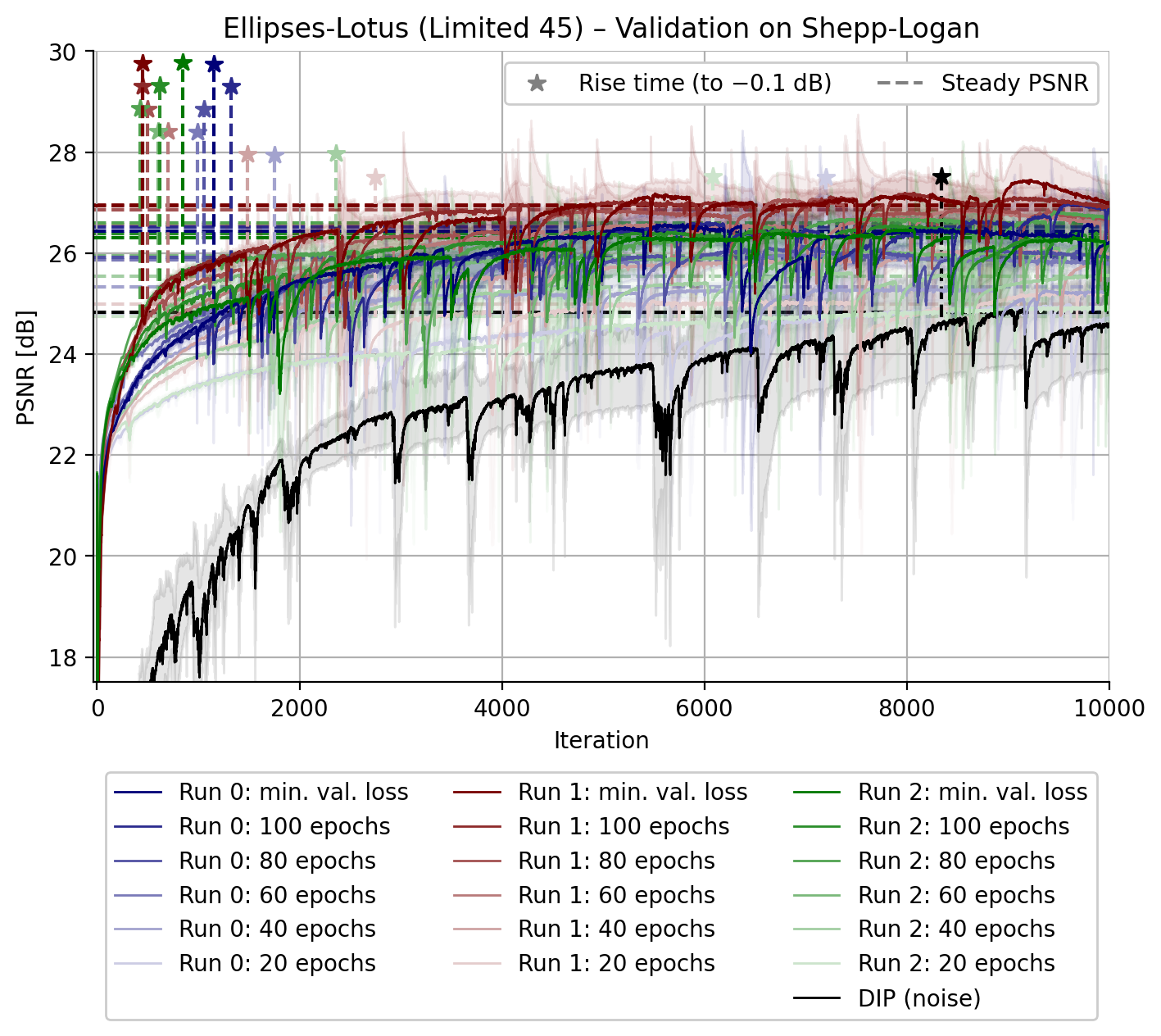}
  \caption{ {Validation runs} on the Shepp-Logan phantom for selecting the initial EDIP (FBP) model parameters for data in the Lotus \texttt{Sparse 20} and \texttt{Limited 45} geometry.  For \texttt{Sparse 20} the model from training run \num{2} after \num{100} epochs is selected because it has the shortest rise time (with a sufficiently high steady PSNR), whilst, for \texttt{Limited 45} run \num{1} after \num{100} epochs is selected. }
  \label{fig:validation_lotus_20}
\end{figure*}

We then investigate whether the selected checkpoints that then are used for the test data --- both the Lotus and the Walnut could be considered an out-of-distribution image class --- are still optimal as we switch from the simulated measurements of the Shepp-Logan phantom to the real-measured test data.
Fig.~\ref{fig:test_checkpoints_lotus_20} and Fig.~\ref{fig:test_checkpoints_ellipses_walnut_120} show the PSNR convergence on the test data using different checkpoints.
While we observe a different behavior between validation and test data, the validation selects one of the best two checkpoint.

\begin{figure}[]
  \centering%
  \includegraphics[
    draft=\draftgraphics,
    width=\linewidth]{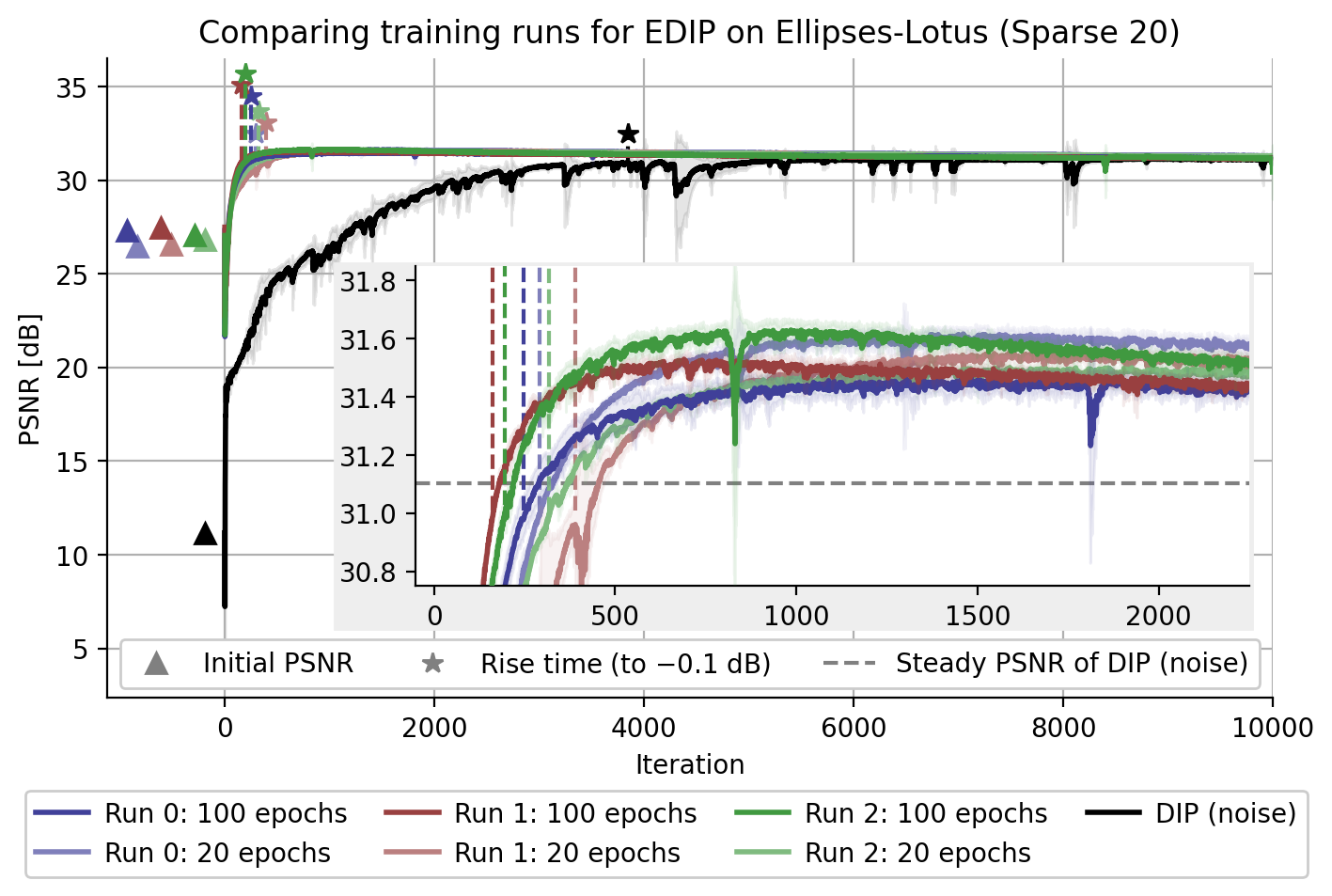}
  \caption{{Optimization of EDIP} using different checkpoints considered during validation (see Fig.~\ref{fig:validation_lotus_20}) for EDIP (FBP) on Lotus \texttt{\texttt{Sparse 20}} data. The parameters from run \num{2} after \num{100} epochs are selected by the validation. The notations $\blacktriangle$ and $\star$ denote the initial PSNR and rise time, respectively, and the horizontal dashed line indicates steady PSNR of DIP (noise).}
  \label{fig:test_checkpoints_lotus_20}
\end{figure}

\begin{figure}[]
  \centering%
  \includegraphics[
    draft=\draftgraphics,
    width=\linewidth]{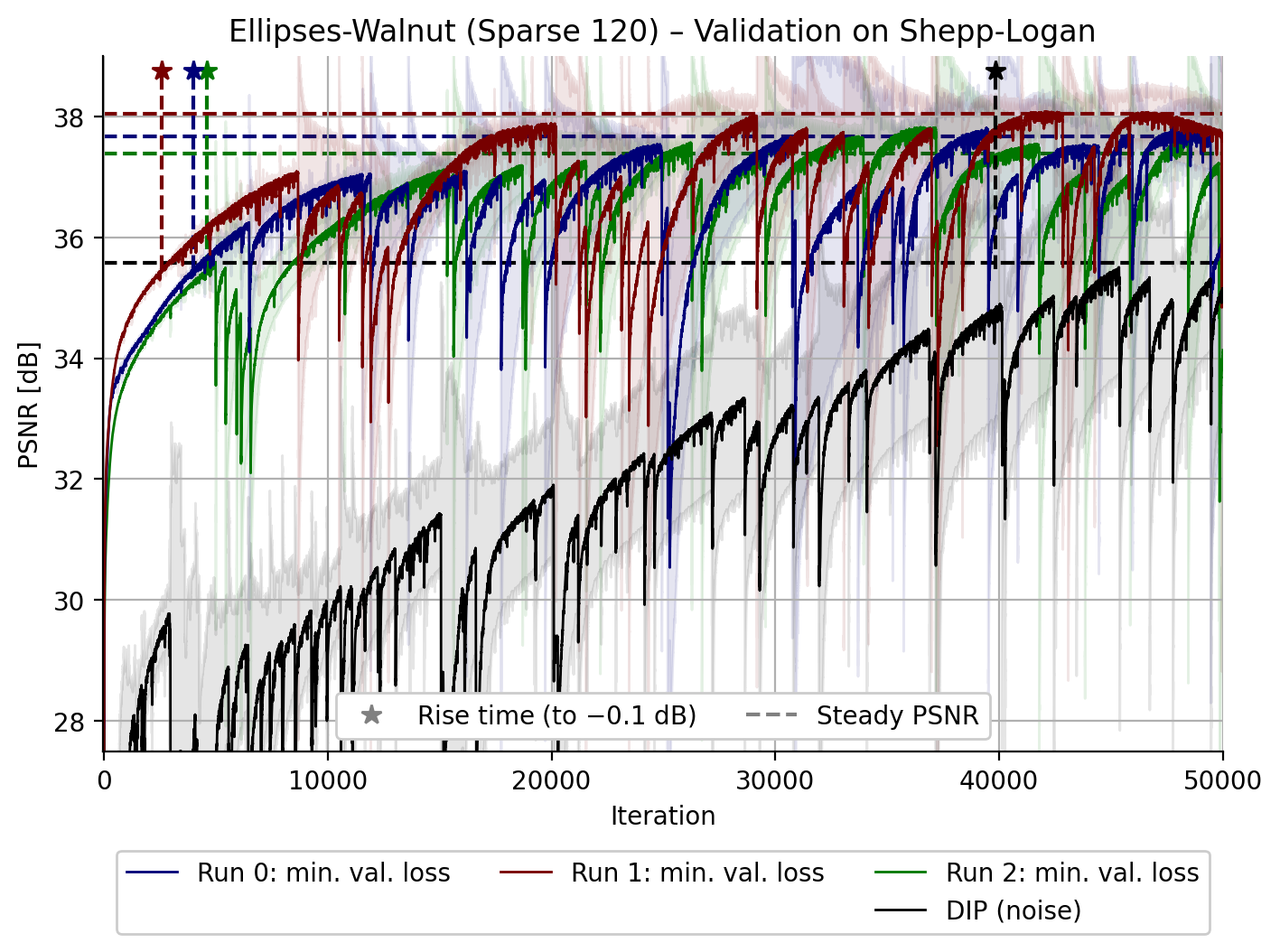}
  \caption{ {Validation runs} on the Shepp-Logan phantom for selecting the initial EDIP (FBP) model parameters for the Walnut \texttt{Sparse 120} geometry. The model from training run \num{1} is selected because it has the shortest rise time (with a high steady PSNR).}
  \label{fig:validation_ellipses_walnut}
\end{figure}

\begin{figure}[]
  \centering%
  \includegraphics[
    draft=\draftgraphics,
    width=\linewidth]{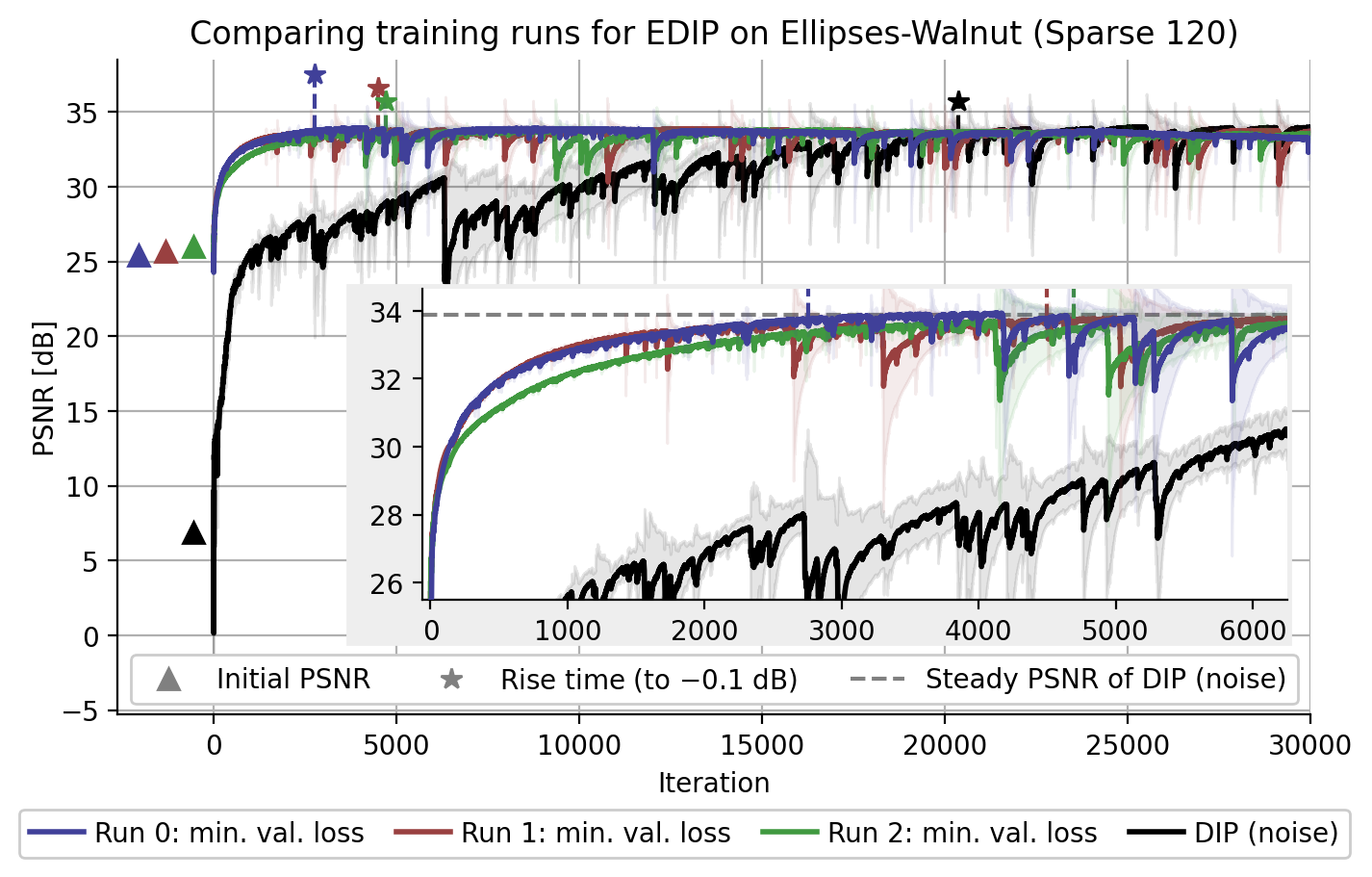}
  \caption{ {Optimization of EDIP} using parameters from different training runs considered during validation (see Fig.~\ref{fig:validation_ellipses_walnut}) for EDIP (FBP), on Walnut \texttt{\texttt{Sparse 120}} data. The parameters from run \num{1} are selected by the validation. The notations $\blacktriangle$ and $\star$ denote the initial PSNR and rise time, respectively, and the horizontal dashed line indicates steady PSNR of DIP (noise).}
  \label{fig:test_checkpoints_ellipses_walnut_120}
\end{figure}

\section{Ablation Study and Limitations}\label{sec:is_not_better}

We showcase one potential pitfall of the ``supervised pretraining + unsupervised fine-tuning'' paradigm for DIP, resorting to a by far too specific and less diverse image class.
Instead of the ellipses dataset, we use human brain images for the supervised learning stage.
We consider MRI images of the human brain from the ACRIN-FMISO-Brain (ACRIN 6684) dataset from \href{https://wiki.cancerimagingarchive.net/x/kQIGAg}{https://wiki.cancerimagingarchive.net/x/kQIGAg}. %
For the synthetic dataset, we normalize the extracted 2D slices and (mis)interpret the values to be X-ray attenuation coefficients.
We use a random data split on patient level, leading to \num{30917} training images and \num{4524} validation images.
Both training and validation images are augmented by random rotations.
Fig.~\ref{fig:pretraining_samples_brain_walnut} shows an exemplary reconstruction of the brain dataset, whilst Fig.~\ref{fig:pretraining_convergences_brain_dataset} reports the pretraining convergence.

\begin{figure}[ht!]
\centering
  Brain-Walnut \texttt{Sparse 120}\\[0.1em]
  \includegraphics[
    draft=\draftgraphics,
    width=\linewidth]{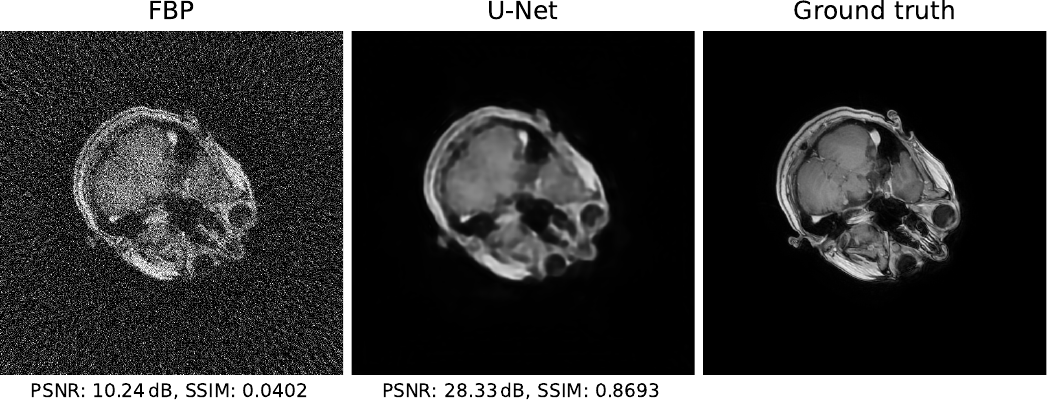}%
  \caption{{Exemplary reconstructions} from the synthetic training dataset of {brain} images for Walnut \texttt{Sparse 120}.}
  \label{fig:pretraining_samples_brain_walnut}
\end{figure}

\begin{figure}[ht!]
    \centering
    \includegraphics[
    draft=\draftgraphics,
    width=\linewidth]{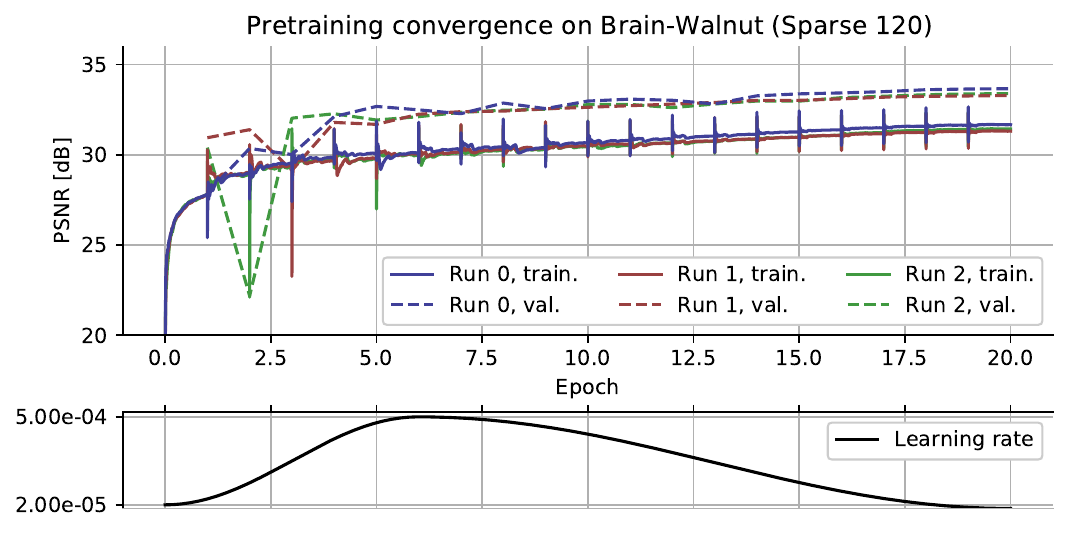}
    \caption{{Pretraining convergence}.}
    \label{fig:pretraining_convergences_brain_dataset}
\end{figure}

In Fig.~\ref{fig:validation_brain_walnut}, we show the validation on the Shepp-Logan.
Fig.~\ref{fig:comp_brain_walnut_120} compares DIP and EDIP trained on the brain dataset.
EDIP performs worse than DIP.

\begin{figure}[ht]
  \centering%
  \includegraphics[
    draft=\draftgraphics,
    width=\linewidth]{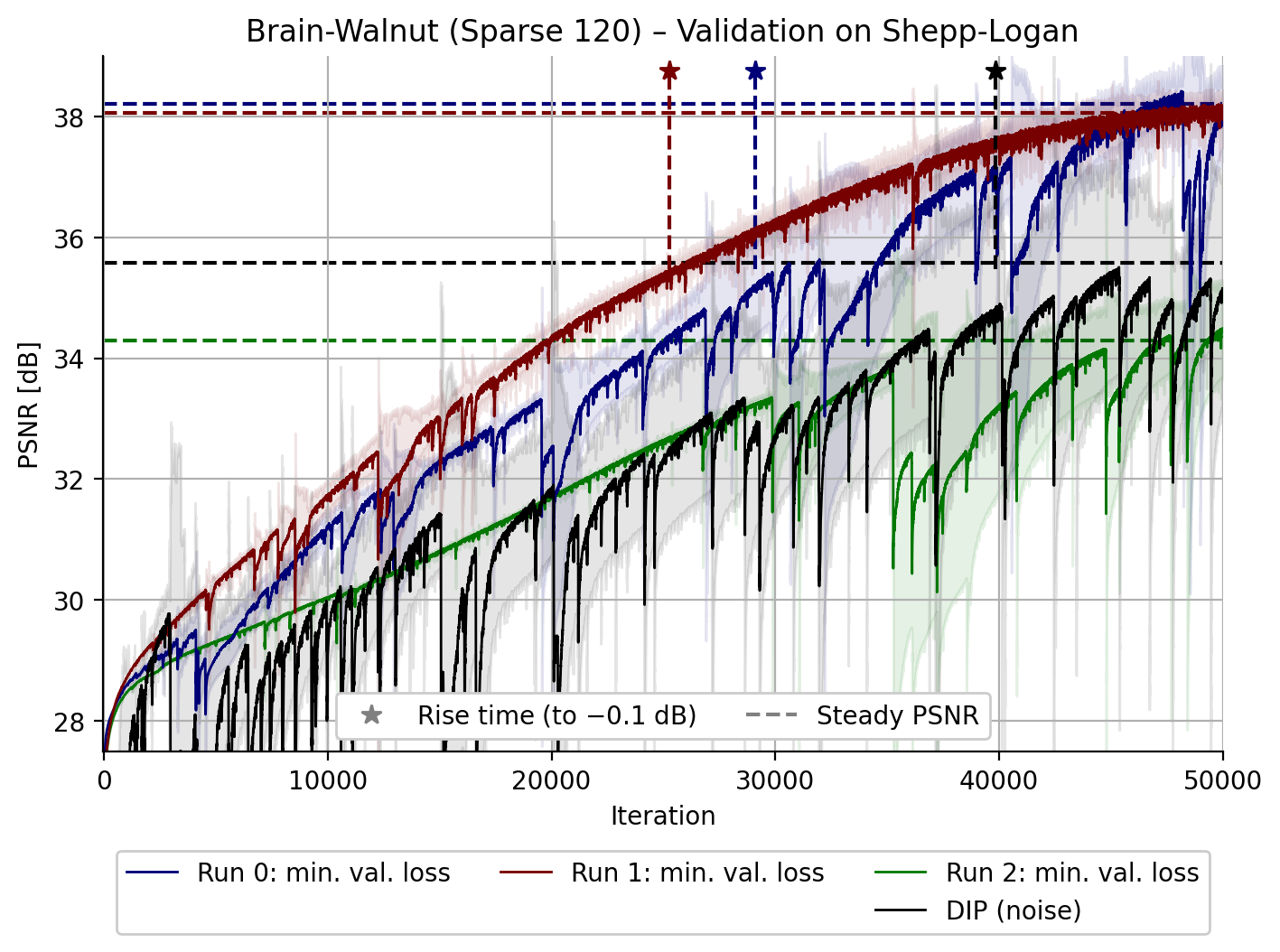}
  \caption{ {Validation runs} on the Shepp-Logan phantom for selecting the initial EDIP (FBP) model parameters pretrained on the {brain} dataset for data in the Walnut \texttt{Sparse 120} geometry. The model from training run \num{1} is selected because it has the shortest rise time. Despite the relatively high number of \num{50}k iterations, the (E)DIP optimizations do not fully converge yet.}
  \label{fig:validation_brain_walnut}
\end{figure}

\begin{figure}[ht]
  \centering%
    \includegraphics[
    draft=\draftgraphics, width=\linewidth]{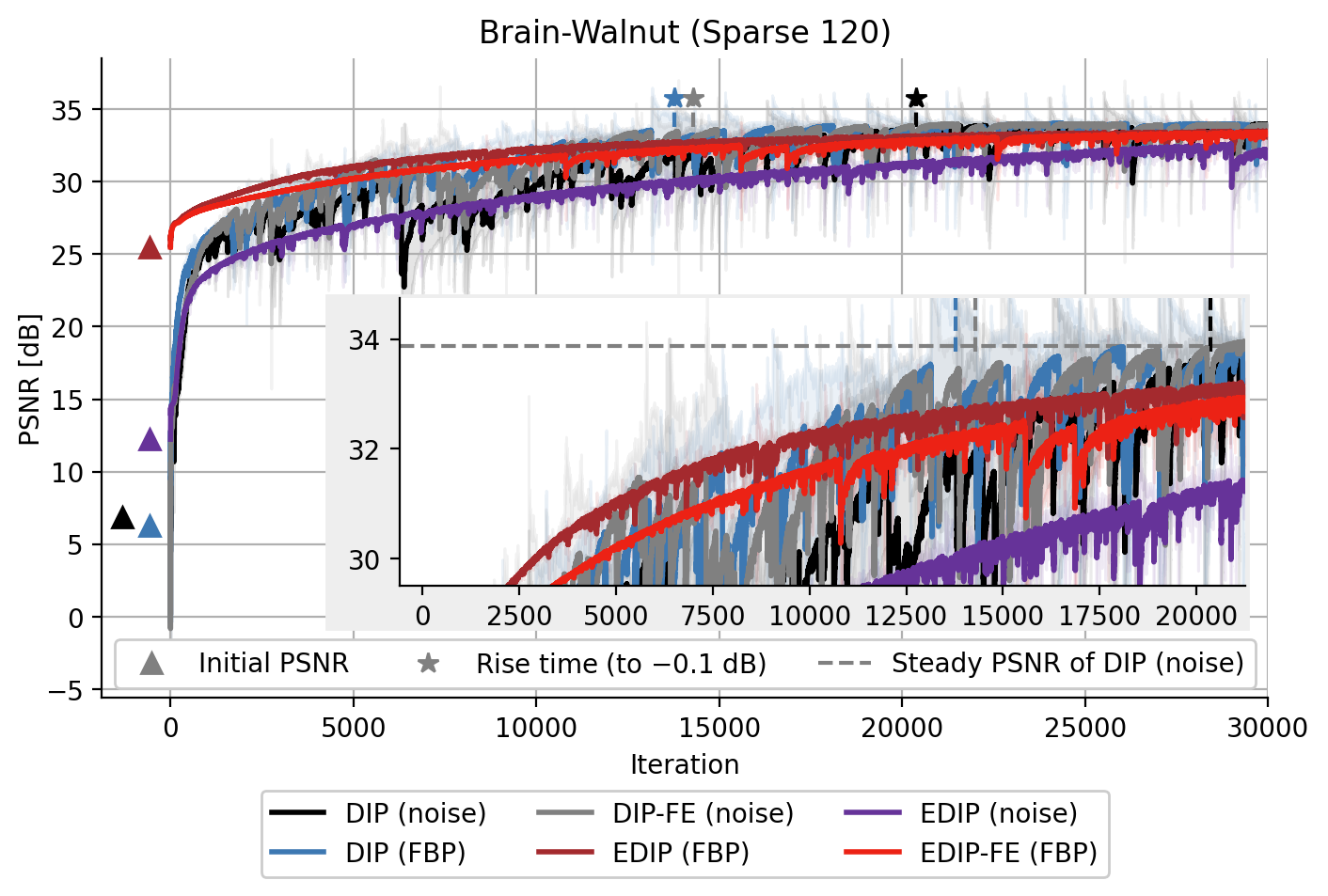}
  \caption{{The optimization of EDIP} versus DIP  pretrained on the {brain} dataset compared to standard DIP on Walnut \texttt{Sparse 120} measurement data. All traces are the mean PSNR of \num{5} repetitions of the same experimental run (varying the random seed). See \cref{tab:brain_numerical_eval_results_walnut} for complementary tabular results. The notations $\blacktriangle$ and $\star$ denote the initial PSNR and rise time, respectively.}
  \label{fig:comp_brain_walnut_120}
\end{figure}

\begin{table}[ht!]
    \caption{{Quantitative evaluation results for EDIP} on Walnut \texttt{Sparse 120} after pretraining on the {brain} dataset for \num{20} epochs. No rise time can be reported, because the PSNR is not reaching the steady PSNR of DIP (noise) minus \SI{0.1}{dB} within the \num{30}k iterations. See Table~\ref{tab:numerical_eval_results_walnut} for the corresponding results from standard DIP and from pretraining on ellipses data.}
  \small%
  \setlength\fboxsep{0pt}%
  \resizebox{\columnwidth}{!}{%
  \begin{tabular}{l@{\extracolsep{4pt}}cccc}
  \strut{} Brain-Walnut \texttt{Sparse 120} --- pretrained for \num{20} epochs \hspace{-20em} & & & & \\
    \cline{1-1}\cline{2-5}\\[-0.7em]
   & \shortstack{\strut{}Rise time} & \shortstack{\strut{} (Max PSNR; iters)} & \shortstack{\strut{}Steady PSNR} & \shortstack{\strut{}Init PSNR} \\\midrule
  EDIP (FBP) & -- & (\num{33.51}; \num{29982}) & \num{33.35} & \num{25.49}\\
  EDIP (noise) & -- & (\num{32.67}; \num{29875}) & \num{32.29} & \num{12.23}\\
  EDIP-FE (FBP) & -- & (\num{33.43}; \num{29862}) & \num{33.24} & \num{25.49}\\
  EDIP-FE (noise) & -- & (\num{31.06}; \num{29989}) & \num{30.39} & \num{12.23}\\
  \bottomrule
  \end{tabular}}
    \label{tab:brain_numerical_eval_results_walnut}
\end{table}

Fig.~\ref{fig:test_checkpoints_brain_walnut_120} suggests that checkpoints from repeated pretraining runs also lead to similar subpar results.
We observe the inadequacy of the brain dataset (of its education!).
Pretraining on the brain dataset induces too dataset specific inductive biases from which EDIP fails to escape, leading to slow convergence and sub-optimal steady PSNR.
Possibly the implicit regularization exerted by the pretraining on the brain dataset essentially restricts the networks from leaving a ``pretrained landscape'' of sub-optimal parameters' configurations.

We then check whether using earlier checkpoints would lead to better transferable performances.
We, indeed, observe that an early-stopping of the pretraining stage on the brain dataset ameliorates EDIP, cf.\ Fig.~\ref{fig:test_checkpoints_epochs_brain_walnut_120}.
The longer we pretrain on the brain dataset, the worse EDIP performs subsequently.

\begin{figure}[ht]
  \centering%
  \includegraphics[
    draft=\draftgraphics,
    width=\linewidth]{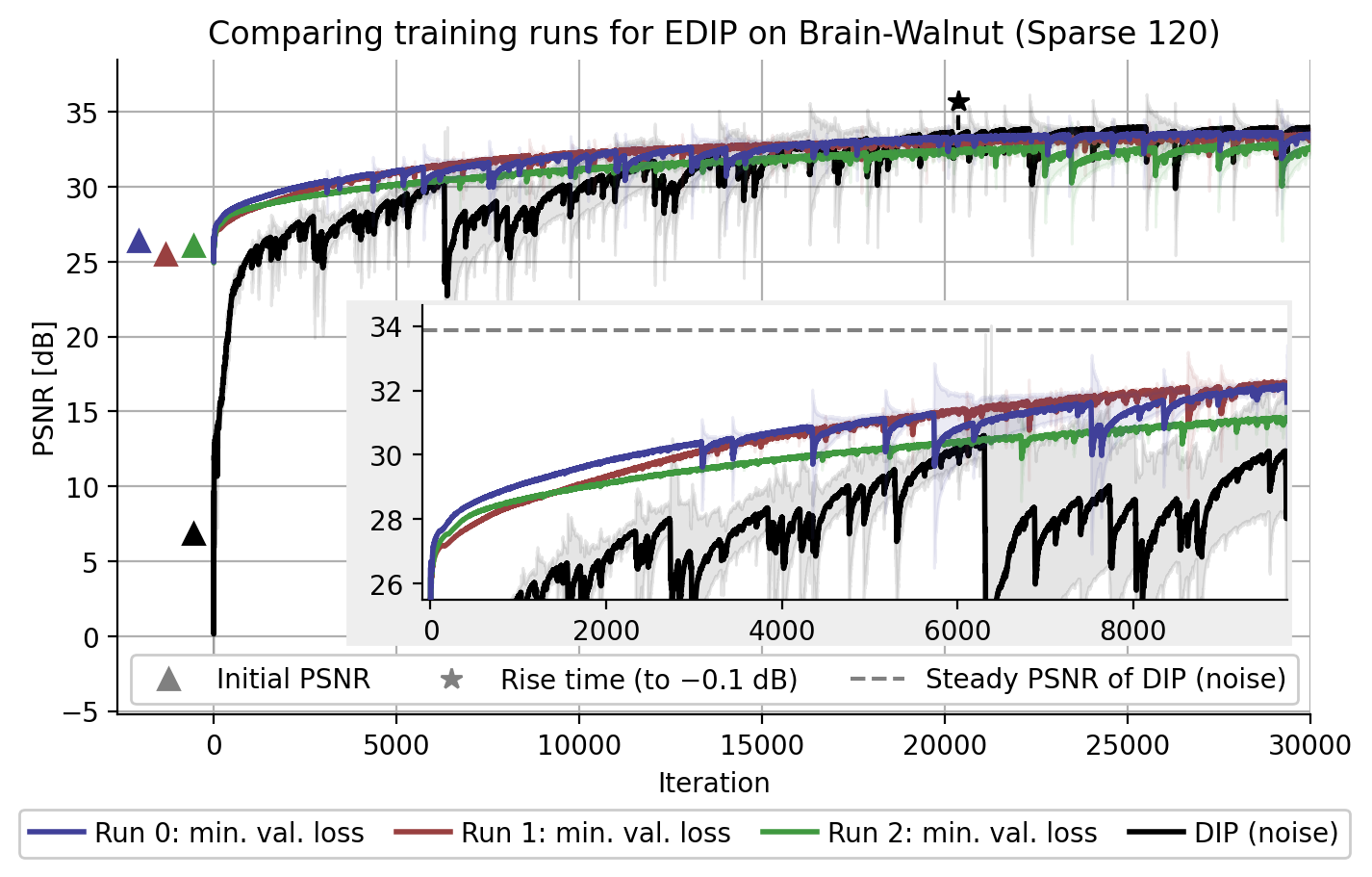}
  \caption{ {The optimization of EDIP} using parameters from different training runs considered during validation (see Fig.~\ref{fig:validation_brain_walnut}) for EDIP (FBP), pretrained on the {brain} dataset, on Walnut \texttt{\texttt{Sparse 120}} measurement data. The parameters from run \num{1} are the ones selected by the validation. The notations $\blacktriangle$ and $\star$ denote the initial PSNR and rise time, respectively, and the horizontal dashed line indicates steady PSNR of DIP (noise).}
  \label{fig:test_checkpoints_brain_walnut_120}
\end{figure}

\begin{figure}[ht]
  \centering%
  \includegraphics[
    draft=\draftgraphics,
    width=\linewidth]{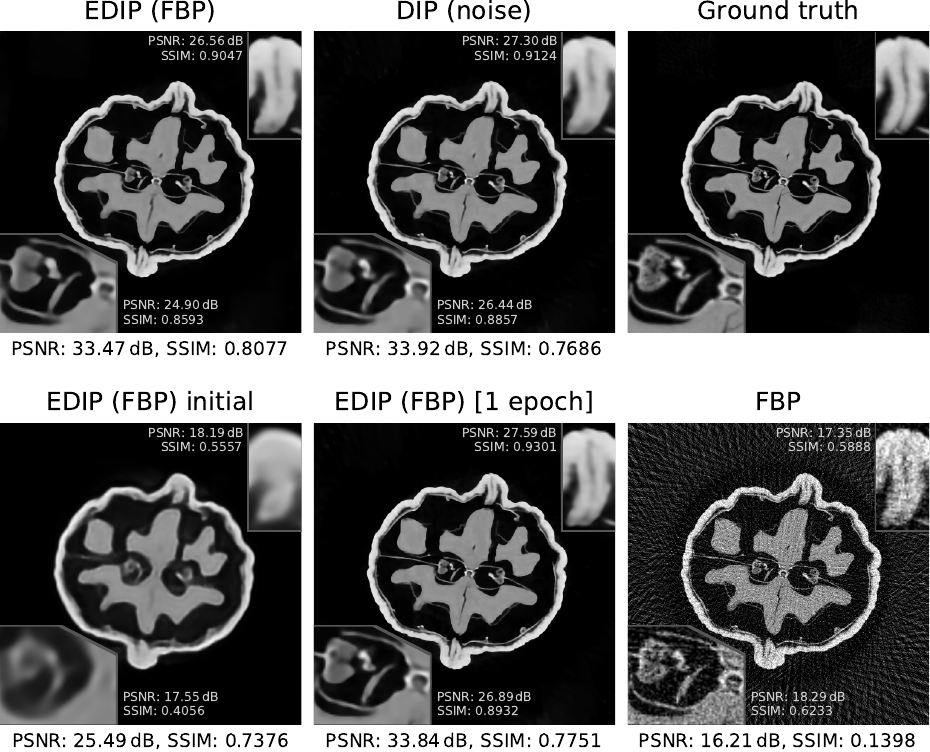}%
  \caption{ {Walnut reconstruction of EDIP} pretrained on {brain} dataset, compared to standard DIP. From the \num{5} runs (varying the seed), the one with the (closest to) median PSNR was selected for each method. See Fig.~\ref{fig:reco_ellipses_walnut_120} for the Walnut reconstruction with EDIP pretrained on the ellipses dataset.}
  \label{fig:reco_brain_walnut_120}
\end{figure}

\begin{figure}[ht]
  \centering%
  \includegraphics[
    draft=\draftgraphics,
    width=\linewidth]{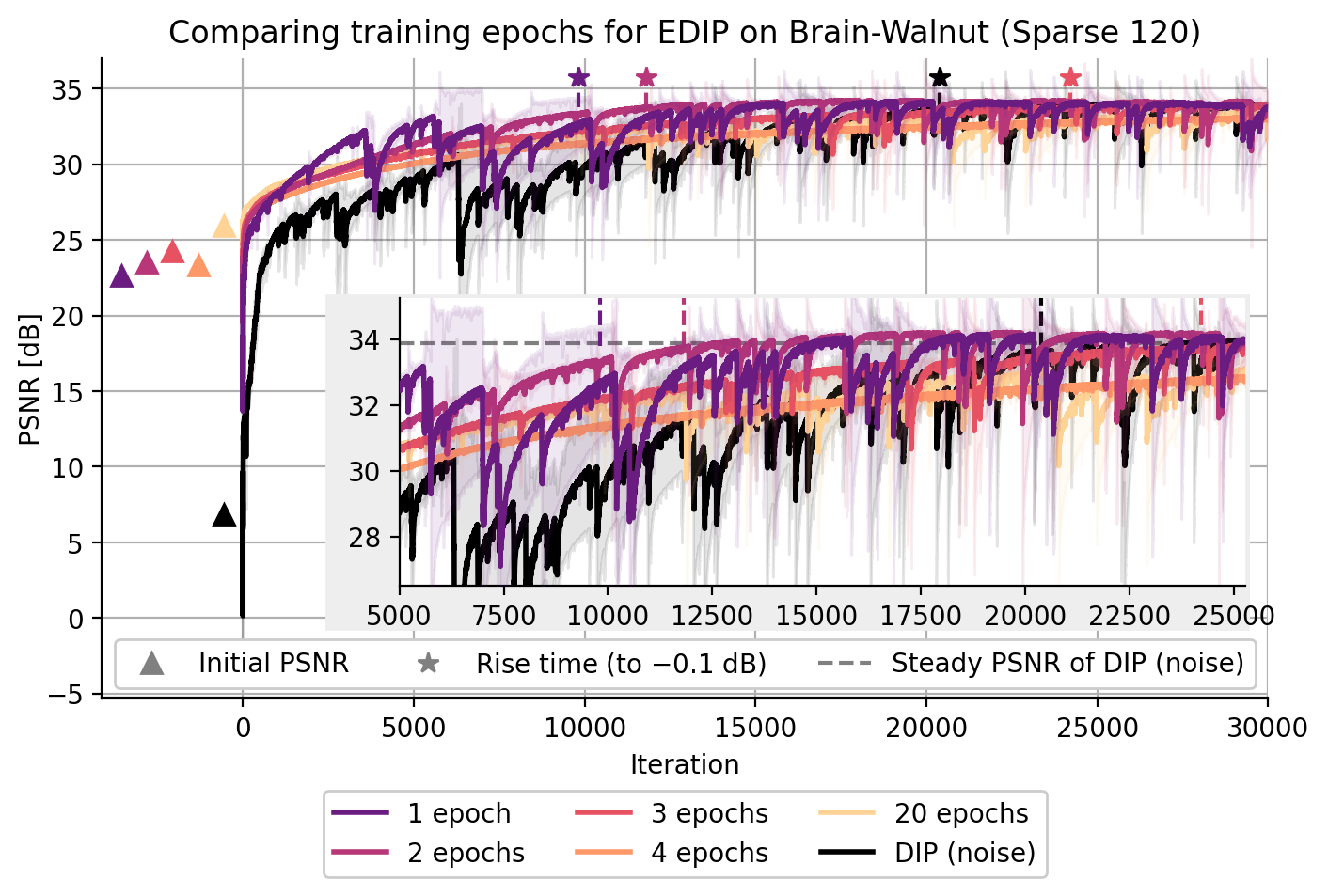}
  \caption{{The optimization of EDIP} using parameters from different epochs for EDIP (FBP) on Walnut \texttt{\texttt{Sparse 120}} measurement data while pretraining on the {brain} dataset. The notations $\blacktriangle$ and $\star$ denote the initial PSNR and rise time, respectively, and the horizontal dashed line indicates steady PSNR of DIP (noise).}
  \label{fig:test_checkpoints_epochs_brain_walnut_120}
\end{figure}

\end{appendices}

\fi

\end{document}